\documentclass[12pt, reqno]{amsart}
\usepackage{amsmath,amssymb,pdflscape,bm,xr}
\usepackage{amsfonts,natbib}
\usepackage{graphicx,epsfig,setspace}
\usepackage[margin=1in]{geometry}

\numberwithin{equation}{section}
\numberwithin{figure}{section}
\numberwithin{table}{section}
\begin{document}

\newcommand{\cov}{\textnormal{Cov}}
\newcommand{\var}{\textnormal{Var}}
\newcommand{\diag}{\textnormal{diag}}
\newcommand{\plim}{\textnormal{plim}_n}
\newcommand{\dum}{1\hspace{-2.5pt}\textnormal{l}}
\newcommand{\ind}{\bot\hspace{-6pt}\bot}
\newcommand{\co}{\textnormal{co}}
\newcommand{\fsgn}{\textnormal{\footnotesize sgn}}
\newcommand{\sgn}{\textnormal{sgn}}
\newcommand{\fatb}{\mathbf{b}}
\newcommand{\fatp}{\mathbf{p}}
\newcommand{\tr}{\textnormal{tr}}
\newcommand{\spn}{\textnormal{span}}
\newcommand{\im}{\textnormal{im}}
\newcommand{\id}{\textnormal{Id}}
\newcommand{\Eta}{\textnormal{H}}

\newtheorem{dfn}{Definition}[section]
\newtheorem{rem}{Remark}[section]
\newtheorem{cor}{Corollary}[section]
\newtheorem{thm}{Theorem}[section]
\newtheorem{lem}{Lemma}[section]
\newtheorem{notn}{Notation}[section]
\newtheorem{con}{Condition}[section]
\newtheorem{prp}{Proposition}[section]
\newtheorem{pty}{Property}[section]
\newtheorem{ass}{Assumption}[section]
\newtheorem{ex}{Example}[section]
\newtheorem*{cst1}{Constraint S}
\newtheorem*{cst2}{Constraint U}
\newtheorem{qn}{Question}[section]

\onehalfspacing

\title[Counterfactual Extrapolation]{Transfer Estimates for Causal Effects across Heterogeneous Sites}
\author[Konrad Menzel]{Konrad Menzel\\New York University}
\date{April 2023, this version: October 2025. Any comments and suggestions will be appreciated.  The author especially thanks Mike Gechter for sharing data used for this paper, and Christoph Rothe and J\"org Stoye for helpful discussions. The author also acknowledges useful feedback from seminar and conference audiences at Cornell, Yale, UCL, Munich, Bonn, Mannheim, Harvard/MIT, Tilburg, Sciences-Po, Amsterdam, BSGE, Tsinghua, EIEF, CIREQ, Y-RISE, and the Africa and East Asia summer meetings of the Econometric Society.}

\begin{abstract} We consider the problem of extrapolating treatment effects across heterogeneous populations (``sites"/``contexts"). We consider an idealized scenario in which the researcher observes cross-sectional data for a large number of units across several ``experimental" sites in which an intervention has already been implemented to a new ``target" site for which a baseline survey of  unit-specific, pre-treatment outcomes and relevant attributes is available. Our approach treats the baseline as functional data, and this choice is motivated by the observation that unobserved site-specific confounders manifest themselves not only in average levels of outcomes, but also how these interact with observed unit-specific attributes. We consider the problem of determining the optimal finite-dimensional feature space in which to solve that prediction problem. We follow a fixed-population approach which evaluates the performance of the predictor given the specific, finite selection of experimental and target sites. Our approach is nonparametric, and our formal results concern the construction of an optimal basis of predictors as well as convergence rates for the estimated conditional average treatment effect relative to the constrained-optimal population predictor for the target site. We quantify the potential gains from adapting experimental estimates to a target location in an application to  conditional cash transfer (CCT) programs using a combined data set from five multi-site randomized controlled trials.\\[4pt]

\end{abstract}

\maketitle

\section{Introduction}

When scaling up an intervention or planning an implementation at a new location, it is often necessary to extrapolate experimental evidence to new sites or contexts. In such settings, average causal effects typically vary across contexts due to environmental factors, only some of which may be observed. We consider a problem in which cross-sectional information on outcomes and covariates is available for both experimental and target sites, and we formalize a process of predicting a causal response that uses disaggregated pre-intervention (baseline) outcome data from the target location to predict such a model shift reflecting site-specific heterogeneity. The underlying premise of such an approach is that the data-generating processes for potential outcomes for pre- and post-intervention outcomes are likely similar, and depend on the same unit- and site specific factors, so that baseline outcomes are in fact predictive for treatment effects. Such an assumption may be particularly plausible when the effect of the intervention is expected to be only incremental rather than fundamentally altering the relationship between unit or site characteristics and the outcome of interest.

It has widely been recognized that pre-intervention outcomes can be useful to predict or control for unobserved heterogeneity at the level of the individual unit (see e.g. \cite{DWa99}). This paper proposes a strategy for doing so at the level of the entire subpopulation to account for shared unobserved heterogeneity at the level of the site rather than the individual. To that end, the relevant baseline information for a given site consists of the full conditional distribution of pre-intervention outcomes given unit covariates, that is we view the baseline as functional data. This choice is motivated by the observation that unobserved site-specific confounders may generally manifest themselves not only in average levels of outcomes, but also how these interact with observed unit-specific attributes. However, in most practically relevant settings, the number of observed sites is not large, forcing the researcher to make pragmatic decisions on how flexibly to model the observable data.\footnote{\cite{All15} considers a setting in which a policy was initially evaluated at 10 sites and eventually scaled up to 111 separate sites. \cite{DPS21} use 142 year/country samples from 61 different countries. The PROGRESA study of conditional cash transfers in Mexico was initially conducted in 506 rural communities across 7 states in Mexico (see \cite{TWo06}). \cite{Mea22} aggregates across seven different RCTs for micro-credit interventions published in 2015.} The corresponding problem of predicting conditional average treatment effects from baseline outcome data can be viewed as functional regression where a realistic implementation can at best achieve a highly regularized solution. Moreover these data constraints also make it all the more important to choose a procedure that makes statistically efficient use of the available data.

Our approach corresponds to a finite-dimensional approximation to that problem, where we determine the optimal feature space in which to solve a linear version of the prediction problem. In our leading application, cross-validation recommends the use of as few as $K=2$ features for prediction, resulting in a highly regularized estimator. Compared to Ridge and other alternative regularization schemes, the resulting transfer estimate can always be interpreted as the best linear predictor given those constructed site-specific features regardless of the degree of regularization. We can furthermore assess whether there exist sites in the experimental population that are similar to a target location in terms of these site characteristics that were determined to be most predictive of conditional average treatment effects. Similar techniques could in principle be developed to predict conditional treatment effects for sites within the experimental sample when treatment assignment was randomized at the site level.

Conditioning on baseline data presumes a statistical framework that defines a joint distribution for pre- and post-intervention outcomes across sites. We choose a fixed-population construction that regards the combined (finite) population of experimental and target sites as fixed, but assumes that the number of cross-sectional units within each cluster is large. Statistical properties of extrapolation estimators are then evaluated under a randomization protocol that assigns experimental versus target status at random among those clusters. In analogy with the literature on conformal prediction, the constructed statistical experiment treats experimental and target locations as finitely exchangeable. We do not necessarily regard this assignment mechanism as factually accurate - e.g. the observed assignment may likely exhibit site-selection effects of the kind documented by \cite{All15}. Rather, this data generating process can alternatively be viewed a device to define a pseudo-true treatment parameter that incorporates the available information on average effects and between cluster heterogeneity. A transfer estimate of this kind would summarize the relevant evidence from the available experimental data and could be subject to additional (qualitative or quantitative) sensitivity analysis with respect to potential violations of the exchangeability assumption.

Rather than imposing strong assumptions necessary for identification of counterfactuals in a target location, our focus is on prediction. Alternatively, we can impose conditions under which that predictor is asymptotically unbiased estimator for a version of the problem in which sites are drawn at random from an infinite superpopulation and consistent for average effects at the target site. Conditions under which the bias from linear interpolation vanishes are discussed in Appendix \ref{sec:linearity_app}.

The empirical application in this paper concerns the effect of conditional cash transfers (CCT) to households on children's school attendance. The effect of CCTs was first evaluated in a large multi-site trial of the PROGRESA/OPORTUNIDADES program in Mexico, which was followed by implementations and additional RCTs in many developing and middle-income countries, often modeled after the PROGRESA study. After applying selection criteria we construct a data set of 640 sites, combining data from five studies in Mexico, Morocco, Indonesia, Kenya, and Ecuador to illustrate our approach. One non-technical contribution of this paper is to exploit cross-site variation within and across studies for extrapolation across populations, where we find that site heterogeneity at baseline predicts cross-study differences in post-intervention responses and conditional average treatment effects.

The problem of adapting empirical findings to new contexts allowing for unobserved heterogeneity is certainly not limited to estimation of discrete treatment contrasts but is also relevant to make more model-based estimates generalizable or comparable across settings. A fully nonparametric approach appears to be well-suited for the particular problem of a binary policy intervention, but can be seen as a stand-in for a more pragmatic estimation approach based on a more explicit model for the outcome of interest. For more ``structural" approaches, it may be preferable to choose low-dimensional models of site heterogeneity that can be directly incorporated into the model, possibly motivated by economic theory or empirical regularities.

\subsection{Literature}

A conceptual framework for the problem of extrapolation of estimated treatment effects across heterogeneous sites was developed in the seminal article by\linebreak \cite{HIM05}. Using their terminology, we assume unconfounded locations, but specifically want to allow for (site-specific) model shifts (``macro effects"), i.e. shared heterogeneity in potential outcomes and treatment effects within each context. We propose a mechanism to incorporate information on pre-treatment outcomes at the cluster/site level when no treated units are observed in the population of interest.

Extrapolation of treatment effects was considered by various studies, including \cite{DPS21}, \cite{Gec23}, \cite{Mea22}, \cite{NIW21}, \cite{ACh22}, and \cite{CSO23}. \cite{DPS21} considered the problem of predicting treatment effects at target sites based on observed site-specific covariates. \cite{Gec23}, \cite{Man20}, and \cite{NIW21} derive bounds that account for selection effects at the individual level, allowing individual heterogeneity to be distributed differently across sites. Our focus is on site-specific heterogeneity, in particular we do not require the support of unobservables $(U_{ig}',V_g')'$ to be shared across sites for the approach to be useful. \cite{ACh22} consider robust extrapolation of treatment rules when there is no separate data on the target site, but the distribution of potential outcomes is in a neighborhood of that for the experimental population.

A separate question concerns the transfer performance of extrapolation methods. \cite{GSDP19} use data from two conditional cash transfer programs to evaluate extrapolation of empirical treatment rules.  \cite{KXCAL18} identify attributes that exhibit a stable predictive relationship to the outcome of interest across environments.  \cite{AFLW22} analyze the problem of assessing transfer performance, where model estimates from data in one domain are transferred to another, whereas this paper optimizes cross-domain model performance within the experimental sample. While our analysis is formally design-based conditional on the experimental sample (rather than assuming i.i.d. draws of contexts from a meta-population), a sampling-based interpretation similar to theirs is also possible. \cite{GHLMMMRS23} discuss optimal selection of experimental locations for extrapolation to other sites.


The work closest to ours is \cite{CSO23} who propose to use the distribution of pre-intervention outcomes for the target site together with post-intervention outcomes from the experimental locations to predict outcomes under a synthetic transferability condition. Their approach predicts policy effects based on the assumption that the policy shift affects outcomes through an index where the supports of pre- and post-intervention index values overlap in the target population. We consider a setting in which the policy intervention is binary and not equivalent to a shift in other observed covariates. Under that scenario the supports for pre- and post-intervention values for that index are disjoint, so that no subpopulation of the target site can be directly matched to post-intervention outcomes in the experimental sample. Our approach predicts counterfactuals conditional on pre-treatment outcomes alone, and is therefore complementary to theirs.

The working assumption of exchangeability between experimental and target sites is shared by conformal prediction methods (see \cite{VGS05} and \cite{LGRTW18}). The focus of the present paper is on a point estimate that is informed by the experimental sample rather than inference, however under an exchangeability assumption our approach could in principle be combined with classical or conformal methods for inference with either asymptotic or finite-sample guarantees. Sensitivity of conformal inference with respect to departures from exchangeability was characterized by \cite{BCRT23}. We do not explore the problems of inference or sensitivity analysis in this paper but leave this for future research.

It is also worth comparing our approach to other conceptual frameworks for aggregation of causal estimates across different populations: \cite{PBa14} gave explicit conditions for transportability of causal estimates across populations in terms of selection diagrams. One interpretation of our approach is the construction of a site-level covariate from baseline outcome data conditional on which potential outcomes are, to an approximation, mean-independent of selection. This paper also differs in the interpretation of transfer estimates, where our focus is on cross-population prediction of causal effects, rather than assuming the idealized conditions that would guarantee transportability in the strict sense.

Conceptually, the extrapolation problem also  has some resemblance with the method of synthetic controls (\cite{AGa03},\cite{ADH10}). However our approach is developed with a setting in mind where we do not have (typically aggregate) time series information on a treated unit and the ``donor pool" of potential controls. Rather we assume that each site/context provides rich cross-sectional information, where a fraction of units is treated in a study population of sites, and we then predict treatment effects for the (yet untreated) target cluster. For that problem, \cite{Gun23} is most similar to our approach in that he proposes to use cross-sectional variation in micro-data to calibrate synthetic weights, however our approach differs in that rather than optimizing weights to match the distribution of baseline outcomes as closely as possible, we construct factors that are optimized to predict post-treatment outcomes based on the conditional distribution given unit-specific attributes. \cite{Shi22} uses a k-means algorithm to model unobserved heterogeneity in a problem with cluster dependence in treatment assignment.

In order to model site-specific conditional mean functions as random objects, we use tools from functional data analysis (see \cite{RSi05} and also \cite{WCM16} for a more recent overview), where function-to-function regression was analyzed by \cite{HMW03} and \cite{HMWY10},\cite{YMS11}, and \cite{BCF17}. Our approach is also related to functional principal components approaches for function completion/reconstruction based on partially observed functional data, where our setting corresponds more closely to that of sparsely sampled functions analyzed in \cite{YMW05}, rather than the dense case considered by \cite{Kra15} and \cite{KLi20}, although we assume that the number of points sampled for each curve (site) grows large. Since our focus is on cases in which only a modest number of trajectories is observed, the basis functions for our approach is constructed in as to be optimal for prediction, using both covariate and outcome data rather than separate principal components for covariate and outcome trajectories.

Generally our problem differs from function reconstruction in that our objective is to predict the \emph{difference} between two curves, corresponding to conditional mean functions for either potential value, rather than the trajectory of the partially observed curve, so that the functional principal components of the conditional mean functions themselves do not generally have the best basis property for this particular task. Our problem differs from that of covariate adaptive reconstruction (\cite{JWa10},\cite{Lie19}) in that we consider unit-specific covariates which correspond to coordinates of the random trajectories, rather than site-specific covariates that shift the distribution of the random curve. Prediction of scalar outcomes based on functional principal components was analyzed by \cite{CHa06} and \cite{HHo07}.

Our focus is on prediction of the conditional average treatment effect as a function of covariates, and we derive a choice of basis functions that is optimal for that prediction task in a sense to be made more specific below. We show that our solution bears some resemblance with, but is distinct from \cite{Hot36}'s classical problem of canonical correlation analysis. For functional data, functional canonical regression has first been proposed by \cite{HMWY10} whose approach differs from the present paper in terms of the approach to regularization. We derive our approach from optimality considerations and establish a (to our knowledge novel) formal optimality result.

Interpreting ``locations" at which random trajectories are evaluated as covariates or causal variables also requires a few subtle adjustments relative to the classical literature on functional data analysis. In particular, the covariate distributions may differ across sites, so nonparametric estimation of moments of the distribution of the random function requires some local reweighting and support conditions.

\subsection{Organization} The remainder of the paper is organized as follows: we first give a formal characterization of transfer estimation as a statistical problem. We then determine the optimal finite-dimensional subspace of features of the baseline data, and propose nonparametric estimators based on the experimental sample. Asymptotic properties of those estimators, assuming the number of experimental sites grows large, are given in Appendix \ref{sec:asymptotics}. The approach is then illustrated using an application to predicting the causal effect of conditional cash transfer programs to new locations.

\section{Problem Description}

The population of interest consists of $G$ sites (``clusters"/``contexts"), where the $g$th site consists of $N_g$ units. Our focus is on the case in which there is a single target site $g^*$ in addition to $G-1$ experimental sites $g\in\{1,\dots,G\}\setminus\{g^*\}$. We also use the dummy variable $R_g\in\{0,1\}$ to indicate whether $g$ is an experimental location ($R_g=1$), or a target site ($R_g=0$).

There is a binary policy variable (``treatment") $D_{gi}\in\{0,1\}$ which acts at the level of the unit $i$ at site $g$, where we assume that the outcome of interest is determined only by the unit's own treatment status (SUTVA). Specifically, the unit is associated with potential outcomes $Y_{gi}(0),Y_{gi}(1)$, where the realized outcome is given by $Y_{gi}:=Y_{gi}(D_{gi})$. Furthermore, each unit is associated with a finite-dimensional vector $X_{gi}$ of attributes whose distribution is given by the p.d.f. $f_g(x)$ for cluster $g$, where we assume that the support $\mathcal{X}$ of $X_{gi}$ is a compact subset of $\mathbb{R}^d$. For the purposes of this paper $N_g$ will be treated as infinite, but the researcher only observes a finite random sample of $n_g$ units for each cluster.

Adapting notation from \cite{NIW21}, we can represent potential outcomes as
\begin{equation}Y_{gi}(d)\equiv y(d;X_{gi},U_{gi},V_g),\hspace{0.5cm}d=0,1\end{equation}
for some unspecified mapping $y(\cdot)$ and potentially multi-dimensional unobserved individual and site-specific heterogeneity $U_{gi}$ and $V_g$. We first define key objects in terms of a superpopulation model in which $V_g$ and $U_{gi}$ are random draws from an unspecified distribution. Our statistical approach will be conditional on a fixed population of $G$ sites with heterogeneity $V_1,\dots,V_G$ without additional restrictions on how those sites were selected.\footnote{Previous work by \cite{Gec23} and \cite{NIW21} proposed strategies to address cross-site differences in the conditional distribution of individual heterogeneity $U_{gi}$, whereas our focus is on site-specific heterogeneity $V_g$. While $V_g$ could be included with the vector $U_{gi}$ as a matter of notation, the approaches in \cite{Gec23} and \cite{NIW21} require $U_{gi}$ to have the same support across sites, which can't be satisfied by variables $V_g$ that are shared by all units at the site. We therefore prefer to keep site-specific heterogeneity explicit in our notation.} 

Using this notation we can write the conditional expectation of post-intervention outcomes at site $g$ for $D_{gi}=d$ as
\[\mu_g(x;d)\equiv\mu(x;1,V_g):=\mathbb{E}[Y_{gi}(d)|X_{gi}=x,V_g]\]
The site-specific conditional average treatment effect is given by
\[\tau_g(x)\equiv\tau(x;V_g):=\mathbb{E}[Y_{gi}(1)-Y_{gi}(0)|X_{gi}=x,V_g]\]
In particular, $\mu_g(x;d)$ and $\tau_g(x)$ are functions of site-specific unobserved heterogeneity $V_g$ and therefore random objects whenever $V_g$ is regarded as stochastic. For a given superpopulation $V_g\sim F_V$, we can also define the  cross-site averages $\mu(x;1):=\mathbb{E}_{F_V}\left[\mu(x;1,V_g)\right]$
and $\tau(x):=\mathbb{E}_{F_V}\left[\tau(x;V_g)\right]$ of the CATE.

Our goal is \emph{prediction} of $\tau_g(x)$ rather than consistent estimation, although under a more restrictive superpopulation framework and a linearity assumption, the prediction problem can also be cast  as estimation of $\tau_g(x)$, see Appendix \ref{sec:linearity_app} for a dicussion. We aim to predict model shifts
\begin{eqnarray}\label{tau_shift}\Delta\tau_g(x)&:=&\tau_g(x) - \tau(x)\\ \label{mu1_shift}\Delta\mu_g(x;1)&:=&\mu_g(x;1)-\mu(x;1).\end{eqnarray}
using the site-specific distribution of pre-intervention outcomes, $Y_{gi}(0)|X_{gi},V_g$.


Prediction of site-specific CATE therefore seeks to account for \emph{model shifts} $\Delta\tau_{g}(x)$. Our method aggregates information on the first two moments of the distribution of conditional expectation functions (pre- and post-intervention) across sites and does not require that we can estimate either conditional mean function consistently for any individual site. In particular, we also discuss a version of our aproach for the case in which treatment assignment was randomized at the site level. In principle, the arguments behind our method can therefore also be extended to imputation of site-specific CATE for \emph{experimental} sites when treatment was randomized at the site level, or the researcher only observes a moderate number of units for each site.

\subsection{Functional Predictors}\label{subsec:functional_pred} Our approach aims to extract predictive information regarding the unobserved site-specific heterogeneity $V_g$ from baseline (pre-intervention) outcome data. Since $V_g$ is shared among all units at the same site, not only the baseline outcome $Y_{gi}(0)$ of a target unit is predictive of the post-intervention out come $Y_{gi}(1)$ of that same unit, but the conditional distribution of pre-intervention outcomes given covariates for that site, $Y_{gi}(0)|X_{gi},V_g$, contains additional information regarding the model shifts (\ref{tau_shift}) and (\ref{mu1_shift}). This is particularly plausible when the effect of the intervention is expected to be only incremental so that pre- and post-intervention outcomes behave similarly and depend on the same unit- and site-specific factors. Under this view of the DGP, unobserved site-specific heterogeneity $V_g$ in expected outcomes is not necessarily separable, but site effects will often manifest themselves in interactions between attributes and outcome variables.

\begin{rem}
For example in the study of conditional cash transfers, school attendance may differ across sites according to whether there is a secondary school in close proximity to the community. If the closest secondary school is difficult to reach, then older age groups will likely have substantially lower attendance at baseline, but also be less responsive to a cash incentive. Also, local price levels may differ across communities, causing shifts between sites in the relationship between attendance and nominal measures of income. Sites may also differ e.g. in terms of how well the site is connected to urban centers, or whether the language of instruction is widely spoken within the community. For the CCT data one might expect a substantial amount of heterogeneity among sites within each study country along these lines, but also for these site-specific factors to play out in similar ways across these countries.
\end{rem}

In practice, the researcher may in addition directly observe site-specific measures e.g. of price variables or the cost of attending school, and our approach could then be viewed as addressing residual site-specific heterogeneity after practically feasible adjustments for observable covariates. We discuss this further in Section \ref{subsec:site_spec_cov_subsec} below.

Our approach models the conditional distribution of $Y_{gi}(0)$ given $X_{gi},V_g$ as functional data, which is then used to extract site-specific factors $m_{g1},\dots,m_{gK}$ (say) to predict a model shift $\Delta\mu_g(x;1)$ or $\Delta\tau_g(x;1)$. While other moments of the conditional distribution of baseline outcomes may reveal additional information regarding $V_g$, in this paper we restrict our attention to the problem of using only the conditional first moment of baseline outcomes
\[\mu_g(x;0)\equiv\mu(x;0,V_g):=\mathbb{E}[Y_{gi}(0)|X_{gi}=x,V_g]\]
as a predictor of $\Delta\tau_g(x)$.\footnote{In our application, the outcome of interest $Y_{gi}$ is a binary indicator whether a school-age child attends school, so that all higher moments of potential outcomes are known functions of $\mu_g(x;0)$, but in general higher-order conditional moments of $Y_{gi}(0)$ given $X_{gi}$ may also be predictive of the CATE at the target site. \cite{IKi21} propose efficient aggregation of noisy site-specific estimates of unconditional ATEs. In our leading scenario, cluster size $n_g$ is large relative to $G$ so that error in estimating $\mu_g(x;0)$ is asymptotically negligible. In the sparsely sampled case in which the number of units per site is not large, estimation error in $\mu_g(x;0)$ gives rise to similar efficiency considerations which we do not address in this paper. We also do not consider the use potentially predictive information form the marginal distribution of covariates $f_{X_g|V_g}(x|V_g)$. It is also possible to incorporate observable site-specific covariates into our approach, as discussed in Section \ref{subsec:site_spec_cov_subsec} below.}


For a target site $g^*$ drawn from a superpopulation, $V_{g^*}\sim F_V$, the best (lowest variance) predictor of $\tau_{g^*}(x)$ given $\mu_{g^*}(\cdot;0)$ is the conditional expectation function,
\begin{equation}\label{cond_exp_pop}\mathbb{E}[\tau(x;V_{g^*})|\mu(:,0,V_{g^*})] = \mathbb{E}\left[\mu(x;1,V_{g^*})-\mu(x;0,V_{g^*})|\mu(\cdot;0,V_{g^*})\right]\end{equation}
Since $\mu(\cdot;0,V_{g^*})$ is generally infinite-dimensional (unless all attributes $X_{gi}$ are discrete), completely flexible interpolation between sites is generally not feasible as a practical matter. Instead, we restrict our attention to predictors that are \emph{linear} in $\mu_{g^*}(x;0)$,
\begin{equation}\label{linear_beta}\Pi\left[\tau(x,V_{g^*})|\mu(\cdot;0,V_{g^*})\right]:= \tau(x) + \int\mu(x_1;0,V_{g^*})\beta(x_1,x)f_0(x_1)dx_1\end{equation}
for a square integrable function $\beta(x_1,x_2)$.  That is, we can view a linear predictor as a regression adjustment over the unconditional CATE $\tau(x)$. Finding the kernel $\beta(x_1,x_2)$ corresponding to the best linear predictor is the classical functional linear regression problem (see \cite{RSi05}, \cite{HMWY10}, and \cite{BCF17}).


\begin{rem} In general, there is no guarantee that a linear projection (\ref{linear_beta}) can extract all relevant information on site heterogeneity from $\mu_g(\cdot;0)$ but will in general result in an interpolation error $e_g(x):=\mathbb{E}\left[\tau(x;V_g)|\mu(\cdot;0,V_g)\right]-\Pi\left[\tau(x;V_{g})|\mu(\cdot;0,V_g)\right]$. While the assumption of linearity is fairly well understood in the finite-dimensional case, we provide stylized examples for the functional prediction case  that are motivated by our leading application in Appendix \ref{sec:linearity_app} in order to illustrate the possibilities and limitations of this approach.
\end{rem}

Estimation of (\ref{linear_beta}) from a modest number of experimental sites requires substantial \emph{regularization.} The dimension of the function generally equals the cardinality of $\mathcal{X}$, and the researcher may choose to work with approximations in an $S$-dimensional sieve space, e.g. using functional principal components or using B-splines, as in our implementation below. We propose to substantially reduce the dimensionality of this problem by constructing a subspace of $K<<S$ predictive features from $\mu_g(x;0)$ in a way that is optimal for prediction in a sense to be made more precise below.

Our approach differs from a conventional application of functional regression techniques in that rather than aiming for consistent estimation, we regard regularization via the choice of $K$ as fixed and instead aim at constructing those predictive features optimally. In our application, cross-validation recommends an approximation using a subspace of dimension as low as $K=2$, a level at which other regularization approaches may be difficult to interpret. Our approach still yields a best linear predictor given those constructed predictive features, whose construction and distributions can be reported and analyzed explicitly in any empirical application.

\subsection{Fixed-Population Approach}


We are interested in solving the functional prediction problem (\ref{linear_beta}) for situations in which the researcher wishes to extrapolate from existing experimental data and therefore has limited control or knowledge on how those sites had been selected. In such a scenario, it is generally implausible to assume a well defined sampling mechanism from a particular superpopulation, however defined. Instead, we follow a fixed-population (design-based) approach to the problem of extrapolating from experimental to target sites, where our statistical theory will regard the combined population of experimental and target sites as fixed, but only the role of the target site $g^*$ is regarded as random.


To be specific, we analyze the statistical properties of a predictor under the distribution defined by the following hypothetical protocol: in a first stage, we select sites to the experimental arm by drawing $G_1$ sites at random from $\{1,\dots,G\}$ uniformly and without replacement. The remaining locations are assigned the role of a target site, and we take $R_g\in\{0,1\}$ to be an indicator variable that equals one if $g$ is an experimental site, and zero otherwise. In a second step, individualized treatments $D_{gi}\in\{0,1\}$ are assigned at random to units, and the intervention is implemented according to that assignment at each experimental site $g$. Our main results concern the case of unit-level randomization at each experimental site, but we also discuss the case of site-level randomization separately. Finally in a third step we sample units uniformly at random at all sites and use the resulting sample to construct extrapolation estimates for the CATE at each target site.

\begin{rem}
This framework treats the combined experimental and target sites as finitely exchangeable (see Assumption \ref{unconf_loc_ass} below), in close analogy with the literature on conformal prediction (see e.g. \cite{VGS05} and \cite{LGRTW18}). This exchangeability condition can be plausible if the researcher ``curates" a sample of experimental sites from available data that is generally comparable \emph{ex ante} to the target site, and potentially discard sites or studies that are known to differ systematically from that site. The resulting prediction may also be interpreted \emph{ex post} in light of possible departures from exchangeability.\footnote{In the literature on conformal prediction such a sensitivity analysis was formally proposed by \cite{BCRT23}.} 
\end{rem}

Under this fixed-population experiment, the cluster-specific conditional average treatment effects $\tau_1(x),\dots,\tau_G(x)$ are nonstochastic, however the assignment $R_g$ of sites to the experimental role as well as the selection $D_{gi}$ of treated units within each experimental cluster are random. In particular, the cross-site average and empirical covariance of the functions $\mu_g(x;d)$ can only be estimated with error since even for units included in the sample, only one of the two potential outcomes $Y_{gi}(0),Y_{gi}(1)$ is observed. For the remainder of the paper we focus on the case in which there is a single target cluster in addition to $G-1$ experimental clusters.

We consider a \textbf{transfer estimate} $\hat{\tau}_{g^*,1,\dots,G}(x)$ for extrapolating from the sites $\{1,\dots,G\}\backslash\{g^*\}$ to $g^*$. Such a transfer estimate combines information on covariates and outcomes from the $G$ experimental and target sites to predict the CATE for the target site, $g^*$. We evaluate the statistical performance of such a transfer estimate in terms of the integrated mean-squared error (IMSE) under the resulting statistical experiment,
\begin{equation}\label{setup_mse}IMSE:=\frac1G\sum_{g=1}^G\mathbb{E}\left[(1-R_g)\int(\hat{\tau}_{g,1\dots,G}(x)-\tau_g(x))^2f_0(x)dx\right]
\end{equation}
with a weight function $f_0(x)$ that has the properties of a p.d.f. and is chosen by the researcher. That function could be e.g. the uniform distribution on a compact set, or an estimate of the covariate distribution across the $G$ sites.

The fixed population approach is therefore used as a way of formalizing the researcher's problem who aims to produce a forecast that performs as well as possible \emph{on average} for prediction \emph{among this fixed population of sites}. The best feasible prediction under those circumstances is a parameter that is specific to the set of observable experimental and target sites. The resulting transfer estimator represents a summary of site-specific unobserved model heterogeneity that can be quantified based on the experimental sample and used to predict the treatment effect at the target site. This is analogous to a situation that would arise when the sample average treatment effect (SATE) is used to predict the treatment effect for an individual participant in an experimental trial on subjects that were not sampled at random from a well-defined population.

We derive theoretical properties of the approach using fixed-population asymptotics (see \cite{AAIW17}), where approximations are derived under a sequence of finite populations along which the number of sites $G$ grows large. We analyze scenarios at which sites are either sampled \emph{densely}, where $n_g\rightarrow\infty$ for each site $g$, or \emph{sparsely}, where $n_g$ remains bounded. While for any given application, $G$ is obviously fixed, embedding the fixed-population prediction problem into such a sequence of statistical experiments allows to establish stochastic orders of magnitude for estimation errors as long as $G$ is sufficiently large for those approximations to be close.

\begin{rem} The arguments in the remainder of the paper could also be directly translated to a sampling based framework by replacing sample with superpopulation moments. However we do not evaluate prediction errors with respect to a (``natural" or constructed) superpopulation, but define transfer estimate as fixed-population, design-based analogs instead. This design-based interpretation of transfer estimation therefore keeps any potential caveats about non-representative sampling of sites explicit. There are also empirical scenarios in which the experimental clusters were in fact chosen at random from the relevant superpopulation, allowing for an alternative, sampling based interpretation. For example, for several of the studies of conditional cash transfers the study population consisted of randomly selected villages or schools in a subset of major administrative regions (states, provinces) of the country in which the study was conducted. In such a setting, a sampling-based approach is well-suited to estimating the anticipated effect of scaling the policy to the remaining sites in those regions. 
\end{rem}

\section{Optimal Predictors for Model Shift}

\label{sec:opt_basis_sec}

This section concerns the optimal choice of basis functions (features) for estimation of the linear projection problem (\ref{linear_beta}). Our approach is based on a representation of the random processes $\mu_{g^*}(x;0)$ and $\tau_{g^*}(x)$ for the target site $g^*$ in terms of orthogonal bases. To be specific, for a given pair of orthogonal bases $\phi_1,\phi_2,\dots$ and $\psi_1,\psi_2,\dots$ of square integrable functions, respectively,  we can write
\begin{eqnarray}
\nonumber \mu_{g}(x;0)&=&\mu(x;0) + \sum_{k=1}^{\infty}m_{gk}\phi_k(x)\\
\label{tau_CATE_exp} \tau_{g}(x)&=&\tau(x) + \sum_{k=1}^{\infty}t_{gk}\psi_k(x)
\end{eqnarray}
for each $g=1,\dots,G$. We use a fixed-population framework in which the target site $g^*$ is a random draw from the deterministic population $\{1,\dots,G\}$, so that $\mu_{g^*}(x;0)$, $\tau_{g^*}(x)$, and the corresponding coefficients $\left\{m_{g^*k},t_{g^*k}\right\}_{k=1}^{\infty}$ are stochastic.

Our approach then estimates a truncated version of this expansion for $\tau_{g^*}(x)$,
\begin{equation}\label{pseudo_true_tau}\tau_{g^*}^K(x):=\tau(x) + \sum_{k=1}^{K}t_{g^*k}\psi_k(x)\end{equation}
to approximate the CATE at site $g$ at a low order $K<<G$. For the scenarios we are envisioning in this paper, the number of experimental clusters is not very large, so $K$ should be thought of as fairly small. In fact, for our empirical application, cross-validation (with respect to cross-site prediction) suggests a value of $K$ equal to 2 or 3, depending on the exact specification. So rather than aiming for consistent estimation of $\tau_g(x)$, we view the use of the first few leading factors in the expansion (\ref{tau_CATE_exp}) as a method of improving over the unconditional forecast $\tau(x)$ in order to account for site-specific heterogeneity.

It is therefore all the more important to have theoretical guidance on how to choose the basis of that expansion optimally so as to prioritize those features in the data that will be most predictive for $\tau_{g^*}(x)$. The need to truncate the expansion for purposes of estimation stems from ill-posedness in the problem of predicting $\tau_{g^*}(x)$ based on trajectories $\mu_{g^*}(x;0)$. While other continuous regularization methods are available (see \cite{CFR07}), an advantage of this finite-dimensional approximation is that it can be interpreted as a linear prediction of the CATE based on the first $K$ factors in an analogous expansion of the function $\mu_{g^*}(x;0)$ for arbitrary fixed values of $K$.

Our approach requires nonparametric estimation of the mean functions
\[\mu(x;d):=\frac1G\sum_{g=1}^G\mu_g(x;d),\hspace{0.5cm}d=0,1\]
and
\[\tau(x;d):=\frac1G\sum_{g=1}^G\tau_g(x;d),\hspace{0.5cm}d=0,1\]
as well as the covariance kernels
\begin{eqnarray}
\nonumber H_{\mu\mu}(x_1,x_2)&:=&\frac1G\sum_{g=1}^G(\mu_g(x_1;0)-\mu(x_1;0))(\mu_g(x_2;0) - \mu(x_2;0))\\
\nonumber H_{\mu\tau}(x_1,x_2)&:=&\frac1G\sum_{g=1}^G(\mu_g(x_1;0)-\mu(x_1;0))(\tau_g(x_2) - \tau(x_2))
\end{eqnarray}
\begin{eqnarray}
\nonumber H_{\tau\tau}(x_1,x_2)&:=&\frac1G\sum_{g=1}^G(\tau_g(x_1;0)-\tau(x_1;0))(\tau_g(x_2) - \tau(x_2))
\end{eqnarray}
These objects can be interpreted as expectations and covariances, respectively, with respect to a random draw of a site $g^*$ from the discrete uniform distribution over $\{1,\dots,G\}$.

A standard representation of the random processes $\mu_g(x;0)$ and $\tau_g(x)$ in (\ref{tau_CATE_exp}) is the Karhunen-Lo\`eve expansion, which chooses the  basis functions $\phi_1,\phi_2,\dots$ and $\psi_1,\psi_2,\dots$ as eigenfunctions of the respective covariance operators $H_{\mu\mu}(\cdot),H_{\tau,\tau}(\cdot)$, see \cite{RSi05} and \cite{RWi06}. These bases of eigenfunctions ordered by their associated eigenvalues are also known as the functional principal components (FPC) of the random functions $\mu_g(x;0)$ and $\tau_g(x)$.  At any finite order, an reconstruction of the function by its leading $K$ FPC is known to be optimal with respect to the mean-square error of approximation. However, our goal is to extract those features of $\mu_g(x;0)$ that are ``most predictive" for the average of $\tau_g(X_{gi})$, which generally do not coincide with the FPC. We show that instead, that optimal choice can be described in terms of a singular value decomposition of an operator characterizing the covariance between $\mu_g(x;0)$ and $\tau_g(x)$.

\subsection{Optimal Basis Functions}

Our main objective is to determine the optimal finite-dimensional feature space for the baseline data in which to solve the prediction problem (\ref{linear_beta}). We regard the conditional mean functions $\mu_{g}(x;d)$ and $\tau_{g}(x)$ as random elements of the Hilbert space $L_2(\mathcal{X},f_0)$ ($L_2(\mathcal{X})$ henceforth) of square integrable functions with norm induced by the scalar product
\[\langle \phi,\psi\rangle = \int\phi(x)\psi(x)f_0(x)dx\]
where $f_0(x)$ denotes the weighting function introduced in (\ref{setup_mse}).

We also define integral operators $T_{\mu\mu},T_{\mu\tau}$ associated with the covariance kernels
\begin{eqnarray}
\nonumber (T_{\mu\mu}\varphi)(x)&:=&\int H_{\mu\mu}(x_1,x)\varphi(x_1)f_0(x_1)dx_1\\
\nonumber (T_{\mu\tau}\varphi)(x)&:=&\int H_{\mu\tau}(x_1,x)\varphi(x_1)f_0(x_1)dx_1\\
\nonumber (T_{\tau\tau}\varphi)(x)&:=&\int H_{\tau\tau}(x_1,x)\varphi(x_1)f_0(x_1)dx_1
\end{eqnarray}
for any square integrable function $\varphi$. The operators $T_{\mu\mu},T_{\tau\tau}$ are self-adjoint, whereas the adjoint of $T_{\mu\tau}$ is given by
\[(T_{\mu\tau}^*\varphi)(x):=\int H_{\mu\tau}(x,x_1)\varphi(x_1)f_0(x_1)dx_1.\]

We now turn to the construction of an optimal $K$-dimensional basis for predicting $\tau_g(x)$ based on $\mu_g(x;0)$. For a collection of $K$ functions $\phi_1,\dots,\phi_K\in L_2(\mathcal{X})$, we let $P_K:L_2(\mathcal{X})\rightarrow\mathcal{H}_K$ denote the operator associated with orthogonal projection onto the closed linear subspace
\[\mathcal{H}_K:=\spn\left(\phi_1,\dots,\phi_K\right):=\left\{\sum_{k=1}^Ka_k\phi_k:a_1,\dots,a_K\in\mathbb{R}\right\}\]
By the classical projection theorem (Theorem 2 on p.51 in \cite{Lue69}) that projection is well-defined.

We then consider the predictors $BP_K\mu_g$ for $\tau_g$ on that subspace corresponding to linear operators $B:L_2(\mathcal{X})\rightarrow L_2(\mathcal{X})$, where we define $B$ via
\[(B h)(x):=\int h(x_1;0)\beta(x_1,x)f_0(x_1)dx_1\]
for any function $h\in\mathcal{H}$. We then let
\begin{eqnarray}\nonumber IMSE_K&\equiv& IMSE_K[\phi_1,\dots,\phi_K]\\ \nonumber&:=&\int\min_{B\in\mathcal{H}_K\times L_2(\mathcal{X})}\mathbb{E}\left[(\Delta\tau_{g^*}(x)-BP_K\mu_{g^*}(x;0))^2\right]f_0(x)dx\end{eqnarray}
denote the integrated mean-square error of prediction, minimized over the set of linear predictors using those $K$ functions. We restrict our attention to basis functions in the closed linear subspace $\mathcal{N}^{\perp}$, the orthogonal complement of the null space of $T_{\mu\mu}$, $\mathcal{N}:=\ker(T_{\mu\mu})$. This restriction is of no practical consequence since for any function $h$ in the null space of $T_{\mu\mu}$, $\var(\langle \mu_g,h\rangle)=\langle h,T_{\mu\mu}h\rangle=0$. Considering any possible choices of $\phi_1,\dots,\phi_K\in L_2(\mathcal{X})$, we first give a lower bound on $IMSE_K$

\begin{lem}\label{IMSE_K_lower_bound} Suppose that $T_{\mu\mu},T_{\mu\tau}$ are compact operators, and define
\begin{equation}\label{IMSE_star_lem}IMSE_K^*:=\inf_{\phi_1,\dots,\phi_K}\left(\int\mathbb{E}[\Delta\tau_{g^*}(x)^2]f_0(x)dx - \sum_{k=1}^K\langle \phi_k,T_{\mu\tau}T_{\mu\tau}^*\phi_k\rangle\right)\end{equation}
where the infimum is taken over $\phi_1,\dots,\phi_K\in L_2(\mathcal{X})$ such that $\langle\phi_k,T_{\mu\mu}\phi_l\rangle=\delta_{kl}$ for all $k,l=1,\dots,K$. Then for an arbitrary choice of $\phi_1(x),\dots,\phi_K(x)$ we have that $IMSE_K\geq IMSE_K^*$.
\end{lem}

See the appendix for a proof. The operators $T_{\mu\mu},T_{\mu\tau}$ are known to be compact if the corresponding covariance kernels are square-integrable, that is if the integrals $\int H_{\mu\mu}(x_1,x_2)^2dx_1dx_2$ and $\int H_{\mu\tau}(x_1,x_2)^2dx_1dx_2$ are finite. Since the operator $T_{\mu\mu}$ in the constraint is compact, there is no guarantee that the infimum will be attained by square integrable functions $\phi_1,\dots,\phi_K$. Intuitively, this ill-posedness stems from the fact that there may be functionals of $\mu_g(x;0)$ that have small variance across sites but are highly predictive with respect to $\tau_{g^*}(x)$. This problem bears some resemblance with functional canonical analysis, where \cite{HMW03} propose high-level conditions on the cross-correlation operator which would also be sufficient to guarantee that the infimum in (\ref{IMSE_star_lem}) is in fact attained at elements in $L_2(\mathcal{X})$.

If such a solution exists, it can be easily seen from the expression for $IMSE_K^*$ that the optimal basis functions for linear prediction are given by the solutions to the generalized eigenvalue problem
\begin{equation}\label{S_op_defn}
T_{\mu\tau}T_{\mu\tau}^*\phi_k^*=\lambda_kT_{\mu\mu}\phi_k^*\hspace{0.5cm}\textnormal{for each }k=1,\dots,K
\end{equation}
where we select the eigenfunctions $\phi_1^*,\dots,\phi_K^*$  associated with the $K$ leading eigenvalues $|\lambda_1|\geq|\lambda_2|\geq\dots$.\footnote{Note that while the self-adjoint operators $T_{\mu\tau}T_{\mu\tau}^*$ and $T_{\mu\mu}$ are both nonnegative, the generalized eigenvalue problem may have solutions associated with a negative eigenvalue.}

Our results allow for multiplicities of eigenvalues rather than requiring the ordering of $\lambda_1,\lambda_2,\dots$ to be strict. In that case, (\ref{S_op_defn}) holds equivalently for any re-ordering or linear combination of eigenfunctions associated with the same eigenvalue. However any such transformations also yield the same minimum in (\ref{IMSE_star_lem}) and are therefore equivalent for the purposes of minimizing the IMSE for prediction.


Rather than imposing conditions for existence, we focus instead on a regularized version of the problem, where we then demonstrate that the solution to that problem is approximately optimal in the sense that they achieve an IMSE that can be arbitrarily close to $IMSE_K^*$ when the regularization parameter is sufficiently small. We discuss conditions for existence of a non-regularized solution to that problem separately in Appendix \ref{subsec:existence_regularization}.

Specifically, we consider the following generalized eigenvalue problem
\begin{equation}\label{S_op_reg}
T_{\mu\tau}T_{\mu\tau}^*\phi_{ka}^*=\lambda_{ka}(T_{\mu\mu}+a \id) \phi_{ka}^*\hspace{0.5cm}\textnormal{for each }k=1,\dots,K
\end{equation}
where $a>0$  is a regularization parameter. We then let $\phi_{1a}^*,\dots,\phi_{Ka}^*$ be the eigenvectors corresponding to the $K$ largest eigenvalues (in absolute value), that is $|\lambda_{1a}|\geq|\lambda_{2a}|\geq\dots|\lambda_{Ka}|\geq|\lambda_{K+sa}|$ for each $s\geq1$, where we impose the normalization $\langle\phi_{ka}^*,T_{\mu\mu}\phi_{ka}^*\rangle=1$ for each $k=1,\dots,K$. In what follows, we also denote the operator $T_{\mu\mu a}:=T_{\mu\mu}+a \id$.

We denote the integrated mean-square error of prediction using the basis from the regularized problem (\ref{S_op_reg}) with
\[IMSE_K^*(a):=\int\min_{B\in\mathcal{H}_K^*\times L_2(\mathcal{X})}\mathbb{E}\left[(\Delta\tau_{g^*}(x)-BP_K^*\mu_{g^*}(x;0))^2\right]f_0(x)dx\]
where $P_K^*$ is the orthogonal projector onto $\mathcal{H}_K^*:=\spn\left(\phi_{1a}^*,\dots,\phi_{Ka}^*\right)$. We show that the solutions to (\ref{S_op_reg}) corresponding to the $K$ largest eigenvalues are approximately optimal as $a\rightarrow0$:

\begin{thm}\label{opt_basis_thm}\textbf{(Optimal Basis for $\mu_g(x;0)$)} Suppose that $T_{\mu\mu}$ and $T_{\mu\tau}$ are compact operators. Then for any $a>0$ and fixed $K$ there exists a solution $\phi_{1a}^*,\dots,\phi_{Ka}^*$ the functions solving the generalized eigenvalue problem (\ref{S_op_reg}), and the resulting IMSE satisfies
\[IMSE_K^*(a) \leq IMSE_K^* + o(1)\]
as $a\rightarrow0$.
\end{thm}

See the appendix for a proof. We can interpret this result as establishing an optimal finite-dimensional feature space for $\mu_g(\cdot;0)$ for predicting the conditional average treatment effect, up to a regularization bias that can be made small in terms of its impact on the IMSE of prediction.

\subsubsection{Prediction of Model Shift}

Given the proposed choice of $\phi_1^*,\dots,\phi_K^*$, we also state the projection of $\tau_g$ onto the optimal basis:

\begin{cor}\label{opt_basis_cor} Suppose the assumptions of Theorem \ref{opt_basis_thm} hold. Then for any $K=1,\dots, K^*$, the projection based on the solution of (\ref{S_op_reg}) takes the form
\[\tau_{g^*K}^*(x):=\tau(x) +\sum_{k=1}^{K}t_{g^*k}\psi_{ka}^*(x)\]
where $t_{g^*k}:=\frac{1+a}{1-a}\langle \mu_{g^*},\phi_{ka}^*\rangle$ and
\begin{equation}\label{psi_star_eqn}\psi_{ka}^*(x):=\left(T_{\mu\tau}^*\phi_{ka}^*\right)(x)\end{equation}
\end{cor}
See the appendix for a proof.  In particular, given the operators $T_{\mu\mu},T_{\mu\tau}$ defined at the population level, the optimal projection depends on the site-specific mean function $\mu_{g^*}(x,0)$ only through $K$ scalar features $(t_{g^*1},\dots,t_{g^*K})$ that can be estimated consistently as the number $n_{g^*}$ of observations in the target cluster grows large.

Incidentally, we can also confirm that each of the functions $\psi_{1a}^*,\dots,\phi_{Ka}^*$ is an eigenfunction of the operator $T_{\mu\tau}^*T_{\mu\mu a}^{-1}T_{\mu\tau}$ at the eigenvalue $\lambda_{ka}$:
\begin{eqnarray}
\nonumber T_{\mu\tau}^*T_{\mu\mu a}^{-1}T_{\mu\tau}\psi_{ka}^* &=&T_{\mu\tau}^*T_{\mu\mu a}^{-1}T_{\mu\tau}T_{\mu\tau}^*\phi_{ka}^*\\
\nonumber &=&\lambda_k T_{\mu\tau}^*\phi_{ka}^* = \lambda_k\psi_{ka}^*
\end{eqnarray}

Hence one interpretation of the approach is as an approximation based on the $K$ leading components of a singular value decomposition of the operator $T_{\mu\mu a}^{-1/2}T_{\mu\tau}^*$ on a suitably chosen linear subspace of $L_2(\mathcal{X})$: Consider the eigensystem $\phi_{1a}^*,\phi_{2a}^*,\dots$ solving (\ref{S_op_reg}) at any nonzero value for the generalized eigenvalue $\lambda_{ka}$, and the corresponding functions $\psi_{1a}^*,\psi_{2a}^*,\dots$. By standard properties of eigenfunctions, these systems form a basis for the orthogonal complements of the null spaces $\ker(T_{\mu\tau}T_{\mu\mu a}^{-1/2})$ and $\ker(T_{\mu\mu a}^{-1/2}T_{\mu\tau}^*)$, respectively. Hence, using these bases as test functions, we can confirm that $\{\phi_{1a}^*,\phi_{2a}^*,\dots\}$, $\{\psi_{1a}^*,\psi_{2a}^*,\dots\}$, and $\{\sqrt{|\lambda_{1a}|},\sqrt{|\lambda_{2a}|},\dots\}$ represent a singular value decomposition of the operator $T_{\mu\mu a}^{-1/2}T_{\mu\tau}^*$ where
\[(T_{\mu\mu a}^{-1/2}T_{\mu\tau}^*h)(s) = \sum_{k=1}^{K^*}\sqrt{|\lambda_{ka}|}\phi_{ka}^*(s)\langle\psi_{ka}^*,h\rangle\]
for any $h\in L_2(\mathcal{X})$.

\subsection{Comparison to Existing Approaches for Functional Regression} We briefly discuss how this approach compares to existing methods in the literature on functional regression with a functional response.

While the basis functions $\phi_{1k},\phi_2,^*,\dots$ in our analysis are derived from optimality considerations, the procedure we arrive at has a close resemblance to canonical correlation analysis which has previously been proposed for functional regression problems by \cite{HMWY10}, see also \cite{LMS93}. Our results differ in that for one the basis $\phi_1^*,\dots,\phi_K^*$ is formally shown to be optimal for the linear prediction problem considered here. Moreover, the canonical variates need not be ordered according to the eigenvalues $\lambda_k$ which we show to be the relevant ordering for the IMSE-optimal choice among the eigenfunctions $\phi_1^*,\phi_2^*,\dots$.

To address the potential non-existence of an unregularized solution to (\ref{S_op_defn}), \linebreak\cite{HMW03} and \cite{HMWY10} impose high-level conditions on the cross-correlation operator to ensure existence (see Proposition 4.2 in \linebreak\cite{HMW03}). Since our focus is on prediction, we focus instead on the achievable IMSE, allowing for the possibility that unregularized canonical variates need not be well-defined. This approach parallels the analysis of \cite{CEGR08} who consider estimation of the largest canonical correlation between two $L_2$ processes and show that this scalar parameter can be approximated arbitrarily closely via regularized canonical correlation analysis.

\cite{YMS11} propose regression based on a singular value decomposition of the operator $T_{\mu\tau}$ rather than $T_{\mu\mu}^{-1/2}T_{\mu\tau}$,
\[(T_{\mu\tau}h)(s):=\sum_{k=1}^K\sqrt{\nu}_k\langle\zeta_k^*,h\rangle\xi_k(s)\]
Similarly, \cite{ROg07} propose functional partial least squares for functional regression. A finite-$K$ expansion based on spectral analysis of $T_{\mu\tau}$ has no known optimality properties but elegantly sidesteps the problem of inverting $T_{\mu\mu}$ and therefore works under weaker conditions and is numerically stable in the absence of regularization.

Another important approach proposed by \cite{BCF17} who directly minimize the mean-square error of prediction in a functional linear regression model, subject to a nuclear norm penalization of the projection operator $B$. The particular appeal of that approach is that it offers a ``one-stop" approach towards regularization with a single tuning parameter, and directly optimizes the in-sample predictive performance subject to that penalty. Their approach assumes that $B$ is a Hilbert-Schmidt (kernel) operator which is not guaranteed under our assumptions. Their approach is also designed towards delivering a consistent estimator for $B$ in a setting where $G$ is large.

Our focus is instead on heavily regularized but interpretable solutions $B_{a,K}$ for moderate values of $G$, where the singular value representation delivers a sparse representation of the operator in terms of a functions of $x$. The estimated scores can then be used to assess whether the target site is comparable to the experimental sample in terms of the most predictive features identified by the method.  The extrapolated CATE can be interpreted as a best linear predictor given the estimated basis functions, and regularization bias results in a potentially suboptimal (with respect to the IMSE), but ultimately valid construction of features from $\mu_g(x;0)$. As \cite{BCF17} point out, ridge regularization also yields more stable predictions in the presence of poorly separated eigenvalues than a truncation of the spectral expansion at a finite dimension, so if the eigenvalue $\lambda_{K}$ at the chosen cutoff is not well separated from $\lambda_{K+1}$, the resulting potential instability of predictions should be flagged when reporting estimation results.


\subsection{Assumptions for Estimation}

We next formalize the identifying conditions which are adapted from \cite{HIM05}. We depart from their main framework in two substantial ways: for one our design-based approach treats experimental and target sites as random draws from a finite population of sites. Moreover, we also consider a version of the problem in which baseline data on pre-treatment outcomes for the target site are available and are to be used to predict site-specific ``macro" effects. We highlight how this affects the interpretation of the assumptions on the assignment mechanism. While our derivation of optimal predictors in section \ref{sec:opt_basis_sec} is directly in terms of high-level properties of covariance operators, the following assumptions are maintained to establish asymptotic rates for estimates in Appendix \ref{sec:asymptotics}.

We assume throughout that for each cluster $g=1,\dots,G$ the researcher observes a sample of $n_g$ units that are drawn independently and uniformly at random from $\{1,\dots,N_g\}$, and also independently of potential values and unit attributes. For notational convenience our results will be stated for the case that the observed number of units is the same for each site, $n_g\equiv n$ for $g=1,\dots,G$. For each experimental site, we assume that selection of units into treatment is based only on observables $X_{gi}$,

\begin{ass}\label{unconf_ass_ass}\textbf{(Unconfounded Assignment)} For all $g$ with $R_g=1$,
\[D_{gi}\ind(Y_{gi}(0),Y_{gi}(1))|X_{gi},R_g=1\]
where $D_{g1},\dots,D_{gN_g}$ are also conditionally independent across units and clusters given attributes and $R_1,\dots,R_G$.
\end{ass}

This condition is met if $D_{gi}$ was assigned at random as part of a randomized controlled trial (RCT) at each experimental site, and it captures the idea of extrapolating from a collection of internally valid estimates of site-specific causal effects to a new site. In a practical application the set of confounders $X_{gi}$ may differ from the conditioning variables chosen by the researcher for define the relevant conditional average treatment effect, however for expositional clarity we only consider the case in which the conditioning variables are the same. It is also possible to adapt our approach to the case of randomization at the cluster level, $D_{gi}\equiv D_g$ for all $i=1,\dots,n_g$, see Appendix \ref{sec:asymptotics} for a brief discussion.

Furthermore, we assume that among the $G$ sites, the $G-1$ experimental locations were selected independently of potential values, conditional on observable covariates:

\begin{ass}\label{unconf_loc_ass}\textbf{(Unconfounded Location)} $g^*$ is drawn uniformly at random from $\{1,\dots,G\}$ independently of $\left\{Y_{gi}(0),Y_{gi}(1),X_{gi}:g=1,\dots G, i=1,\dots,N_g\right\}$.
\end{ass}

This assumption is strengthened version of Assumption 2 in \cite{HIM05} and describes an idealized observational protocol that rules out systematic ex-ante site selection bias. It can be seen immediately that under this condition, for a randomly selected experimental site $\tilde{g}$ with $R_{\tilde{g}}=1$, $\left(Y_{\tilde{g}i}(0),Y_{\tilde{g}i}(1),X_{\tilde{g}i}\right)\stackrel{d}{=}\left(Y_{g^*i}(0),Y_{g^*i}(1),X_{g^*i}\right)$, where ``$\stackrel{d}{=}$" denotes equality in marginal distributions. Therefore, Assumption \ref{unconf_loc_ass} implies that experimental and target sites are exchangeable, the fundamental assumption in the literature on conformal prediction (\cite{VGS05} and \cite{LGRTW18}).

In practice, we do not expect that assumption to be an accurate description on how experimental (study) and target sites were selected. Rather, in the absence of additional knowledge regarding site selection, this auxiliary assumption defines a pseudo-true parameter, which aggregates estimates from experimental sites into a ``best" prediction for the target population. The resulting transfer estimate should therefore be interpreted as a summary of the directly quantifiable relevant information from previous experiments, which could be subject to additional (qualitative or quantitative) sensitivity analysis with respect to suspected violations of that exchangeability condition (see e.g. \cite{BCRT23} for the problem of conformal prediction).

For the next assumption, we define the site-specific propensity score as
\[p_g(x):=\mathbb{P}(D_{gi}=1|X_{gi}=x)\]
We require that the supports of covariates overlap, both between treated and control units, as well as across the sites $g=1,\dots,G$.

\begin{ass}\label{support_ass}\textbf{(Support Conditions)} There exists $\delta$, $0<\delta<1$ such that
\[\delta<p_g(x)<1-\delta\hspace{0.3cm}\textnormal{ and }\delta<f_g(x)/f_0(x)<1/\delta\]
for all $g\in\{1,\dots,G\}\backslash\{g^*\}$ and $x$ in the support of $f(x)$. Furthermore, the support $\mathcal{X}$ of $X_{gi}$ is a compact subset of $\mathbb{R}^d$, without loss of generality $\mathcal{X}=[0,1]^d$, and we assume $\inf_{x\in[0,1]^d}f_g(x)\geq \kappa>0$ for all $g=1,\dots,G$.
\end{ass}

The role of this assumption is to ensure that conditional moments of either potential value are identified and can be estimated consistently across sites. While Assumption \ref{unconf_loc_ass} does allow for experimental and target sites to differ in terms of the distribution of observables, we require that the site-specific supports overlap, potentially after trimming non-overlapping regions in the covariate space as suggested in \cite{HIM05}. This assumption also does not cover site-specific aggregate covariates that may serve as additional predictors as analyzed by \cite{HIM05} and \linebreak\cite{DPS21}. Randomization at the level of the site would not satisfy the support condition on the site-specific propensity score and therefore requires a different approach which is discussed in Appendix \ref{sec:asymptotics}. Additional adjustments for site-specific variables may be possible, but would also be severely constrained by the small number of observable sites. While our focus is on the optimal use of cross-sectional information for extrapolation, we briefly discuss how to incorporate site-level covariates in Section \ref{subsec:site_spec_cov_subsec}.

Nonparametric estimation of the first two conditional moments of potential values $Y_{gi}(d)$ given attributes $X_{gi}$ requires additional moment and smoothness conditions, where we specifically assume the following:

\begin{ass}\label{loc_lin_ass} \textbf{(Distribution and Moments)} For $g=1,\dots,G$, (a) $X_{gi}$ is continuously distributed on $[0,1]^d$ with marginal p.d.f. that is bounded from above $\sup_{x\in[0,1]^d}f_g(x)\leq B_0<\infty$. (b) The site-specific density $f_g(x)$, propensity score $p_g(x)$, and conditional mean functions $\mu_g(x;d)$ are twice continuously differentiable in $x$ with uniformly bounded derivatives. We also assume that (c) potential outcomes have bounded moments $\mathbb{E}|Y_{gi}(d)|^s<\infty$ for $d=0,1$ and some $s>3$, (d) there exist finite constants $B_0,B_1$ such that $\sup_x f(x)\leq B_0$ and $\sup_x\mathbb{E}[|Y_{gi}(d)|^s|X_{gi}=x]f_g(x)\leq B_1$ for all $g$.
\end{ass}

To avoid additional notation, we do not explicitly discuss the case in which some components of $X_{gi}$ may be discrete. With the exception of part (c), the conditions in Assumption \ref{loc_lin_ass} are commonly assumed for nonparametric estimation of conditional moments, see e.g. \cite{Han08}. Notice also that we effectively need to be able to estimate conditional moments separately for each site, and therefore require these conditions to hold uniformly over $g$. In the absence of covariate shifts, i.e. if the distribution of covariates $f_g(x)$ or propensity score $p_g(x)$ did not vary over $g$, this issue could be avoided (see \cite{YMW05}), however we do not find such an assumption plausible for the problem considered here.

\subsection{Implementation for Densely Sampled Clusters}

The representation in Corollary \ref{opt_basis_cor} motivates an estimator of the form
\[\hat{\tau}_{g^*}(x):=\hat{\tau}(x) + \sum_{k=1}^{K}\hat{\tilde{t}}_{g^*k}\hat{\psi}_{ka}(x)\]
where $\hat{\tau}(x):=\hat{\mu}(x;1)-\hat{\mu}(x;0)$, $\hat{\tilde{t}}_{g^*k}=\langle\hat{\mu}_{g^*},\hat{\phi}_{ka}\rangle$ for a nonparametric estimator $\hat{\mu}_g$ of $\mu_g(x;0)$, and the basis functions $\hat{\phi}_{1a},\dots,\hat{\phi}_{Ka}$ are obtained by solving an empirical analog of the generalized eigenvalue problem (\ref{S_op_reg}).

Here we develop our approach for the case of densely sampled clusters, $n\rightarrow\infty$, separate results for the setting with sparse samples are given in Appendix \ref{sec:asymptotics}. In contrast to the densely sampled case, that approach requires that site-specific covariate distributions $f_g(x)$ are either known or can be estimated consistently, which does in general not allow those distributions to be fully nonparametric.

We estimate $\mu(x;d):=\mathbb{E}[\mu_{g^*}(x;d)]$ and $H(x_1,x_2;d_1,d_2):=\cov(\mu_{g^*}(x_1;d_1),\mu_{g^*}(x_2;d_2))$ using nonparametric estimators $\hat{\mu}(x;d)$ and $\hat{H}(x_1,x_2;d_1,d_2)$. While our theory is not restricted to one particular choice of nonparametric estimators, following \cite{YMW05} we give results for local linear estimators: For each experimental cluster, let
\begin{equation}\label{mu_hat_loclin}\hat{\mu}_g(x;d):=\arg_{b_0}\min_{b_0,b_1}\sum_{i=1}^{n_g}w_{gi}(x;d)(Y_{gi} - b_0 - b_1(x-X_{gi}))^2\end{equation}
with nonparametric weights \[w_{gi}(x;d):=\dum\{D_{gi}=d\}K\left(\frac{X_{gi}-x}{h}\right).\]
Here, the notation ``$\arg_{b_0}\min_{b_0,b_1}$" corresponds to the first component vector of the joint argmax of a function with respect to $b_0,b_1$.

Here, $K(u)$ is a kernel function with standard properties (see Assumption \ref{kernel_ass} in the Appendix for formal conditions on $K(\cdot)$), and the bandwidth $h>0$ is chosen according to sample size $G,n$, the dimension of $X_{gi}$ and assumed smoothness of the estimands. We also let
\begin{equation}\label{H_hat_loclin} \hat{M}_g(x_1,x_2;d_1,d_2):=\arg_{b_{0}^{(g)}}\min_{b_{0}^{(g)},b_{11}^{(g)},b_{12}^{(g)}}\sum_{j\neq i}
H_{gij}(x_1,x_2,\mathbf{b})w_{gi}(x_1;d)w_{gj}(x_2;d)\end{equation}
where
\[H_{gij}(x_1,x_2,\mathbf{b}):= \left(Y_{gi}Y_{gj}-b_0^{(g)}-b_{11}^{(g)}(X_{gi}-x_1) - b_{12}^{(g)}(X_{gj}-x_2)\right)^2.\]
We then construct
\begin{eqnarray}
\nonumber \hat{\mu}(x;d)&:=&\frac1{G-1}\sum_{g=1}^GR_g\hat{\mu}_g(x;d)\\
\nonumber \hat{H}(x_1,x_2;d_1,d_2)&:=&\frac1{G-1}\sum_{g=1}^GR_g\hat{M}_g(x_1,x_2;d_1,d_2)-\hat{\mu}(x_1;d_1)\hat{\mu}(x_2;d_2)
\end{eqnarray}

In principle, the bandwidth could be chosen differently for estimation of $\hat{\mu}(x;d)$ and $\hat{H}(x_1,x_2;d_1,d_2)$, however in our theory in Appendix \ref{sec:asymptotics}, the optimal rate turns out to be the same for either estimator in the densely sampled case. Apart from kernel-based approaches, other possible methods include series estimators, random forests, or neural networks. The choice of nonparametric estimator will generally depend on the support of the covariates and other practical considerations.

This estimator is an average of separate local linear estimators for each of the $G-1$ experimental clusters, in a departure from the approach in \cite{YMW05} who propose a local linear estimator based on the pooled data from all $G-1$ clusters. There are two reasons for a different approach in the densely sampled case: for one we do not assume that attributes (``positions") are sampled from the same distribution in all clusters, but sites may differ in the distribution of $X_{gi}$. We furthermore assume ``dense" samples from a small number of clusters, whereas they consider scenarios in which $n$ is small, but $G$ grows large. In our setup, cluster-specific moments can be estimated consistently, whereas between-cluster variation is the dominant source of estimation noise due to small $G$. That source of estimation error would be amplified in a nonparametric regression step, so our approach seeks to avoid that potential problem.


To describe the estimator for the basis functions $\hat{\phi}_1,\dots,\hat{\phi}_K$ let
\[\hat{H}_{\mu\mu}(x_1,x_2):=\hat{H}(x_1,x_2;0,0)\hspace{0.5cm}\textnormal{and }\hat{H}_{\mu\tau}(x_1,x_2):=\hat{H}(x_1,x_2;1,0)-\hat{H}(x_1,x_2;0,0).\]
In analogy to the definition for the operators $T_{\mu\mu}$ and $T_{\mu\tau}$, we can construct the estimators
\begin{eqnarray}
\nonumber(\hat{T}_{\mu\mu}h)(x)&=&\int\hat{H}_{\mu\mu}(x,s)h(x)f_0(s)ds\\
\nonumber(\hat{T}_{\mu\tau}h)(x)&=&\int\hat{H}_{\mu\tau}(x,s)h(x)f_0(s)ds\\
\nonumber(\hat{T}_{\mu\tau}^*h)(x)&=&\int\hat{H}_{\mu\tau}(s,x)h(x)f_0(s)ds\\
\end{eqnarray}
for any square integrable function $h$, and let $\hat{T}_{\mu\mu a}:=\hat{T}_{\mu\mu} + a\textnormal{Id}$.

In order to estimate the eigenfunctions $\phi_{1a}^*,\phi_{2a}^*,\dots$, we solve the generalized eigenvalue problem (\ref{S_op_reg}) after replacing the operators $T_{\mu\tau},T_{\mu\mu}$ with their estimates as defined above. Specifically, we can find the functions $\hat{\xi}_{1a},\dots,\hat{\xi}_{Ka}$ solving the eigenvalue problem
\begin{equation}\label{eig_estimator_1}\hat{T}_{\mu\mu a}^{-1/2}\hat{T}_{\mu\tau}\hat{T}_{\mu\tau}^*\hat{T}_{\mu\mu a}^{-1/2}\hat{\xi}_{ka}=\hat{\lambda}_k\hat{\xi}_{ka}\end{equation}
and that are associated with the $K$ largest eigenvalues in $\hat{\lambda}_1\geq\hat{\lambda}_2\geq\dots$. We then solve for
\begin{equation}
\label{eig_estimator_2}\hat{\phi}_{ka}:=\hat{T}_{\mu\mu a}^{-1/2}\hat{\xi}_{ka}.\end{equation}
Since $\hat{T}_{\mu\mu}$ is a nonnegative (nonnegative definite) operator and $a>0$, the operator on the left-hand side of (\ref{eig_estimator_1}) is Hermitian and compact, and the inverse problem (\ref{eig_estimator_2}) is well-posed. To implement the procedure we use linear sieve approximations to the eigenfunctions, which converts (\ref{eig_estimator_1}) into a finite-dimensional eigenvalue problem.\footnote{See e.g. \cite{RSi05}, chapter 8.4.2.}

We then construct $\hat{\psi}_{ka}$ by applying the estimator of $T_{\mu\tau}^*$ to the estimated eigenfunction $\hat{\phi}_{ka}$,
\[\hat{\psi}_{ka} (x):=\left(\hat{T}_{\mu\tau}^*\hat{\phi}_{ka}\right)(x)\equiv\int\hat{H}_{\mu\tau}(s,x)\hat{\phi}_{ka}(s)f_0(s)ds\]
for $k=1,\dots,K$. Using these estimates, we then obtain
\[\widehat{\tilde{t}}_{g^*k}:=\langle \hat{\mu}_{g^*},\hat{\phi}_{ka}\rangle\]
Substituting this into the formula from Corollary \ref{opt_basis_cor}, our estimate of the conditional ATE $\tau_{g^*}(x)$ is
\[\hat{\tau}_{g^*}(x)= \hat{\tau}(x) +  \sum_{k=1}^{K}\widehat{\tilde{t}}_{g^*k}\hat{\psi}_{ka}(x)\]

Appendix \ref{sec:asymptotics} gives convergence rates for these estimators both for densely and sparsely sampled sites. Specifically, assuming equal numbers of cross-sectional observations for each site, $n_g\equiv n$, Theorem \ref{loc_lin_consistency_thm} gives the rate
\[r_{Gn} = \frac1{G} + h^2 + \left(\frac{\log n}{Gnh^d}\right)^{1/2}\]
for the preliminary nonparametric estimators of mean and covariance functions if sites are \textbf{densely sampled} ($n_G\rightarrow\infty$ as $G\rightarrow\infty$) and treatment is randomized among units in each site. If treatment is instead randomized at the site level, the approach to estimating the covariance function $H(x_1,x_2;d_1,d_2)$ has to be modified as discussed in the appendix, resulting in a rate
\[r_{Gn} = \frac1{\sqrt{G}} + h^2 + \left(\frac{\log n}{Gnh^d}\right)^{1/2}\]
for the densely sampled case. For \textbf{sparsely sampled} sites ($n_G$ bounded), $H(x_1,x_2;d_1,d_2)$ can still be estimated consistently as $G\rightarrow\infty$ by pooling observation pairs across sites. Convergence for eigenfunctions and the IMSE of prediction depends on the asymptotic rate of estimation of the covariance function $H(x_1,x_2;d_1,d_2)$,
\[r_{Gn} = h^2 + \left(\frac{\log G}{Gh^{2d}}\right)^{1/2}\]
whereas the rate for estimating the conditional mean function $\mu(x_1;d_1)$ is faster for reasonable bandwidth choices.

Given these preliminary rates, Theorem \ref{eig_cons_thm} gives a rate
\[\|\hat{\phi}_{k}-\phi_{ka}\|=O_p\left(a^{-3/2}r_{Gn}\right)\]
for estimation of the eigenfunctions, and Corollary \ref{IMSE_conv_rate} shows that the IMSE of prediction using the estimated basis function is
\begin{equation}\label{imse_hat_eq}|IMSE_K[\hat{\phi}_1,\dots,\hat{\phi}_K]-IMSE_K^*|=O_P\left(a + a^{-3/2}r_{Gn}\right)\end{equation}
Appendix \ref{sec:asymptotics} also provides comparable rates for nonparametric estimation of mean and covariance functions using B-splines instead of local linear regression. When the eigenvalues of the corresponding population problem (\ref{S_op_reg}) are not simple, the  eigenfunctions $\phi_k$ are only estimated up to a data-dependent rotation of each eigenspace associated with multiple eigenvalues. Since any such transformation yields the same value for the problem of minimizing the IMSE of prediction (\ref{IMSE_K_lower_bound}), this does not affect the rate at which the IMSE is minimized in (\ref{imse_hat_eq}).

Since the transfer estimate is always a linear projection on the constructed features $\hat{\phi}_{k}$, these rates illustrate how fast the quality of the prediction improves as we approximate the optimal basis functions $\phi_k^*$ more closely. In general, that approximation requires the number of sites $G$ to be not too small, especially if treatment was not randomized within each site. The difference in rates between the densely and sparsely sampled cases also illustrates how a larger number of cross-sectional observations $n_g$ within each site can be leveraged to retrieve the optimal predictors more accurately, although in practice typically the number of sites $G$ is the main limiting factor.

\subsection{Site-Specific Covariates}

\label{subsec:site_spec_cov_subsec}

A natural extension of the main framework concerns site-specific covariates $W_g$ which may be observed in addition to the unit-level attributes $X_{gi}$. In this section we sketch a conceptual extension to our approach under the assumption that these covariates satisfy unconfoundedness conditions analogous to those for $X_{gi}$. When the number of experimental sites is not very large, controlling nonparametrically for a significant number of site covariates is generally not feasible in practice, so we consider this extension to be primarily of theoretical interest. For the purposes of this section, we also regard the $G$ sites as random draws from a superpopulation in order to be able to define conditional expectations given the covariate $W_g$ in a meaningful way.

To be specific, we consider a version of the original problem, where Assumption \ref{unconf_ass_ass} is changed to
\[D_{gi}\ind(Y_{gi}(0),Y_{gi}(1))|X_{gi},W_g,R_g=1\]
and Assumption \ref{unconf_loc_ass} is strengthened to assume that $g^*$ is drawn independently of\linebreak $Y_{g^*i}(0),Y_{g^*i}(1),X_{g^*i}$, and $W_{g^*}$. Assuming that the $g$th cluster represents a random draw from a superpopulation, we can define the conditional expectation
\[\mu(x,w;d):=\mathbb{E}[Y_{g^*i}(d)|X_{g^*i}=x,W_{g^*}=w]\]
and covariance function
\begin{eqnarray}\nonumber H(x_1,x_2,w;d_1,d_2)&:=&\mathbb{E}\left[\left.\frac{}{}(Y_{g^*1}(d_1)-\mu(x_1,w;d_1))\right.\right.\\
\nonumber&&\textnormal{  }\left.\left.\frac{}{}\times(Y_{g^*2}(d_2)-\mu(x_2,w;d_2))
\right|X_{g^*1}=x_1,X_{g^*2}=x_2,W_{g^*}=w\right]\end{eqnarray}
where expectations are with respect to the joint distribution of potential values, attributes, and $W_g$ in that superpopulation.

We can then apply the previous method conditional on $W_{g^*}=w_{g^*}$, where we replace the unconditional mean function $\mu(x;d)$ with an estimate of estimate $\mu(x;w_{g^*};d)$, and form the analogs of the covariance operators $T_{\mu\mu}$ and $T_{\mu\tau}$ from estimates of the conditional covariance function $H_{d_1d_2}(x_1,x_2;w_{g^*})$. The conditionally optimal basis functions $\phi_1^*,\dots,\phi_K^*$ are then obtained from an eigenanalysis of the conditional covariance operators given $W_{g^*}$. Such an approach would effectively amount to a regression adjustment for the mean and covariance functions for $\mu_{g^*}(x;d)$ with respect to $W_{g^*}$.

For modest values of $G$, the scope for fully nonparametric adjustments to site-specific covariates is fairly limited for practical purposes, in contrast to ``micro" (unit-specific) covariates where our approach can leverage the size of the cross-sectional sample for each site to construct approximately optimal adjustments to estimates for the CATE. \cite{DPS21} used machine learning methods to adjust (unconditional) ATE estimates for site-specific covariates, however a fully nonparametric site-specific adjustment to the estimated CATE poses greater challenges given realistic sample sizes.

\section{Empirical Application}


We illustrate our approach with an empirical application to the estimation of the effect of conditional cash transfers on children's school attendance. In this literature, a conditional cash transfer is a recurring grant paid to an eligible household that is explicitly linked to a child attending school or other household decisions the policy maker wants to encourage, with transfer amounts of the order of 5-20 percent of average household consumption in the target population. In 1998-99 the government of Mexico conducted a large-scale randomized trial during the roll-out of the \emph{PROGRESA/OPORTUNIDADES} program (\cite{Sch04} and \cite{TWo06}), and similar programs have subsequently been implemented in over 50 other countries (see \cite{BHKO17} for a recent summary).

\subsection{Data}

We combine samples from  \emph{PROGRESA} with four additional randomized studies that were conducted in Indonesia (\emph{Program Keluarga Harapan (PKH)}, see \cite{Ala11} and \cite{CHOPSS20}), Morrocco (\emph{Tayssir}, see \cite{BDDDP16}), Kenya (\emph{Kenya CT-OVC}, \cite{CT_OVC12}), and Ecuador (\emph{Bono de Desarrollo Humano (BDH)}, \cite{ESch12}).\footnote{These studies were selected according to ease of access to the underlying microdata, where we excluded one additional study from Colombia (\emph{Subsidios Condicionados a la Asistencia Escolar}, \cite{BBLP11}) due to our inability to reconstruct baseline attendance data from the replication package.} Each of these field trials was a multi-site study conducted by the national government, where participants were recruited from a previously selected sample of clusters (schools, villages, or other comparable unit). In each study, clusters were drawn from a subset of the major administrative regions in each of these countries.

It should be noted that there were substantial differences in the exact design of the incentive between these five studies. In particular, Progresa and PKH explicitly make part of the transfer dependent on school attendance, whereas Tayssir experimented with a nudge rather than a strictly conditional transfer. For the remaining two studies in Kenya and Ecuador, cash transfers were unconditional. We deliberately pool the sites to replicate a realistic scenario for which a policy as been adapted to local circumstances, due to practical constraints and the policymaker's preferences.

Our main focus is on leveraging cross-site variation within each multi-site trial to extract predictive information on site-specific heterogeneity in the CATE. The five study populations in Mexico, Indonesia, Morocco, Ecuador, and Kenya are likely systematically different in terms of many factors that cannot be modeled explicitly, such as the local educational system, the chosen target population within the geographic reach of the study, the specific manner in which the transfer scheme was implemented, etc. Nevertheless, sites also vary substantially within each study, e.g. according to travel distance to urban centers or secondary school, or whether the language of instruction is widely spoken in the community. Hence, some communities in the heterogeneous pool of clusters in, say, Mexico, may still be sufficiently similar to a target location in Morocco or Indonesia in terms of the predictive attributes, as determined by our method. We will assess to what extent between-study variation can be predicted from between site variation on a more disaggregated level.

We retain all observations of households that met the eligibility criteria for the program, and for whom we can reconstruct measures of school attendance and per capita household expenditure at baseline and follow-up, along with children's age and gender, and the household head's level of education. For school attendance we use self-reports from baseline and follow up household surveys rather than data from school records or random checks which were only collected for some of the studies used in our analysis. After dropping households with incomplete data and locations with fewer than 15 school-aged children, we obtain a sample of 640 clusters (sites) with average cluster sizes ranging from 18 (PKH, Indonesia) to 47 (PROGRESA, Mexico) and 51 (BDH, Ecuador). PROGRESA and TAYSSIR (Morocco) contribute the largest number of clusters (297 and 238, respectively) compared to 50 for PKH (Indonesia), 31 for BDH (Ecuador), and 24 for CT-OVC (Kenya). For the purposes of this analysis we assign equal weight to each cluster. Of those clusters, 434 were treated, the remaining clusters were in the control group.

\subsection{Implementation}

We compare our approach across three different prediction tasks - as a benchmark, we report some results for the in-sample fit, with $\mu(\cdot)$ and $\Eta(\cdot)$ and resulting basis functions $\phi_k,\psi_k$ estimated from the full data set. We then consider cross-site prediction where for a given target site $g^*$, the basis functions are estimated from the remaining $G-1$ sites, and the transfer estimate is obtained by estimating the principal scores $m_{g^*1},\dots,m_{g^*K}$ from the baseline for the target site. Finally, we perform cross-study extrapolation, with the predictive model estimated from data excluding all other sites from the study that included the target site, for example predicting the outcome at a Progresa site using only data from sites in the remaining four studies.

Given the small to moderate cluster sizes, we choose an estimation approach suited to sparsely sampled functional data, see also Appendix \ref{sec:asymptotics}. The main difference to the densely sampled case is that the cluster-specific covariate distribution $f_g(x)$ for the weights in (\ref{mu_hat_loclin}) and (\ref{H_hat_loclin}) cannot be estimated nonparametrically. We make the simplifying assumptions that gender and age are independent of location and household per capita expenditure, and per capita expenditure follows a log-normal distribution within each cluster, which we then estimate parametrically.

The setting also differs from the idealized setup discussed in the theoretical sections of the paper in that there is baseline data available for each experimental cluster. Furthermore, in each of theses studies, treatment was randomized at the cluster level. We therefore construct predictors from the observed baseline data for $\mu_{gt}(x;0):=\mathbb{E}[Y_{git}(0)|X_{git}=x]$ at $t=0$, which are then used to predict conditional expectations $\mu_{gt}(x;1):=\mathbb{E}[Y_{git}(d)|X_{git}=x]$ for $d\in\{0,1\}$ and $t=1$. The covariance operators between $\mu_{g0}(x;0)$
and $\mu_{g1}(x;1)$ or $\mu_{g1}(x;0)$ are then estimated using the treatment and control clusters in the experimental population. We first consider the problem of predicting post-treatment outcomes from baseline outcomes in the treated clusters, where we can validate predictions directly against the observed data at the site level. We then implement the algorithm for predicting conditional average treatment effects, which are not directly observed at the cluster level for any of the experimental sites.

Given the limited number of distinct sites, and also in order to apply consistent variable definitions across studies we restrict the unit-specific covariates $X_{gi}$ to four variables, the child's gender, the child's age in years, enrollment status at baseline, and log per-capita household expenditure. We also restrict the estimators for $\mu(\cdot)$ and $\Eta(\cdot)$ to be additively separable in covariates, where we flexibly dummy out gender and age in years, and use B-splines of degree 2 to model variation with respect to log expenditure. Tuning parameters are chosen using cross-validation across clusters, where we separately target the integrated mean-square error of estimating the mean and covariance functions to determine the bandwidths for local linear regression, and the mean-square error for cross-cluster prediction for the regularization parameter $a$ in (\ref{S_op_reg}).

\subsection{Results}

We first report results for prediction of the \textbf{model shift in post-intervention outcomes} $\Delta\mu_{g}(x;1):=\mu_g(x;1)-\mu(x;1)$ using the estimated IMSE-optimal predictors from (\ref{S_op_reg}), which were estimated using only the 434 treated sites. We assess their performance as predictors at the level of the individual site as well as after aggregating sites within each study. The number of knots for B-spline approximations was determined using (leave-one-site-out) cross-validation, targeting the mean function $\mu(x;d)$ and covariance function $H(x_1,x_2;d_1,d_2)$, respectively. The ridge parameter $a$ was chosen based on estimated cross-site predictive performance, and cross-validation also suggests that for this application the optimal number of basis functions is $K=2$.

\begin{table}\label{fig:mu1_pred_corr}\footnotesize
\begin{tabular}{lcrcrrrrrrrrrr}
&&\multicolumn{8}{c}{Prediction using IMSE-Optimal Basis Functions}\\
\hline\hline\\[4pt]
&&baseline mean&&K=1&K=2&K=3&K=4&K=5&K=6\\[4pt]
\hline\\[2pt]
in-sample fit &&  0.1379  &&  0.4398  &  0.5039 &   0.5038 &   0.5167 &   0.5191 &   0.4835 \\
cross-site prediction &&   0.0136 &&   0.3600  &  0.4484 &   0.4514 &   0.4524  &  0.4603 &   0.4252 \\
cross-study prediction&& -0.0400 &&   0.2164 &   0.3198 &   0.2987 &   0.3282  &  0.3028 &   0.3267 \\[4pt]
\hline\\[8pt]

&&\multicolumn{8}{c}{Prediction using Functional PC}\\
\hline\hline\\[4pt]
&&baseline mean&&K=1&K=2&K=3&K=4&K=5&K=6\\[4pt]
\hline\\[2pt]
in-sample fit && 0.1379  &&  0.1377 &   0.3871  &  0.3776 &   0.4051  &  0.4221 &   0.4224 \\
cross-site prediction  &&  0.0136 &&   0.0896 &   0.3743 &   0.3544 &   0.3706 &   0.3878 &   0.3852 \\
cross-study prediction  &&  -0.0400 &&   0.1519  &  0.2414  &  0.2988 &   0.2868 &   0.3134  &  0.3356 \\[4pt]
\hline\\[8pt]
\end{tabular}
\caption{Prediction of Post-Intervention Outcomes $\mu_{g1}(x;1)-\mu(x;1)$ using the leading $K$ IMSE-optimal basis functions (top panel) and functional PC (bottom panel) as predictors. Entries correspond to correlation coefficients between actual mean at the site level and the baseline average (first column) and the prediction based on the leading $K$ basis functions (remaining columns).}
\normalsize
\end{table}

Table \ref{fig:mu1_pred_corr} reports the correlation coefficient between the predicted model shift for the average effect at site $g$, $\widehat{\Delta\mu_{g1K}}:=\frac1{n_g}\sum_{i=1}^{n_g}\sum_{k=1}^K\hat{t}_{gk}\hat{\psi}_k(x_{gi})$ with its post-hoc empirical counterpart, $\widehat{\Delta\mu_{g1}}:=\frac1{n_g}\sum_{i=1}^{n_g}(Y_{gi1}-\hat{\mu}_1(X_{gi}))$. A natural alternative strategy would be to predict post-intervention outcomes using separate regression estimates stratified by average pre-intervention outcomes. In the first column we therefore report correlation coefficients with the corresponding predictors as a benchmark, where sites were binned into three groups of equal size (terciles) according to average enrollment at baseline.

According to our results, optimal basis functions result in substantially more precise predictions relative to binned estimates and standard FPC, where gains are largest for the first two basis functions, and then plateau for 3 or more components. For example for cross-site prediction, we find a correlation coefficient of around $0.36$ (corresponding to an R-square of $0.13$) after using only the leading baseline function ($K=1$), which still gradually improves as additional terms are included. For $K$ larger than 5 or 6, terms are fairly noisily estimated and therefore do not lead to substantial additional improvements. As expected, the strength of correlation for cross-study extrapolation is lower than for cross-site prediction, but still substantial. Stratified estimation by pre-intervention \emph{levels} of outcomes does not appear to extract much predictive information at all, suggesting that the gains observed for our estimator exploit information on how outcomes vary together with covariates at each site.

\begin{table}\label{fig:mu1_pred_studies_opt}
\footnotesize
\begin{tabular}{lcrcrrrrrrrrrr}
&&\multicolumn{8}{c}{Cross-Site Prediction}\\
\hline\hline\\[4pt]
Study&\hspace{0.5cm}&actual&&K=1&K=2&K=3&K=4&K=5&K=6\\[4pt]
\hline\\[2pt]
MEX&&   -0.0077 &&  -0.0006 &  -0.0033 &  -0.0015 &  -0.0018 &  -0.0018 &  -0.0033 \\
MOR&&    0.0220 &&  -0.0016 &   0.0259 &   0.0231 &   0.0228 &   0.0210 &   0.0266 \\
IDN&&    0.0037 &&  -0.0003 &   0.0053 &   0.0094 &   0.0086 &   0.0059 &   0.0133 \\
KEN&&    0.0300 &&   0.0353 &   0.0356 &   0.0432 &   0.0591 &   0.0580 &   0.0694 \\
ECU&&   -0.0495 &&  -0.0172 &  -0.0170 &  -0.0230 &  -0.0198 &  -0.0217 &  -0.0252 \\[4pt]
\hline\\[8pt]
&&\multicolumn{8}{c}{Cross-Study Prediction}\\
\hline\hline\\[4pt]
MEX&&   -0.0077 &&  -0.0064 &  -0.0014 &   0.0007 &   0.0009 &   0.0004 &   0.0012 \\
MOR&&    0.0220 &&  -0.0008 &   0.0108 &   0.0213 &   0.0161 &   0.0238 &   0.0290 \\
IDN&&    0.0037 &&  -0.0004 &  -0.0006 &   0.0022 &   0.0014 &  -0.0002 &   0.0016 \\
KEN&&    0.0300 &&   0.0230 &   0.0200 &   0.0283 &   0.0441 &   0.0405 &   0.0412 \\
ECU&&   -0.0495 &&  -0.0165 &  -0.0150 &  -0.0201 &  -0.0186 &  -0.0128 &  -0.0144 \\[4pt]
\hline\\[8pt]
\end{tabular}

\caption{Predictions of post-intervention outcomes $\mu_{g1}(x;1)-\mu(x;1)$ using the leading $K$ IMSE optimal basis functions as predictors. Rows correspond to averages across sites for each of the five studies (Progresa, Tayssir, PKH, CT-OVC, BDH), ``actual" corresponds to the empirical mean of $\mu_{g1}(x;1)-\mu(x;1)$.}

\normalsize

\end{table}

In Table \ref{fig:mu1_pred_studies_opt}, we compare cross-site averages of predictions, where we let $\mathcal{G}_s$ denote the subset of $\{1,\dots,G\}$ corresponding to sites that were part of study $s=1,\dots,5$. For each study $s$ we then compare $\frac1{|\mathcal{G}_s|}\sum_{g\in\mathcal{G}_s}(\hat{\mu}_{g1K}-\hat{\mu}_1)$ to their ``realized" empirical counterparts, $\frac1{|\mathcal{G}_s|}\sum_{g\in\mathcal{G}_s}(\hat{\mu}_{g1}-\hat{\mu}_1)$. We find that the predicted average outcomes reflect some of the systematic differences, although especially for BDH and CT-OVC, the numbers and sizes of clusters are smaller, so results are likely noisier than for the first three studies. It should also be noted that the baseline outcome $Y_{gi0}$ is already included as a control for post-intervention outcomes in the specification of $\mu_1$. Without controlling for state-dependence at the individual level (not reported here), the correlation between pre- and post-intervention outcomes at the site-level is substantially stronger, but the relative comparison between using baseline averages as the ``naive" predictor and prediction using $K$ estimated basis functions is qualitatively similar.

\begin{table}\label{fig:mu1_pred_studies_pc}
\footnotesize
\begin{tabular}{lcrcrrrrrrrrrr}
&&\multicolumn{8}{c}{Cross-Site Prediction}\\
\hline\hline\\[4pt]
Study&\hspace{0.5cm}&actual&&K=1&K=2&K=3&K=4&K=5&K=6\\[4pt]
\hline\\[2pt]
MEX&&   -0.0077 &&   0.0089 &  -0.0049 &  -0.0051 &  -0.0040 &  -0.0036 &  -0.0034 \\
MOR&&    0.0220 &&  -0.0042 &   0.0258 &   0.0285 &   0.0271 &   0.0271 &   0.0281 \\
IDN&&    0.0037 &&   0.0043 &   0.0090 &   0.0221 &   0.0166 &   0.0255 &   0.0213 \\
KEN&&    0.0300 &&   0.0074 &   0.0550 &   0.0943 &   0.0921 &   0.0912 &   0.0961 \\
ECU&&   -0.0495 &&   0.0047 &  -0.0138 &  -0.0182 &  -0.0213 &  -0.0251 &  -0.0371 \\[4pt]
\hline\\[8pt]
&&\multicolumn{8}{c}{Cross-Study Prediction}\\
\hline\hline\\[4pt]
MEX&&   -0.0077 &&   0.0005 &  -0.0032 &  -0.0066 &  -0.0013 &  -0.0021 &  -0.0018 \\
MOR&&    0.0220 &&  -0.0011 &  -0.0043 &   0.0286 &   0.0373 &   0.0364 &   0.0364 \\
IDN&&    0.0037 &&   0.0035 &   0.0077 &   0.0174 &   0.0105 &   0.0168 &   0.0125 \\
KEN&&    0.0300 &&   0.0041 &   0.0340 &   0.0604 &   0.0543 &   0.0505 &   0.0507 \\
ECU&&   -0.0495 &&   0.0053 &  -0.0118 &  -0.0152 &  -0.0192 &  -0.0366 &  -0.0370 \\[4pt]
\hline\\[8pt]
\end{tabular}
\caption{Predictions of post-intervention outcomes $\mu_{g1}(x;1)-\mu(x;1)$ using the leading $K$ functional PC as predictors. Rows correspond to averages across sites for each of the five studies (Progresa, Tayssir, PKH, CT-OVC, BDH), ``actual" corresponds to the empirical mean of $\mu_{g1}(x;1)-\mu(x;1)$.}

\normalsize

\end{table}

We next repeat the same analysis using the respective functional PC for $\hat{\mu}_{g0}(x)$ and $\hat{\mu}_{g1}(x)$, see Table \ref{fig:mu1_pred_studies_pc}. Since the general patterns of school attendance as a function of child and household attributes were unlikely to have shifted fundamentally between baseline and follow-up, and the effect of the intervention was sizeable but incremental, we should expect the functional PC for the baseline to be fairly closely aligned with those at follow-up, and therefore perform very well as predictors for post-intervention outcomes. This is confirmed by the quantitative results, where performance is very similar to the IMSE optimal predictors, likely within or close to the margin of error, although we do not formally quantify estimation error for these results.

\begin{figure}\centering\label{fig:mu1_pred_IMSE}
\includegraphics[scale=0.6]{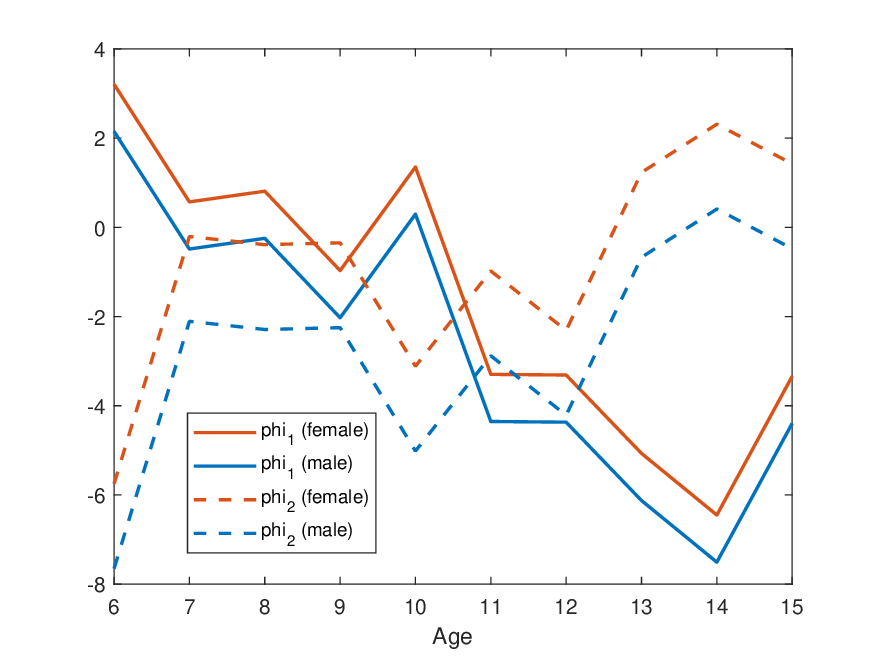}\includegraphics[scale=0.6]{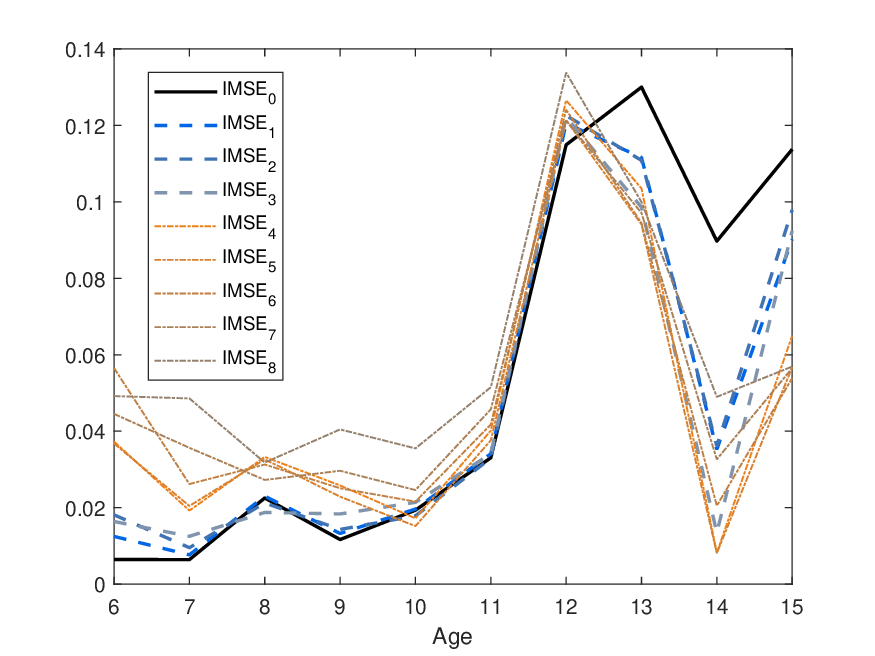}
\caption{Optimal basis functions for predicting $\mu_{g1}(x)$ (left), and conditional MSE of prediction given age using 0 to 8 basis functions (right).}
\end{figure}
In Figure \ref{fig:mu1_pred_IMSE}, we report estimates of the leading two leading optimal basis functions for predicting conditional post-intervention outcomes. These basis functions do not appear to vary much with income, so we plot $\phi_1,\phi_2$ only as functions of gender and age alone. Since post-intervention outcomes are also observed at all treated sites, we also plot the conditional mean square error for predicting the post-intervention response for those sites, where $IMSE_0$ corresponds to the case in which we use the unadjusted cross-site average as a predictor, and $IMSE_k$ for the prediction using the first $k$ basis functions $\phi_1,\dots,\phi_k$ as predictors. While the predictors appear to be responsive to differences in enrollments at young and old ages, most of the improvement in the forecast is for enrollment at ages 12 and above, where (within and across site) variation is generally highest. Most of the improvement in the conditional forecast results from including the first two factors, whereas additional predictors lead to a significant deterioration of the forecast at lower ages. This is in line with the number $K=2$ of factors selected by cross-validation.

\begin{figure}\centering\label{fig:mu1_pred_scores}
\includegraphics[scale=0.8]{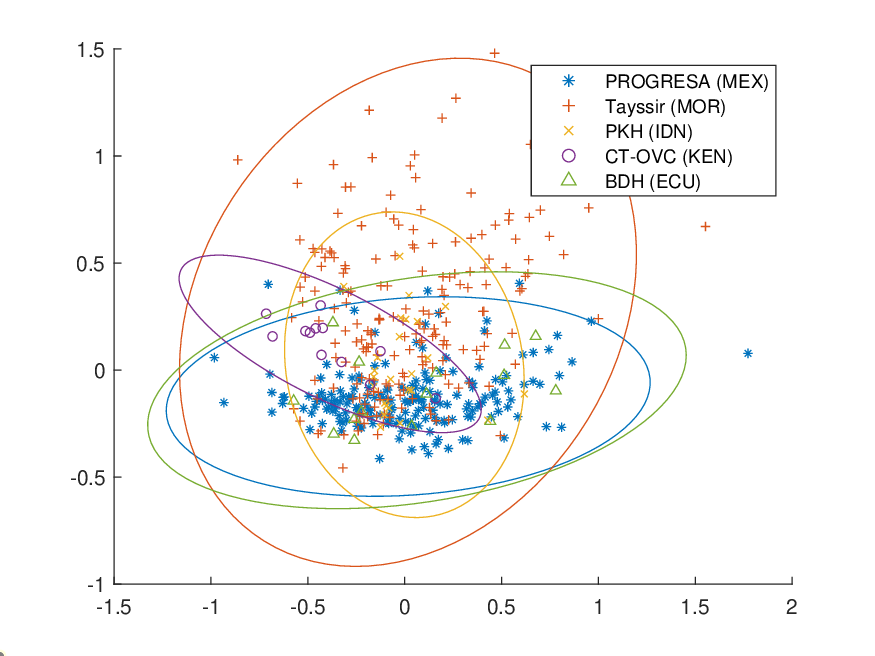}
\caption{Estimated scores for leading $K=2$ IMSE-optimal predictors of $\mu_{g1}(x)$.}
\end{figure}

One important question is whether these features constructed based on their predictive power capture systematic differences between the five different study countries (Mexico, Morocco, Indonesia, Kenya, Ecuador), but also whether there is substantial overlap between those populations. The latter is especially important since we use extrapolation that is linear in those features $\phi_k$. Figure \ref{fig:mu1_pred_scores} plots the estimated scores corresponding to the leading two basis functions, $\hat{m}_{g1},\hat{m}_{g2}$, for each site. To visualize differences in the factor loadings between the five countries included in our analysis, we also plot study-specific variance ellipses corresponding to a 80 percent confidence set for jointly normal variates. We can see that while there is substantial overlap in the support, their distributions vary substantially across the five studies, with especially some sites in the BDH and CT-OVC differing quite substantially from those in the other three studies.

\begin{table}\label{fig:tau_pred_study_opt}
\footnotesize
\begin{tabular}{lcrcrrrrrrrrrr}
&&\multicolumn{8}{c}{Cross-Site Prediction}\\
\hline\hline\\[4pt]
Study&\hspace{0.5cm}&actual&&K=1&K=2&K=3&K=4&K=5&K=6\\[4pt]
\hline\\[2pt]
MEX&&    0.0011 &&  -0.0006 &  -0.0019 &  -0.0011 &  -0.0028 &  -0.0029 &  -0.0031  \\
MOR&&    0.0029 &&  -0.0002 &  -0.0021 &   0.0161 &   0.0198 &   0.0213 &   0.0178  \\
IDN&&   -0.0173 &&  -0.0022 &  -0.0135 &  -0.0455 &  -0.0570 &  -0.0564 &  -0.0577  \\
KEN&&   -0.0506 &&  -0.0394 &  -0.0301 &  -0.0253 &  -0.0245 &  -0.0197 &  -0.0198  \\
ECU&&   -0.0264 &&  -0.0005 &  -0.0053 &  -0.0057 &  -0.0076 &  -0.0083 &  -0.0107  \\[4pt]
\hline\\[8pt]
&&\multicolumn{8}{c}{Cross-Study Prediction}\\
\hline\hline\\[4pt]
MEX&&    0.0005 &&  -0.0088 &  -0.0046 &  -0.0039 &  -0.0013 &   0.0040 &   0.0036 \\
MOR&&    0.0021 &&  -0.0015 &   0.0005 &   0.0309 &   0.0306 &   0.0236 &   0.0154 \\
IDN&&   -0.0170 &&  -0.0009 &  -0.0050 &  -0.0169 &  -0.0286 &  -0.0344 &  -0.0247 \\
KEN&&   -0.0522 &&  -0.0540 &  -0.0470 &  -0.0376 &  -0.0273 &  -0.0199 &  -0.0191 \\
ECU&&   -0.0258 &&   0.0023 &  -0.0062 &  -0.0081 &   0.0073 &   0.0033 &   0.0055 \\[4pt]

\hline\\[8pt]
\end{tabular}
\caption{Prediction of Conditional ATE $\tau_{g}(x)-\tau(x)$ averaging across sites for each of the five studies (Progresa, Tayssir, PKH, CT-OVC, BDH), using the leading $K$ IMSE optimal basis functions as predictors. ``Actual" corresponds to the empirical mean of $\tau_g(x)-\tau(x)$.}
\normalsize
\end{table}

\begin{figure}\label{fig:tau_pred_scores}\centering
\includegraphics[scale=0.8]{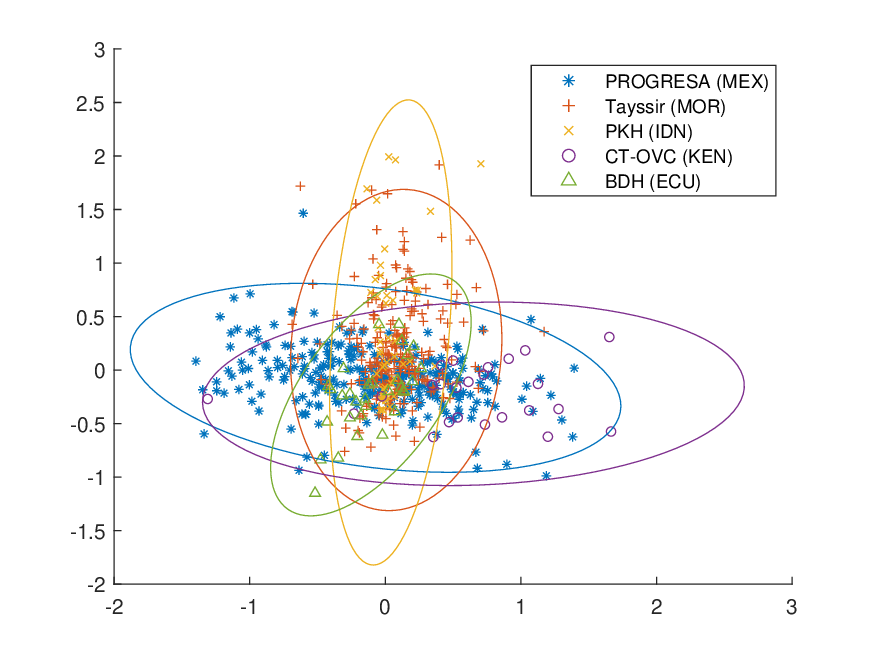}
\caption{Estimated scores for leading $K=2$ IMSE-optimal predictors of $\tau_g(x)$.}
\end{figure}

We next repeat the analysis for prediction of \textbf{model shifts in site-specific treatment effects} $\tau_g(x)-\tau(x)$, both using IMSE-optimal basis functions and functional PC as predictor (Tables \ref{fig:tau_pred_study_opt}). These results were obtained combining the data from the 434 treated and 206 control sites, and since the covariance operator of $\mu_g(x;0)$ with the baseline is estimated using only data from the substantially smaller control group, we should expect the resulting estimates to be less precise than for predicting post-intervention average outcomes.

In all five RCTs, treatment was randomized at the cluster level, so we can't directly assess the performance of either type of predictor at that level, but we can still aggregate actual and predicted effects at the level of the study. Here, average predictions based on the IMSE optimal basis functions match the sign and approximate magnitudes of post-hoc realized effects for all values of $K$, whereas at least 3 or 4 functional principal components appear to be necessary to match at least some qualitative aspects of study-level averages. We also report the estimated scores for predicting conditional ATEs plotted in Figure \ref{fig:tau_pred_scores}.

For any of these comparisons, it should also be noted that both types of predictions are based on the unconfounded location assumption (Assumption \ref{unconf_loc_ass}), whereas realized conditional effects also reflect systematic differences between studies that can't be predicted by extrapolating intra-study variation among sites. Most importantly, the five studies differ in terms of the exact implementation of the incentive, and also country specific factors. Most importantly, cash transfer for CT-OVC in Kenya and BDH in Ecuador were unconditional, whereas transfers under Progresa, PKH, and Tayssir were conditioned on, or connected to, the child's enrollment in school. While cross-site average treatment effects for those two studies were indeed substantially lower than the cross-study average (first column in Table \ref{fig:tau_pred_study_opt}), our method appears to replicate most of that difference for the Kenyan sites, whereas it fails to reproduce the deviation from the cross-study average only for Ecuador.

\begin{table}
\footnotesize
\begin{tabular}{lcrcrrrrrrrrrr}
&&\multicolumn{8}{c}{Cross-Site Prediction}\\
\hline\hline\\[4pt]
Study&\hspace{0.5cm}&actual&&K=1&K=2&K=3&K=4&K=5&K=6\\[4pt]
\hline\\[2pt]
MEX&&    0.0011 &&  -0.0008 &  -0.0014 &  -0.0015 &  -0.0018 &  -0.0020 &  -0.0027 \\
MOR&&    0.0029 &&   0.0004 &   0.0153 &   0.0150 &   0.0157 &   0.0160 &   0.0164 \\
IDN&&   -0.0173 &&   0.0003 &  -0.0002 &  -0.0160 &  -0.0169 &  -0.0167 &  -0.0245 \\
KEN&&   -0.0506 &&  -0.0321 &  -0.0308 &  -0.0202 &  -0.0152 &  -0.0151 &  -0.0132 \\
ECU&&   -0.0264 &&   0.0018 &  -0.0025 &  -0.0018 &  -0.0095 &  -0.0182 &  -0.0185 \\[4pt]
\hline\\[8pt]
&&\multicolumn{8}{c}{Cross-Study Prediction}\\
\hline\hline\\[4pt]
MEX&&    0.0005 &&  -0.0005 &  -0.0010 &  -0.0027 &  -0.0022 &  -0.0027 &  -0.0026 \\
MOR&&    0.0021 &&  -0.0003 &  -0.0073 &  -0.0288 &   0.0028 &   0.0034 &   0.0031 \\
IDN&&   -0.0170 &&  -0.0001 &   0.0018 &  -0.0022 &   0.0072 &   0.0070 &   0.0075 \\
KEN&&   -0.0522 &&  -0.0299 &  -0.0297 &  -0.0021 &   0.0031 &   0.0074 &   0.0052 \\
ECU&&   -0.0258 &&   0.0047 &  -0.0005 &   0.0047 &   0.0073 &  -0.0039 &  -0.0038 \\[4pt]
\hline\\[8pt]
\end{tabular}
\caption{\label{fig:tau_pred_study_pc}Prediction of Conditional ATE $\tau_{g}(x)-\tau(x)$ averaging across sites for each of the five studies (Progresa, Tayssir, PKH, CT-OVC, BDH), using the leading $K$ functional PC as predictors. ``Actual" corresponds to the empirical mean of $\tau_g(x)-\tau(x)$.}
\normalsize
\end{table}

\section{Conclusion}

We investigate how to exploit observed between-site variation within one or several studies to predict outcomes using baseline data for new ``target" sites. The premise of our approach is that agent responses at the micro level follow some universal patterns across study populations. These responses are generally confounded by site-specific factors of an unknown structure, but cross-sectional patterns of attributes and outcome at baseline for each site typically contain useful information regarding those environmental factors in a target site, and may help identify ``comparable" sites in the experimental sample. We chose to focus on a nonparametric, linear version of the problem primarily for clarity and ease of implementation, and nonseparable or structural models with sufficiently flexible specifications of site-specific heterogeneity may be another fruitful approach to this problem.

We give a finite-population formulation for the statistical problem of evaluating out of sample forecast performance. We define the target for the transfer estimate as a pseudo-true parameter which reflects the relevant information regarding likely outcomes at the target site that may be learned from previously observed contexts. The corresponding prediction problem is equivalent to functional regression, but given the limited number of sites can only estimate heavily regularized version of the problem. We therefore choose a regularization approach that targets a small number of ``most predictive" features of the distribution of outcomes in the baseline. Those optimal predictors are solutions to a generalized eigenvalue problem in terms of the covariance operators of $\mu_g$ and $\tau_g$. The approach can be adapted to sparsely or densely sampled sites, as well as randomization within or between clusters, resulting in different convergence rates.

\appendix

\section{Asymptotic Theory}
\label{sec:asymptotics}

This section gives convergence rates for the estimated eigenfunctions $\hat{\phi}_k,\hat{\psi}_k$ and the resulting integrated mean square error relative to the optimal benchmark $IMSE_K^*$. We consider rates as both $G$ and $n_g\equiv n$ grow to infinity, possibly at different rates. We first state results for the case of densely sampled clusters, $n\rightarrow\infty$ based on local linear estimators for the mean and covariance functions. We then discuss estimation using B-splines, and rates for the sparsely sampled case as well as site-based randomization. 

\subsection{Nonparametric Estimation of Covariance Functions}

We first derive convergence rates for the local linear estimator of the covariance function $H(x_1,x_2;d_1,d_2)$. Following \cite{Han08}, we assume the following regarding the kernel function

\begin{ass}\label{kernel_ass}\textbf{(Kernel Function)} The multivariate kernel function $K:\mathbb{R}^d\rightarrow\mathbb{R}$ satisfies (a) $|K(u)|\leq\bar{K}<\infty$ and $\int|K(u)|du\leq\mu<\infty$. Furthermore, (b) $K(u)$ is differentiable and there exist $\Lambda_1,L<\infty$ and $\nu>1$ such that $\left\|\nabla_uK(u)\right\|\leq\Lambda_1\min\left\{1,\|u\|^{-\nu}\right\}$ for $\|u\|>L$. (c) The first two moments of the kernel satisfy the conditions $\int K(u)du = 1$, $\int uK(u)du=0$, and $\int uu'K(u)du=\Omega$, a finite, positive definite matrix.
\end{ass}

Parts (a) and (c) are fairly standard in the literature, the bound in part (b) is important for the uniformity arguments. These assumptions are satisfied by commonly used kernel functions such as the Gaussian or Epanechnikov kernel. \cite{Han08} and \cite{GNP21} consider an alternative set of conditions for kernel functions with bounded support that need not be differentiable, which could be used to replace part (b).

We consider uniform convergence over a compact subset of the support of $X$, without loss of generality $[0,1]^d$. We can now state our main result regarding the rate of consistency for the local linear estimator for the conditional mean functions and their covariance kernel:

\begin{thm}\label{loc_lin_consistency_thm} Suppose that Assumptions \ref{unconf_ass_ass}-\ref{loc_lin_ass} and \ref{kernel_ass} hold. Furthermore, the bandwidth $h$ satisfies $\left(\frac{n}{\log n}\right)^{1/3} h^d\rightarrow\infty$ and $\left(\frac{\log n}{Gnh^d}\right)^{1/2}\rightarrow0.$ Then  the local linear estimators for $\mu(x;\mathbf{d})$ and the covariance operator $H_{d_1d_2}(x_1,x_2)$ are consistent at the rate
\begin{eqnarray}
\nonumber\sup_{d,x_1}\left|\hat{\mu}(x_1,d)-\mu(x_1,d)\right|&=&O_p\left(r_{Gn}\right)\\
\nonumber\sup_{d_1,d_2,x_1,x_2}\left| \hat{H}(x_1,x_2;d_1,d_2)-H(x_1,x_2;d_1,d_2)\right|&=&O_p\left(r_{Gn}\right)
\end{eqnarray}
where $r_{Gn} = \frac1{G} + h^2 + \left(\frac{\log n}{Gnh^d}\right)^{1/2}$, and the suprema are for $x_1,x_2\in[0,1]^d$. The rate optimal bandwidth minimizing the second term of either error is of the order $h^*=O\left(\left(\frac{\log n}{Gn}\right)^{\frac1{4+d}}\right)$, resulting in a rate $\frac1G + \left(\frac{\log n}{Gn}\right)^{\frac2{4+d}}$.
\end{thm}

The proof uses uniform consistency results by \cite{Han08} and \cite{GNP21} for nonparametric regression using cross-sectional and dyadic data and is provided in the appendix. The convergence rate consists of two separate contributions: the first component represents the error from estimating moments from the $G-1$ experimental clusters, which excludes the single target cluster. The rate for this component is of the order $O_p(G^{-1})$ rather than its square root, owing to the fact that the estimands are the mean and covariance function for the sample of $G$ sites, $G-1$ out of which are used for estimation. If instead the population of interest consisted of a greater number additional target clusters growing at least at the order of $G$ or an infinite meta-population, the rate on that leading term would instead be $O(G^{-1/2})$.

The second component represents the sampling error in nonparametric estimation of conditional mean functions in each cluster, where the optimal bandwidth sequence balances the respective rates for the smoothing bias and sampling error. It is also instructive to compare the rate to that in Theorem 1 of \cite{YMW05} who consider the case of sparse (finite-$n$) rather than dense samples from each site. For one the effective dimension for nonparametric estimation of the covariance kernel is only $d$ in our problem rather than $2d$, reflecting the fact that the contribution for each cluster is a U-statistic averaging over $\binom{n}{2}$ terms, so that allowing $n$ to grow results in a more favorable rate. A similar phenomenon was pointed out for nonparametric regression with dyadic data in \cite{GNP21}.

\subsection{Estimation of Basis Functions $\phi_k,\psi_k$}
We next consider convergence rates for the estimated eigenfunctions given a preliminary estimator for mean and covariance function for the conditional average treatment effect function $\mu_g(x)$. We distinguish between settings where the optimal predictors $\phi_k^*$ are well-defined in the absence of regularization according to (\ref{S_op_defn}) and the general case in which we consider estimation of $\phi_{ka}^*$ for the regularized version of the problem (\ref{S_op_reg}), and $\psi_{ka}^*$ given by Corollary \ref{opt_basis_cor}.

We first state result in terms of a generic nonparametric estimator and its convergence rate, both under the inner product norm $\|\cdot\|$ as well as the sup norm. Asymptotic rates based on the local linear estimator are given further below.

\begin{thm}\label{eig_cons_thm} Suppose the estimator $\hat{H}(x_1,x_2;d_1,d_2)$ is consistent with rate  \[\sup_{x_1,x_2,d_1,d_2}\left|\hat{H}(x_1,x_2;d_1,d_2)-H(x_1,x_2;d_1,d_2)\right|=O_p(r_{Gn}).\] Then the estimators for the generalized eigenvalue problem (\ref{S_op_reg}) with regularization parameter $a>0$ are consistent at respective rates $|\hat{\lambda}_{k}-\lambda_{ka}| = O_p\left(a^{-3/2}r_{Gn}\right)$. Furthermore, if the eigenvalue $\lambda_k$ is simple,
\begin{eqnarray}
\nonumber \|\hat{\psi}_{k}-\psi_{ka}\|&=&O_p\left(a^{-3/2}r_{Gn}\right)\\
\nonumber \|\hat{\phi}_{k}-\phi_{ka}\|&=&O_p\left(a^{-3/2}r_{Gn}\right)
\end{eqnarray}
If on the other hand the multiplicity of some $\lambda_k>0$ is $m+1$, i.e. $\lambda_{k-1}>\lambda_{k}=\dots=\lambda_{k+m}>\lambda_{k+m+1}$, then there exist orthonormal basis $\phi_{ka}^*,\dots,\phi_{(k+m)a}^*$ for $\textnormal{span}\{\phi_{ka},\dots,\phi_{(k+m)a}\}$ and
$\psi_{ka}^*,\dots,\psi_{(k+m)a}^*$ for $\textnormal{span}\{\psi_{ka},\dots,\psi_{(k+m)a}\}$ such that
\begin{eqnarray}
\|\hat{\psi}_{k}-\psi_{ka}^*\|=O_p\left(a^{-3/2}r_{Gn}\right)\\
\|\hat{\phi}_{k}-\phi_{ka}&(\|=O_p\left(a^{-3/2}r_{Gn}\right)
\end{eqnarray}
for each fixed $k=1,\dots,K$.
\end{thm}

The proof has a similar overall structure as that for results for functional principal components in \cite{GHR04} and \cite{YMW05} (Proposition 4.2 and Theorem 2, respectively), but requires some major adjustments. For the problem considered here, estimation of the operator itself requires regularization, and furthermore the rank of all covariance operators is less than or equal to $G$, treating the population of sites as fixed. We also allow the dimension of $x$ to be greater than one and some of the relevant eigenvalues need not be unique. While each of these adjustments is incremental and leaves the general structure of the argument unchanged, we provide a self-contained proof in Appendix \ref{sec:app_asymptotics_proofs} below.

For the local linear estimator we can immediately obtain the following from Theorems \ref{loc_lin_consistency_thm} and \ref{eig_cons_thm}:

\begin{cor} For each $k=1,\dots,K$, let $\hat{\psi}_k,\hat{\phi}_k$ and $\hat{\nu}_k,\hat{\lambda}_k$  the estimators for eigenfunctions and eigenvalues using local linear estimators for $\mu(x)$ and $H_{d_1d_2}(x_1,x_2)$ with a bandwidth sequence $h$ satisfying $\left(\frac{n}{\log n}\right)^{1/3} h^d\rightarrow\infty$ and $\left(\frac{\log n}{nh^d}\right)^{1/2}\rightarrow0.$ Under the assumptions for Theorem \ref{loc_lin_consistency_thm}, the conclusions of Theorem \ref{eig_cons_thm} hold for these estimators with the rate
\[r_{Gn}=\frac1{G} + h^2 + \left(\frac{\log n}{Gnh^d}\right)^{1/2}\]
\end{cor}

Similarly, we can give the asymptotic rate for estimating the best linear predictor for the conditional average treatment effect as defined in (\ref{pseudo_true_tau}), where the rates of individual components follow immediately from Theorem \ref{eig_cons_thm}.

\begin{cor}\label{cate_cons_thm} Suppose the estimators $\hat{\mu}(x;d)$ and  $\hat{H}_{d_1d_2}(x_1,x_2)$ are consistent with rates $\sup_{x_1,d_1}\left|\hat{\mu}(x_1;d_1)-\mu(x_1;d_1)\right|=O_p(r_{Gn})$ and  $\sup_{x_1,x_2,d_1,d_2}\left|\hat{H}(x_1,x_2;d_1,d_2)-H(x_1,x_2;d_1,d_2)\right|=O_p(r_{Gn})$. Then for any fixed value of $K$, the estimator $\hat{\tau}_g^K(x)$ based on (\ref{pseudo_true_tau}) is consistent at the rate
\[\sup_{x\in\mathcal{X}}\left|\hat{\tau}_g(x)-\tau_g^K(x)\right|=O_p\left(a^{-3/2}r_{Gn}\right)\]
\end{cor}

At present we do not derive the asymptotic distribution for functionals of $\hat{\tau}_g(x)$. Asymptotic normality of certain functionals of estimated eigenfunctions were derived in a different site by \cite{Chr17}. Whether his strategy of proof can be adapted to derive distributional results for our setup is left for future research.

We finally consider the case in which condition (\ref{kh_bound_condition}) holds and the solution to the unregularized problem (\ref{S_op_defn}) is well defined. It can be seen from the proof of Theorem \ref{eig_cons_thm} that regularization is generally still necessary for estimation of the optimal functions $\phi_1^*,\dots,\phi_K^*$, but we can give a rate with which the IMSE given the estimated functions converges to the lower bound in Lemma \ref{IMSE_star_lem}. Specifically, for any $K$ square-integrable functions $\phi_1,\dots,\phi_K$, we define
\[IMSE_K[\phi_1,\dots,\phi_K]:=\int\min_{B\in\hat{\mathcal{H}}_K\times L_2(\mathcal{X})}\mathbb{E}\left[(\tau_g(x)-BP_K\mu_g(x))^2\right]f_0(x)dx\]
where  $P_K:L_2(\mathcal{X})\rightarrow\mathcal{H}_K$ denotes the operator associated with orthogonal projection onto the closed linear subspace $\mathcal{H}_K:=\spn\left(\phi_1,\dots,\phi_K\right)$. Combining the rate results in Proposition \ref{C1_IMSE_rate_prp} and Theorem \ref{eig_cons_thm}, we can then give the following

\begin{cor}\label{IMSE_conv_rate} Suppose that the Assumptions of Theorems \ref{loc_lin_consistency_thm} and \ref{eig_cons_thm} hold along with Condition (\ref{kh_bound_condition}). For the local linear estimator with bandwidth $h^*=O\left(\left(\frac{\log n}{Gn}\right)^{\frac1{4+d}}\right)$ and regularization parameter $a>0$,
\[|IMSE_K[\hat{\phi}_1,\dots,\hat{\phi}_K]-IMSE_K^*|=O_P\left(a + a^{-3/2}r_{Gn}\right)\]
where $r_{Gn}:=\frac1G + \left(\frac{\log n}{Gn}\right)^{\frac2{4+d}}$.
\end{cor}
The rate for the IMSE in $a$ and $r_{Gn}$ suggests that this upper bound is minimized at a rate $a=r_{nG}^{2/5}$ for the regularization parameter. For methods of functional data analysis, it is common to determine regularization parameters by cross-validation (see e.g. \linebreak\cite{HMWY10}), however we do not formally establish consistency of such a method.

\subsection{Sparsely Sampled Clusters}

If the number of observed units in each cluster $n_g$ is bounded or grows at a slow rate relative to $G$, we have to adapt the strategy for estimating the mean and covariance functions of $\mu_g(x;d)$ along the lines proposed by \cite{YMW05} for the problem of functional principal components.

A challenge relative to the densely sampled case is that the site-specific distribution of attributes $f_g(x)$ can no longer be estimated consistently. Instead, we assume that the cluster-specific distribution of attributes $X_{gi}$ is either known or can be estimated parametrically with sufficient precision from the $n_g$ observations in each cluster, so that individual observations can be reweighted accordingly to match the target distribution $f_0(x)$ in each cluster.\footnote{For the combined studies of conditional cash transfers in the empirical application, this is unproblematic for some of the demographic variables, including the child's age and gender. On the other hand, the means and variances of log per capita household expenditure vary between sites, but separate histograms for each study suggest that the log-normal distribution provides a plausible parametric approximation to the marginal distribution.}

Specifically, let
\[\hat{\mu}(x;d):=\arg_{b_0}\min_{b_0,b_1}\sum_{g=1}^G\sum_{i=1}^{n_g}w_{gi}(x;d,h_{\mu})
(Y_{gi} - b_0 - b_1(x-X_{gi}))^2\]
for $\mu(x;d)$. As before, $K(u)$ is a kernel function satisfying Assumption \ref{kernel_ass}, and the bandwidth $h>0$ is chosen according to sample size $G,n_g$, the dimension of $X_{gi}$, and assumed smoothness of the estimands. The weights $w_{gi}(x;d,h):=\dum\{D_{gi}=d\}\frac{f_0(x)}{f_g(x)}K\left(\frac{X_{gi}-x}{h}\right)$ are assumed to be known.
We also let
\[\hspace*{-0.5cm}\hat{H}(x_1,x_2;d_1,d_2):=\arg_{b_{0}^{(g)}}\min_{b_{0}^{(g)},b_{11}^{(g)},b_{12}^{(g)}}\sum_{g=1}^G\sum_{j\neq i}H_{gij}(x_1,x_2,\mathbf{b})
w_{gi}(x_1;d,h_H)w_{gj}(x_2;d,h_H)\]
for an appropriately chosen bandwidth $h_H$, where \[H_{gij}(x_1,x_2,\mathbf{b}):=\left(Y_{gi}Y_{gj}-b_0^{(g)}-b_{11}^{(g)}(X_{gi}-x_1) - b_{12}^{(g)}(X_{gj}-x_2)\right)^2.\]

Then using arguments parallel to the proof of Theorem \ref{loc_lin_consistency_thm}, the convergence rates of the local linear estimators for $\mu(x;\mathbf{d})$ and the covariance operator $H_{d_1d_2}(x_1,x_2)$ are given by
\begin{eqnarray}
\nonumber\sup_{d,x_1}\left|\hat{\mu}(x_1,d)-\mu(x_1,d)\right|&=&O_p\left(r_{G\mu}\right)\\
\nonumber\sup_{d_1,d_2,x_1,x_2}\left| \hat{H}(x_1,x_2;d_1,d_2)-H(x_1,x_2;d_1,d_2)\right|&=&O_p\left(r_{GH}\right)
\end{eqnarray}
where the sup is taken over $x_1,x_2\in[0,1]^d$ and $r_{G\mu} = h_{\mu}^2 + \left(\frac{\log G}{Gh_{\mu}^d}\right)^{1/2}$ and
$r_{GH} = h_H^2 + \left(\frac{\log G}{Gh_H^{2d}}\right)^{1/2}$. A similar result was proven under slightly different conditions in Theorem 1 by \cite{YMW05}.

Comparing these rates to those for the densely sampled case in Theorem \ref{loc_lin_consistency_thm}, the contributions of order $\frac1G$ are now unambiguously dominated by the remaining errors. Furthermore, in the rate for the covariance kernel, the bandwidth $h_H$ now appears at the power $h_H^{2d}$ (compared to $h_H^d$ in the dense case). This is a consequence of the fact that the number of unit pairs in each cluster no longer increases to infinity under the sparsely sampled case. As a result, the optimal bandwidths for estimating $\mu(x;d)$ and $H(x_1,x_2;d_1,d_2)$ are generally at different rates and should therefore also be chosen separately in this case. As in the densely sampled case, we can then use Theorem \ref{eig_cons_thm} to determine rates for eigenfunctions and eigenvalues. Unbiased estimators for the scores $m_{gk},t_{gk}$ are also available, however consistent estimation requires $n_g$ growing large for the sites of interest.

\subsection{Cluster-Based Randomization}

We can also adapt the approach to the scenario in which treatment assignment is randomized at the cluster level, but a baseline survey of pre-intervention outcomes is available. Specifically, we let $D_{gi}\equiv D_g$ be the assigned treatment for all units in cluster $g$, and $Y_{git}(d)$ denote the potential value for the policy variable $D_{git}=d$ unit $i$ at site $g$ in period $t=0,1$. We then consider the problem of predicting post-intervention conditional average treatment effects
\[\tau_{g1}(x):=\mathbb{E}\left[Y_{gi1}(1)-Y_{gi1}(0)|X_{gi}=x\right]\]
based on $\mu_{g0}(x):=\mathbb{E}\left[Y_{gi0}(0)|X_{gi}=x\right]$, where the covariates $X_{gi}$ are also assumed to be measured at baseline.

With cluster-level randomization, we only observe potential values $Y_{gi1}(1)$ for all units in a cluster $g$ assigned to treatment, $D_g=1$, or only $Y_{gi1}(0)$ for all units in a control cluster with $D_g=0$. However under ignorable assignment,
\[D_g \ind Y_{gi0}(0),Y_{gi1}(0),Y_{gi1}(1)|X_{gi}\]
and independent assignment across clusters, we can estimate the covariance function\linebreak $H_{\mu\tau}(x_1,x_2):=\cov(\mu_{g0}(x_1),\tau_g(x_2))$ consistently as $G\rightarrow\infty$ via
\[\hat{H}_{\mu\tau}(x_1,x_2):=\hat{H}(x_1,x_2;0,1)-\hat{H}(x_1,x_2;0,0).\]
where $\hat{H}(x_1,x_2;0,1)$ and $\hat{H}(x_1,x_2;0,0)$ are nonparametric estimators for $\cov(\mu_{g0}(x_1),\mu_{g1}(x_2;1))$ and\linebreak  $\cov(\mu_{g0}(x_1),\mu_{g1}(x_2;0))$, respectively, obtained separately from the treatment and control clusters.

With minor modifications of the proofs of Theorem \ref{loc_lin_consistency_thm} and Theorem 1 in \cite{YMW05} we can then obtain the convergence rates $r_{Gn} = \frac1{\sqrt{G}} + \delta_{Gn}^{-1}\left(h^2 + \left(\frac{\log n}{Gnh^d}\right)^{1/2}\right)$
for the densely sampled case where $n_g\equiv n\rightarrow\infty$ for each cluster. For the sparsely sampled case where $n_g$ remains fixed, we obtain $r_{G\mu} = h_{\mu}^2 + \left(\frac{\log G}{Gh_{\mu}^d}\right)^{1/2}$ and $r_{GH} =h_H^2 + \left(\frac{\log G}{Gh_H^{2d}}\right)^{1/2}$, so that the rates coincide with the case of within cluster-randomization. The eigenfunctions in (\ref{S_op_reg}) can then be obtained from Theorem \ref{eig_cons_thm} as before.

\subsection{Estimation using B-Splines}

This section contains additional results for nonparametric estimation of the mean and covariance functions using B-splines as a convenient alternative to kernel estimation. As before, we consider estimation at covariate values on a compact subset $\mathcal{X}^*$ of $\mathbb{R}^d$, without loss of generality, $\mathcal{X}^*\equiv [0,1]^d$. We first define the B-spline basis functions, following the exposition in \cite{CCh15}.

We first partition the support of the $d'$th dimension of the continuously distributed components of $X_{gi}$ with $M_d$ knots $0\equiv t_{0d'}<...<t_{M_{d'}d'}\equiv 1$ for each $d'=1,\dots,d$. For the B-spline of order $r>1$ we also set $t_{-(r-1)d'}=t_{-(r-2)d'}=\dots\equiv t_{0d'}$, and for notational simplicity we only consider the case $M_{d'}\equiv M$ and $t_{jd'}\equiv t_j$ for all $d'$ and assume that the mesh ratios for successive spline bases, $\max_{0\leq j\leq M}\{t_{j+1}-t_j\}/\min_{0\leq j\leq M}\{t_{j+1}-t_j\}$ are uniformly bounded for all $M=1,2,\dots$.

The univariate basis functions are then defined according to
\begin{eqnarray}
\nonumber N_{j,1}(x)&:=&\dum\{t_{j}\leq x<t_{j+1}\}\\
\nonumber N_{j,r'}(x)&:=&\frac{x-t_j}{t_{j+r'-1}-t_j}N_{j,r'-1}(x) + \frac{t_{j+1}-x}{t_{j+r'}-t_{j+1}}N_{j+1,r'-1}(x)
\end{eqnarray}
recursively for $r'=1,\dots,r$ and all $j$ and we set $\frac10\equiv 0$. The resulting function $N_{j,r}(x)$ are piecewise polynomial of order $r$ and continuously differentiable up to order $r-1$. After rescaling we denote the basis functions for a particular choice of $r$ and $M$ with
\[b_{j}^M(x):=\sqrt{M+r}N_{j,r}(x)\]
and denote
\begin{eqnarray}\nonumber\mathbf{b}^{M}(x;d_1)&:=&\left(\frac{}{}b_{-(r-1)}(x)\dum\{d_1=0\},\dots,b_{M}(x)\dum\{d_1=0\},\right.\\
\nonumber&&\left.\frac{}{}b_{-(r-1)}(x)\dum\{d_1=1\},\dots,b_{M}(x)\dum\{d_1=1\}\right).
\end{eqnarray}
where the argument $d_1\in\{0,1\}$ corresponds to the treatment indicator.

Noting that we chose the same univariate basis for each dimension, we then define the multivariate spline bases
\[\mathbf{b}^{M,1}(\mathbf{x}_1;d_1):=\bigotimes_{d'=1}^d\mathbf{b}^{M}(x_{1d'};d_1)\]
and
\[\mathbf{b}{M,2}(x_1,x_2;d_1,d_2):=\mathbf{b}^{M}(x_1;d_1)\otimes \mathbf{b}^{M}(x_2;d_2)\]
by forming the tensor product of the univariate spline bases, so the resulting dimension of $\mathbf{b}^{M,1}$ is $(M+r)^d$. As for any linear sieve, it is also straightforward to impose additive separability among dimensions of $X_{gi}$ by omitting all interaction terms among the corresponding univariate basis functions from this tensor product.

Given the sample $X_{g1},\dots,X_{gn_g}$ for the $g$th site, we construct the matrices \[\mathbf{B}_{g1}:=(\mathbf{b}_1^M(X_{g1};D_{g1}),\dots,\mathbf{b}_1^M(X_{gn_g};D_{gn_g}))\] and \[\mathbf{B}_{g2}:=(\mathbf{b}_2^{M,2}(X_{g1},X_{g2};D_{g1},D_{g2}),\dots,\mathbf{b}_2^{M,2}(X_{gn_g-1},X_{gn_g};D_{gn_g-1},D_{gn_g})).\] We can then define the site-specific estimates
\begin{eqnarray}
\nonumber\hat{\mu}_g(x_1;d_1)&:=&\mathbf{b}_w^{M,1}(x_1;d_1)\left(\mathbf{B}_{g1}'\mathbf{B}_{g1}\right)^{-1}\mathbf{B}_{g1}'\mathbf{Y}_g\\
\nonumber&=:&\frac1{n_g}\sum_{i=1}^nm_{gi}(x_1;d_1)\\
\nonumber\hat{\Eta}_g(x_1,x_2;d_1,d_2)&:=&\mathbf{b}_w^{M,2}(x_1,x_2;d_1,d_2)\left(\mathbf{B}_{g2}'\mathbf{B}_{g2}\right)^{-1}\mathbf{B}_{g2}'
((\mathbf{Y}_g-\mu_g)\otimes(\mathbf{Y}_g-\mu_g))\\
\nonumber&=:&\binom{n_g}2^{-1}\sum_{1\leq i<j\leq n_g}H_{n_gij}(x_1,x_2;d_1,d_2)
\end{eqnarray}
The resulting B-spline estimators for the mean and covariance functions are then given by
\[\hat{\mu}(x_1;d_1) \equiv \frac1G\sum_{g=1}^G \hat{\mu}_g(x_1;d_1)\] and \[\hat{\Eta}(x_1,x_2;d_1,d_2)\equiv\frac1G\sum_{g=1}^G\hat{\Eta}_g(x_1,x_2;d_1,d_2).\]
We can give the following convergence rates for B-spline estimators using this construction:

\begin{prp}\label{b_spline_consistency_thm} Suppose that Assumptions \ref{unconf_ass_ass}-\ref{loc_lin_ass} hold, and that the number $M$ of knots satisfies $M\rightarrow\infty$ and $\frac{M^d\log n}{n}\rightarrow0$. Then the B-spline estimators for $\mu(x;\mathbf{d})$ and the covariance operator $H_{d_1d_2}(x_1,x_2)$ are consistent at the rate
\begin{eqnarray}
\nonumber\sup_{d,x_1}\left|\hat{\mu}(x_1,d)-\mu(x_1,d)\right|&=&O_p\left(r_{Gn}\right)\\
\nonumber\sup_{d_1,d_2,x_1,x_2}\left| \hat{H}(x_1,x_2;d_1,d_2)-H(x_1,x_2;d_1,d_2)\right|&=&O_p\left(r_{Gn}\right)
\end{eqnarray}
where $r_{Gn} = \frac1{G} + \left(M/c_n\right)^{-2} + \left(\frac{M^d\log n}{Gn}\right)^{1/2}$, and the suprema are for $x_1,x_2\in[0,1]^d$. The rate optimal number of knots minimizing the second term of either error is of the order $M^*=O\left(\left(\frac{Gn}{\log n}\right)^{\frac1{4+d}}\right)$, resulting in a rate $\frac1G + \left(\frac{\log n}{Gn}\right)^{\frac2{4+d}}$.
\end{prp}

This result takes the role of Theorem \ref{loc_lin_consistency_thm}, and convergence rates for estimation of eigenfunctions and the transfer estimator follow by plugging that rate into Theorem \ref{eig_cons_thm} and Corollaries \ref{cate_cons_thm} and \ref{IMSE_conv_rate}. The proof follows closely that of Theorem 2.1 in \cite{CCh15}, with only minor modifications to allow for expanding support and two-way averages in estimation of the covariance function. We therefore only provide a brief outline of the formal argument below.

\subsection{Digression: Existence of Solution without Regularization}\label{subsec:existence_regularization}

We next give results on the behavior of $IMSE_K(a)$ under conditions for which a solution to the non-regularized problem (\ref{S_op_defn}) exists. Following \cite{HMWY10}, we state those sufficient conditions in terms of separate Karhunen-Lo\`eve expansions for $\mu_g(x;0)$ and $\tau_g(x;0)$,
\begin{eqnarray}
\nonumber \mu_{g^*}(x;0)&=&\mu(x;0) + \sum_{k=1}^{\infty} \alpha_{g^*k}\xi_k(x)\\
\label{mu_tau_kl} \tau_{g^*}(x)&=&\tau(x) + \sum_{k=1}^{\infty} \beta_{g^*k}\zeta_k(x)
\end{eqnarray}
where $\xi_1,\xi_2,\dots$ and $\zeta_1,\zeta_2,\dots$ are orthonormal systems corresponding to the eigenfunctions of the covariance operators for $\mu_{g^*}(\cdot;0)$ and $\tau_{g^*}(\cdot)$ associated with eigenvalues $\mathbb{E}[\alpha_{g^*k}^2]$ and $\mathbb{E}[\beta_{g^*k}^2]$, respectively. By the usual properties of the Karhunen-Lo\`eve representation, the coefficients satisfy $\mathbb{E}[\alpha_{g^*k}]=\mathbb{E}[\beta_{g^*k}]=0$ and $\mathbb{E}[\alpha_{g^*k}\alpha_{g^*l}]=\mathbb{E}[\beta_{g^*k}\beta_{g^*l}]=0$ for all $k=1,2,\dots$, and $l\neq k$.

We then analyze regularization bias under the following condition:
\begin{equation}
\label{kh_bound_condition}\sum_{k=1}^{\infty}\sum_{l=1}^{\infty}\frac{\mathbb{E}[\alpha_{g^*k}\beta_{g^*l}]^2}{\mathbb{E}[\alpha_{g^*k}^2]^{3/2}}<\infty
\end{equation}
Note that by compactness of $T_{\mu\mu}$, $\mathbb{E}[\alpha_{g^*k}^2]\rightarrow0$ as $k$ grows, so that this condition is slightly weaker than the sufficient condition given by \cite{HMWY10} for the existence of a functional linear regression model representing the linear projection of $\tau_{g^*}$ on $\mu_{g^*}$ (see their Proposition 2.4).

While the squared correlation coefficient $r_{kl}^2:=\frac{\mathbb{E}[\alpha_{g^*k}\beta_{g^*l}]^2}{\mathbb{E}[\alpha_{g^*k}^2]\mathbb{E}[\beta_{g^*l}^2]}$ is bounded by 1 for any fixed $k,l$ and $\mathbb{E}[\beta_{g^*l}^2]$ are square summable in $l$, (\ref{kh_bound_condition}) may fail when there are many features of the conditional mean of baseline outcomes $Y_{g^*i}(0)$ given $X_{gi}$ that have low variability in $\alpha_{g^*k}$ but are highly predictive of $\tau_{g^*}$.

\begin{prp}\label{C1_IMSE_rate_prp}Suppose that $T_{\mu\mu}$ and $T_{\mu\tau}$ are compact and that $\mu_{g^*}(x;0)$ and $\tau_{g^*}(x)$ have a Karhunen-Lo\`eve representation (\ref{mu_tau_kl}) with coefficients satisfying (\ref{kh_bound_condition}). Then (a) the solutions to the problem (\ref{S_op_defn}) are well-defined, and the conclusion of Theorem \ref{opt_basis_thm} holds with
\[IMSE_K^*(a) \leq IMSE_K^* + O(a)\]
\end{prp}

See Appendix \ref{sec:proofs} for a proof. This result establishes not only that the optimal IMSE can be achieved at a linear rate in $a$, but also that the condition (\ref{kh_bound_condition}) is sufficient to guarantee that the optimal predictors defined in (\ref{S_op_defn}) exist independently of a particular regularization scheme, and can also be estimated consistently.


\section{Linearity of $\mathbb{E}[\tau(\cdot;V_{g^*})|\mu(\cdot;V_{g^*})]$}
\label{sec:linearity_app}

As discussed in the main text, our approach uses the linear projection (\ref{linear_beta}) rather than the conditional expectation $\mathbb{E}[\tau(\cdot;V_{g^*})|\mu(\cdot;V_{g^*})]$ as a predictor for $\tau(\cdot;V_{g^*})$. If the conditional expectations are in fact linear in $\mu(\cdot;V_{g^*})$, then the two coincide, in which case focussing on linear predictors is not restrictive. This appendix analyzes stylized examples to discuss the plausibility of that linearity assumption.

\begin{ex} \textbf{Location-Scale Model.} $\mu_g(x)= v_{g11} + v_{g12}m(x)$ and $\tau_g(x) = v_{g21} + v_{g22}t(x)$ for functions $m(x),t(x)$, where $m(x)$ takes at least two different values in the support of $X_{gi}$ for all $g$, and the conditional expectations $\mathbb{E}[v_{g2s}|v_{g11},v_{g12}]$ are linear in $v_{g11},v_{g12}$ for $s=1,2$. Then $\mathbb{E}[\tau_g(x)|\mu_g(\cdot;0)]$ is linear in $\mu_g(\cdot;0)$, so that in particular the interpolation error $e_g(x)\equiv0$.

Specifically, without loss of generality assume that the support of $X_{gi}$ equals $[0,1]$, and that there exists $0<\kappa<1$ such that  $w_0\equiv\frac1{\kappa}\int_0^{\kappa}m(s)ds\neq \frac1{1-\kappa}\int_{\kappa}^1m(s)ds\equiv w_1$. Then, for $w_{g0}:=\frac1{\kappa}\int_0^{\kappa}\mu_g(s)ds$ and $w_{g1}:=\frac1{1-\kappa}\int_{\kappa}^1\mu_g(s)ds$ we can write $v_{g12}=\frac{w_{g1}-w_{g0}}{w_1-w_0}$ and $v_{g11} = \int_0^1\mu_g(s)ds - v_{g12}\int_0^1\mu(s)ds$. Moreover,
\[\mathbb{E}[\tau_g(x)|\mu_g(\cdot)]=\mathbb{E}[\tau_g(x)|v_{g11},v_{g12}]=\mathbb{E}[v_{g21}|v_{g11},v_{g12}] + \mathbb{E}[v_{g22}|v_{g11},v_{g12}]t(x)\]
is linear in $v_{g11},v_{g12}$ by assumption, and therefore also linear in $\mu_g(x)$, so that recursive substitution yields a bounded expression for the kernel $\beta(x_1,x_2)$ in (\ref{linear_beta}) as long as $\kappa$ is bounded from zero and one, and $t(x)$ and the conditional expectations of $v_{g2s}$ are bounded.
\end{ex}

We next give an example where the distribution of the outcome variable is discontinuous at a known value of a covariate. The main empirical motivation for this setting concerns school attendance according to the child's age, where the cost of attending secondary school at age 12 or above may be different from that for primary school, and both costs may also vary across sites. For example, many sites may have access to a primary school in close proximity, but the nearest secondary school may be more difficult to reach at some sites, leading to lower attendance pre- and post-intervention.

\begin{ex} \textbf{Common Structural Break.} Suppose that $\mu_g(x)= v_{g11}\underline{m}(x)\dum\{x\leq x_0\} + v_{g12}\bar{m}(x)\dum\{x>x_0\}$ and $\tau_g(x)= v_{g21}\underline{t}(x)\dum\{x\leq x_0\} + v_{g22}\bar{t}(x)\dum\{x>x_0\}$, where the conditional expectations $\mathbb{E}[v_{g2s}|v_{g11},v_{g12}]$ are linear in $v_{g11},v_{g12}$ for $s=1,2$. Then $\mathbb{E}[\tau_g(x)|\mu_g(\cdot;0)]$ is linear in $\mu_g(\cdot;0)$ and $e_g(x)\equiv0$. The structure of this example closely parallels the previous one, we therefore omit a separate proof.
\end{ex}

The following example concerns the case in which a covariate may be measured at different units at each site. For example, agents choices may be determined by income and costs in terms of site-specific purchasing power, whereas recorded amounts are in terms local currency units, typically expressed in US Dollar equivalents according to purchasing power parity at the national level.

\begin{ex} \textbf{Heterogeneous Measurement Units I} Suppose that sites are heterogeneous with respect the scale at which a characteristic is measured, where $\mu_g(x) = m(v_g x)$ and $\tau_g(x)=t(v_gx)$, and that $v_g$ only takes finitely many values $v_1,\dots,v_J$. If for some collection of intervals $I_1,\dots,I_J$, the matrix $M:=\left(\int_{I_i}m(v_j s)f_0(s)ds\right)_{i,j}$ has full rank then $\tau_g(x)\equiv \int \mu_g(s)\beta(s,x)f_0(s)ds$ where $\beta(s,x):=(\dum\{s\in I_j\})_{j=1}^J M^{-1}\left(t(v_jx)\right)_{j=1}^J$ and we take $(c_j)_{j=1}^J$ to be a column vector with entries $c_1,\dots,c_J$. In particular, the resulting interpolation error is zero, $e_g(x)\equiv0$.
\end{ex}

Alternatively, under global smoothness of $m(x)$, the relevant information on $v_g$ may be recovered as a limit of linear functionals in $\mu_g(\cdot)$ even when the site-specific scaling factor may take infinitely many values.

\begin{ex} \textbf{Heterogeneous Measurement Units II} Consider again the setting from the previous example, where this time, $v_g$ may be continuously supported but $m(x)$ is differentiable at any order at $x=0$, and $[0,\varepsilon)$ is contained in the support of $X_{gi}$. Furthermore suppose that $t(x)$ is the limit of a convergent power series on the support of $X_{gi}$ where the polynomial coefficients $k_1,k_2,\dots$ are nonzero. Also,  Then, if the corresponding derivatives $m^{(k_s)}(0):=\left.\frac{d^{k_s}}{dx^{k_s}}m(x)\right|_{x=0}$  are all nonvanishing at zero, the linear projection of $\tau_g(x)$ on $\mu_g(x;0)$ equals $\tau_g(x)$, in particular $\mathbb{E}[\tau_g(x)|\mu_g(\cdot;0)]$ is linear in $\mu_g(\cdot;0)$.

To see why this is the case, note first that the derivatives $\mu_g^{(k_s)}(0):=\frac{d^{k_s}}{dx^{k_s}}\mu_g(x;0)\equiv v_g^{k_s}\frac{d^{k_s}}{dx^{k_s}}m(x)$ are limits of finite differences of $\mu_g(x;0)$ and therefore linear in values of the function at points in the neighborhood of $x$. By assumption we can then represent
\[\tau_g(x) = \sum_{s\geq1} v_g^{k_s}\frac{t^{(k_s)}(0)x^{k_s}}{k_s!} = \sum_{s\geq1}\frac{\mu_g^{(k_s)}(0)}{m^{(k_s)}(0)}\frac{t^{(k_s)}(0)x^{k_s}}{k_s!}\]
which is linear in $\mu_g(\cdot,0)$ since $m^{(k_s)}(0)$ and $t^{(k_s)}(0)$ are constant across sites.

Using this argument, $\tau_g(x)$ is recovered only \emph{in the limit} along sequences of Hilbert-Schmidt operators acting on $\mu_g(\cdot)$. In particular there is no guarantee that the limit itself will be Hilbert-Schmidt, so the function $\tau_g(x)$ may only be recovered as an approximation.
\end{ex}

The preceding examples suggest that even in the presence of interpolation error, linear projection can be responsive to the patterns of unobserved site heterogeneity used to motivate our approach although conditional expectations are linear only under admittedly fairly stylized assumptions. While our arguments are given in terms of $m(x),t(x)$, an implementation of linear projection obviously does not require knowledge of these functions.

Finally, even under linearity, there may still remain relevant site-specific unobserved heterogeneity that does not express itself in $\mu_g(\cdot;0)$ and therefore cannot be predicted from baseline outcomes. For example, modeling pre-intervention outcomes $\mu_g(x;0)$ as a function of household income could help predict aggregate heterogeneity in income effects in the response to a conditional cash transfer program. At the same time, a predictor based on income alone may not be responsive to site-specific (rather than only individual-specific) heterogeneity in substitution effects which may manifest themselves more clearly if in addition some measure of the cost of school attendance were taken into account.

\section{Proofs for Section \ref{sec:opt_basis_sec}}\label{sec:proofs}

Before proving the main result, we give a characterization of the integrated mean square error of projection into closed linear subspaces of $L_2(\mathcal{X})$. Specifically, we consider the mean square error of projection of $\tau_{g^*}$ with respect to a general linear transformation $\mu_{g^*}$.

We consider the problem where the domain of the operator $B$ is restricted to $\mathcal{H}$, a closed linear subspace $\mathcal{H}$ of $\mathcal{N}^{\perp})$, which was defined as the orthogonal complement of the null space of $T_{\mu\mu}$, $\mathcal{N}:=\ker(T_{\mu\mu})$. It is known that for any linear operator $T:L_2(\mathcal{X})\rightarrow L_2(\mathcal{X})$, $\im(T)$ and $\ker(T)$ are linear subspaces of $L_2(\mathcal{X})$, and that $\ker(T)$ is closed if the linear operator $T$ is bounded. Since $\mathcal{H}$ is the orthogonal complement of the null space $\ker(T_{\mu\mu})$, it is a closed linear subspace of $L_2(\mathcal{X})$, so this allows for the choice $\mathcal{H}\equiv \mathcal{N}^{\perp}$. We also let the operator $P:L_2(\mathcal{X})\rightarrow\mathcal{H}$ to denote the orthogonal projection on $\mathcal{H}$. Since $\mathcal{H}$ is closed by assumption, that projection is well-defined by the Classical Projection Theorem (Theorem 2 on p.51 in \cite{Lue69}).

We can then characterize the integrated mean square error of projection as follows:

\begin{thm}\label{min_mse_updated_thm}\textbf{(Integrated MSE of Projection)} There exists a best linear predictor for $\tau_{g^*}$ of the form (\ref{linear_beta}) based on $\mu_{g^*}$ that is of the form
\[\tau_{g^*}^*(x):=\int\mu_{g^*}(s)\beta(s,x)f_0(s)ds\]
If furthermore the operator $PT_{\mu\mu}A^*P:\mathcal{H}\rightarrow\mathcal{H}$ possesses an inverse, the minimized mean square error of prediction satisfies
\begin{equation}\label{min_mse_updated_eq}
\int\mathbb{E}\left[(\tau_{g^*}(x)-\tau_{g^*}^*(x))^2\right]f_0(x)dx = \mathbb{E}\|\tau_{g^*}\|^2 - \tr\left(T_{\mu\tau}^*P(PT_{\mu\mu}P)^{-1}PT_{\mu\tau}\right)
\end{equation}
\end{thm}

For the definition of the inverse in (\ref{min_mse_updated_eq}), note that the operator and therefore its inverse are understood to be restricted to the subspace $\mathcal{H}$.

\textsc{Proof:} We first consider the case $P=\id$. The set $\mathcal{M}$ of linear predictors of the form (\ref{linear_beta}) can be identified with the Hilbert space $L_2(\mathcal{X}\times\mathcal{X})$ endowed with the scalar product $\langle T_1,T_2\rangle=tr(T_1^*T_2) =\int\int g(x_2,x_1)h(x_1,x_2)f_0(x_1)f_0(x_2)dx_1dx_2$ and the trace norm induced by that scalar product.

To obtain a representation of the minimized projection error, we derive a projection analog for the regression model developed in \cite{HMW00}. We first show that $\mathcal{M}$ is a closed linear subspace of $L_2(\mathcal{X})$: Consider a random element $\phi_{g^*}\in L_2(\mathcal{X})$ and define $H_{\phi_{g^*},\mu_{g^*}}(x_1,x_2):=\mathbb{E}[\phi_{g^*}(x_1)\mu_{g^*}(x_2;0)]$ and let $T_{\phi\mu}$ denote the corresponding Hilbert-Schmidt integral operator. We then have that $\phi_{g^*}$ is orthogonal to $B\mu_{g^*}$ if
\begin{eqnarray}\nonumber 0&=&\mathbb{E}\langle\phi_{g^*},B\mu_{g^*}\rangle\\
\nonumber&=&\mathbb{E}\left[\int \phi_{g^*}(x_1)\mu_{g^*}(x_2)\beta(x_1,x_2)f_0(x_1)f_0(x_2)dx_1dx_2\right]\\
\nonumber&=&\int\mathbb{E}\left[\phi_{g^*}(x_1)\mu_{g^*}(x_2)\right]\beta(x_1,x_2)f_0(x_1)f_0(x_2)dx_1dx_2\\
\nonumber&=&\int H_{\phi_{g^*},\mu_{g^*}}(x_1,x_2) \beta(x_1,x_2)f_0(x_1)f_0(s)dx_1dx_2=tr\left(T_{\phi\mu}B\right)
\end{eqnarray}
Since $B$ may in particular include the identity, $\phi_{g^*}$ is orthogonal to $\mathcal{M}$ for any $B\in L_2(\mathcal{X}\times\mathcal{X})$ if and only if $H_{\phi_{g^*},\mu_{g^*}}(x_1,x_2)\equiv0$. Since $\mathcal{M}$ is the orthogonal complement of the set of all such vectors $\phi_{g^*}\in L_2(\mathcal{X})$, it is in particular a closed linear subspace of $L_2(\mathcal{X})$.

By the classical projection theorem (Theorem 2 in \cite{Lue69}, p.51), it then follows that there exists a unique minimizer $\tau_0^*\in \mathcal{M}$. Furthermore, that minimizer also satisfies the orthogonality condition $\langle m,\tau_{g^*}-\tau_0^*\rangle$ for any $m\in \mathcal{M}$. It follows that
\begin{eqnarray}\nonumber\min_{B\in L_2(\mathcal{X}\times\mathcal{X})}\mathbb{E}\left\|\tau_{g^*} - B\mu_{g^*}\right\|^2&=&\mathbb{E}\|\tau_{g^*}\|^2-2\mathbb{E}\langle\tau_{g^*},\tau_0^*\rangle + \mathbb{E}\|\tau_0^*\|^2\\
\nonumber&=&\mathbb{E}\|\tau_{g^*}\|^2 - 2(\mathbb{E}\|\tau_{g^*}^*\|^2+\mathbb{E}\langle\tau_{g^*}-\tau_{g^*}^*,\tau_{g^*}^*\rangle) + \mathbb{E}\|\tau_{g^*}^*\|^2\\
\label{TSS_ESS}&=&\mathbb{E}\left\|\tau_{g^*}\right\|^2 - \mathbb{E}\left\|\tau_{g^*}^*\right\|^2
\end{eqnarray}


We next characterize the optimal solution $\tau_{g^*}^*$ in terms of the operators $T_{\mu\mu}$ and $T_{\mu\tau}$. Suppose that $B_0$ with kernel $\beta_0^*(x_1,x_2)$ is such that $\tau_{g^*}^*=B_0\mu_g$. Then, we have for any $h\in L_2(\mathcal{X})$ that
\begin{eqnarray}
\nonumber (T_{\mu\tau}h)(x_1)&=&\int \mathbb{E}[\mu_{g^*}(x_1)\tau_{g^*}(x_2)]h(x_2)f_0(x_2)dx_2\\
\nonumber &=&\int\mathbb{E}\left[\mu_{g^*}(x_1)\left[(B_0\mu_{g^*})(x_2) + \left\{\tau_{g^*}(x_2)-(B_0\mu)(x_2)\right\}\right]\right]h(x_2)f_0(x_2)dx_2\\
\nonumber &=&\int \mathbb{E}[\mu_{g^*}(x_1)\mu_{g^*}(x_2)]\beta_0^*(x_2,x_3)f_0(x_2)dx_2h(x_3)f_0(x_3)dx_3\\
\nonumber &&+\int\mathbb{E}\left[\mu_{g^*}(x_1)\left\{\tau_g(x_3)-(B_0\mu)(x_3)\right\}\right]h(x_3)f_0(x_3)dx_3\\
\label{B_star_normal_eqn} &=&\left(T_{\mu\mu}B_0h\right)(x_1) +0
\end{eqnarray}
where the last equality follows from orthogonality of the projection error, noting that $B\mu_{g^*}:=\textnormal{Id}\mu_{g^*}$ is in $\mathcal{M}$.

Now suppose that the minimum is attained at both $B_0\mu_{g^*}\in \mathcal{M}$ and $B_1\mu_{g^*}\in \mathcal{M}$. Then by orthogonality of the projection error,
\begin{eqnarray}
\nonumber \mathbb{E}\|\tau_{g^*}-B_1\mu_{g^*}\|^2&=&\mathbb{E}\|\tau_{g^*}-B_0\mu_{g^*} + (B_0-B_1)\mu_{g^*}\|^2\\
\nonumber&=&\mathbb{E}\|\tau_{g^*}-B_0\mu_{g^*}\|^2 +2\mathbb{E}\langle\tau_{g^*}-B_0\mu_{g^*} ,(B_0-B_1)\mu_{g^*}\rangle + \mathbb{E}\|(B_0-B_1)\mu_{g^*}\|^2\\
\nonumber&=&\mathbb{E}\|\tau_{g^*}-B_0\mu_{g^*}\|^2 + \mathbb{E}\|(B_0-B_1)\mu_{g^*}\|^2
\end{eqnarray}
Hence the minimum can be attained at both $B_1$ and $B_0$ iff
\begin{eqnarray}
\nonumber0&=& \mathbb{E}\|(B_0-B_1)\mu_{g^*}\|^2\\
\nonumber&=&\int(\beta_0(x_1,x_2)-\beta_1(x_1,x_2))\mathbb{E}[\mu_{g^*}(x_1)\mu_{g^*}(x_3)](\beta_0(x_3,x_2)-\beta_1(x_3,x_2))\\
\nonumber&&\times f_0(x_1)f_0(x_3)f_0(x_2)dx_1dx_2dx_3\\
\nonumber&=&\tr\left((B_0-B_1)T_{\mu\mu}(B_0^*-B_1^*)\right)
\end{eqnarray}
or equivalently iff $(B_0-B_1)\in \ker(T_{\mu\mu})$. In particular, the orthogonal projection $B_0^*$ of any solution $B\in L_2(\mathcal{X}\times\mathcal{X})$ onto  the closed subspace $\mathcal{N}^{\perp}\times L_2(\mathcal{X})$ exists and is a solution to the same minimization problem. So without loss of generality we can restrict our attention to the minimization problem
\begin{equation}\label{restricted_proj_problem}\min_{B\in \mathcal{N}^{\perp}\times L_2(\mathcal{X})}\mathbb{E}\left\|\tau_{g^*} - B\mu_{g^*}\right\|\end{equation}

Noticing that restricted to its range $\mathcal{N}^{\perp}$, the operator $T_{\mu\mu}$ possesses an inverse, we can solve the normal equations (\ref{B_star_normal_eqn}) to obtain
\[B_0^*h = T_{\mu\tau}^*T_{\mu\mu}^{-1}h\]
for any $h\in\mathcal{N}^{\perp}$. Substituting this expression into (\ref{TSS_ESS}), we therefore obtain
\[\min_{B\in L_2(\mathcal{X}\times\mathcal{X})}\mathbb{E}\left\|\tau_{g^*} - B\mu_{g^*}\right\|^2 = \mathbb{E}\|\tau_{g^*}\|^2 - \tr\left(T_{\mu\tau}^*T_{\mu\mu}^{-1}T_{\mu\tau}\right)\]
establishing the claim for $P_1=\id$. For the general case, notice that $\mathcal{H}$ was a closed linear subspace of $\mathcal{N}^{\perp}$ and $A$ was assumed to be invertible on $\mathcal{H}$, so that the argument continues to apply when restricted to the subspace of linear predictors based on $P\mu_{g^*}$\qed

\subsection{Proof of Lemma \ref{IMSE_K_lower_bound}} We use the formula from (\ref{min_mse_updated_eq}) with $\tilde{\mathcal{H}}_K:=\spn\left(\tilde{\phi}_1,\dots,\tilde{\phi}_K\right)$ as the linear subspace of $L_1(\mathcal{X})$ spanned by the $K$ orthonormal functions $\tilde{\phi}_1(x),\dots,\tilde{\phi}_K(x)$.

We first consider the operator $\tilde{S}_K:=\tilde{P}_KT_{\mu\mu}\tilde{P}_K$. Since $\tilde{\mathcal{H}}_K$ is finite-dimensional, we can identify $\tilde{S}_K$ with a $K\times K$ matrix of coefficients with respect to the basis $\tilde{\phi}_1,\dots,\tilde{\phi}_K$ of $\tilde{\mathcal{H}}_K$,
\[\tilde{S}_K:=\left(\langle\tilde{\phi}_k,T_{\mu\mu}\tilde{\phi}_l\rangle\right)_{k,l}\]
Noting that $T_{\mu\mu}$ is injective on $\mathcal{N}^{\perp}$ we will impose the normalization
\begin{equation}
\label{opt_bas_proof_normalization}\langle\tilde{\phi}_k,T_{\mu\mu}\tilde{\phi}_l\rangle=\delta_{k,l}\end{equation} so that $\tilde{S}_K=I_K$, the $K$-dimensional identity matrix.

Furthermore, evaluating the trace using an arbitrarily chosen orthonormal system for $L_2(\mathcal{X})$, we can also verify that
\begin{eqnarray}
\nonumber \tr(T_{\mu\tau}^*\tilde{P}_K\tilde{P}_KT_{\mu\tau})&=&\int\sum_{k=1}^K \left(\int H_{\mu\tau}(x_1,x_2)H_{\mu\tau}(x_3,x_1)\tilde{\phi}_k(x_2)\tilde{\phi}_k(x_3)f_0(x_2)f_0(x_3)dx_2dx_3\right)\\
\nonumber&&\times f_0(x_1)dx_1\\
\nonumber&=&\sum_{k=1}^K\langle \tilde{\phi}_k,T_{\mu\tau}T_{\mu\tau}^*\tilde{\phi}_k\rangle
\end{eqnarray}

Hence, using the formula from Theorem \ref{min_mse_updated_thm},
\[\int\mathbb{E}\left[(\tau_{g^*}(x)-\tau_{g^*}^*(x))^2\right]f_0(x)dx = \sum_{k=1}^K\langle \tilde{\phi}_k,T_{\mu\tau}T_{\mu\tau}^*\tilde{\phi}_k\rangle\]
for any collection of functions $\tilde{\phi}_1,\dots,\tilde{\phi}_K$ satisfying the constraint (\ref{opt_bas_proof_normalization}).

Hence, the mean-square error optimal basis functions are determined by the quadratic program
\begin{equation}\label{opt_proof_const_opt_problem}\inf_{\phi_1,\dots,\phi_K}\sum_{k=1}^K\langle \phi_k,T_{\mu\tau}T_{\mu\tau}^*\phi_k\rangle\hspace{0.8cm}\textnormal{s.t. }
\langle\phi_k,T_{\mu\mu}\phi_l\rangle=\delta_{kl}\end{equation}
which establishes the claim\qed

\subsection{Proof of Theorem \ref{opt_basis_thm}}

From the definition of $IMSE_K^*$, there exist $\tilde{\phi}_{1\varepsilon},\dots,\tilde{\phi}_{K\varepsilon}\in L_2(\mathcal{X})$, such that $\langle\tilde{\phi}_{k\varepsilon},T_{\mu\mu}\phi_{l\varepsilon}\rangle=\delta_{kl}$ and
\[IMSE_K\left[\tilde{\phi}_{1\varepsilon},\dots,\tilde{\phi}_{K\varepsilon}\right]\leq IMSE_K^* + \frac{\varepsilon}2\]
Since  $\tilde{\phi}_{1\varepsilon},\dots,\tilde{\phi}_{K\varepsilon}\in L_2(\mathcal{X})$, we can find $L_{\varepsilon}<\infty$ such that $\|\tilde{\phi}_{k\varepsilon}\|\leq L_{\varepsilon}$.

These $K$ functions do not necessarily satisfy the regularized orthogonality constraint, rather we find ``close" alternative functions
$\bar{\phi}_{1\varepsilon},\dots,\bar{\phi}_{K\varepsilon}$ such that $\langle\check{\phi}_{k\varepsilon},T_{\mu\mu a}\check{\phi}_{l\varepsilon}\rangle=\delta_{kl}$. Since the operator $T_{\mu\mu a}$ is Hermitian and positive, we can define a scalar product
$\langle u,v\rangle_a:=\langle u,T_{\mu\mu a}v\rangle$. We first obtain $K$ functions $\check{\phi}_{1\varepsilon},\dots,\check{\phi}_{K\varepsilon}$ that satisfy the orthogonality constraints $\langle\check{\phi}_{k\varepsilon},\check{\phi}_{l\varepsilon}\rangle=0$ for all $k\neq l$. To that end, we use the Gram-Schmidt procedure with respect to the scalar product $\langle\cdot,\cdot\rangle_a$, where $\check{\phi}_{1\varepsilon}:=\tilde{\phi}_{1\varepsilon}$, and
\[\check{\phi}_{k\varepsilon}:=\tilde{\phi}_{k\varepsilon} - \sum_{l=1}^{k-1}
\frac{\langle\tilde{\phi}_{k\varepsilon},\check{\phi}_{l\varepsilon}\rangle_a}
{\langle\check{\phi}_{l\varepsilon},\check{\phi}_{l\varepsilon}\rangle_a}\check{\phi}_{l\varepsilon}\]

Defining $c_{kk}=1$ and $c_{lk}:=-\frac{\langle\tilde{\phi}_{k\varepsilon},\check{\phi}_{l\varepsilon}\rangle_a}
{\langle\check{\phi}_{l\varepsilon},\check{\phi}_{l\varepsilon}\rangle_a}$ for $k\neq l$, we can confirm by induction that
\[\check{\phi}_{k\varepsilon} = \check{\phi}_{k\varepsilon} - \sum_{l=1}^{k-1} d_{lk}\tilde{\phi}_{l\varepsilon}\]
where $d_{lk} = \sum_{\mathcal{J}(l,k)}\prod_{s=2}^{|k-l|}c_{j_sj_{s-1}}$, and $\mathcal{J}(l,k)$ is the set of all $(j_1,\dots,j_s)$ such that $s\leq|k-l|$ and $j_1< j_2\dots< j_{s}$. For notational convenience, we also define $d_{kk}=1$.

We now recursively bound $c_{kl}$ and $d_{kl}$ for $k\neq l$. Due to the constraint on $\tilde{\phi}_{1\varepsilon},\dots,\tilde{\phi}_{K\varepsilon}$ it follows for any $k\neq l$,
\[\langle\tilde{\phi}_{k\varepsilon},\tilde{\phi}_{l\varepsilon}\rangle_a=\langle\tilde{\phi}_{k\varepsilon},T_{\mu\mu}\tilde{\phi}_{l\varepsilon}\rangle
+a\langle\tilde{\phi}_{k\varepsilon},\tilde{\phi}_{l\varepsilon}\rangle=a\langle\tilde{\phi}_{k\varepsilon},\tilde{\phi}_{l\varepsilon}\rangle\]
Hence we can calculate the scalar products
\begin{eqnarray}
\nonumber \langle\tilde{\phi}_{k\varepsilon},\check{\phi}_{l\varepsilon}\rangle_a&=&a\sum_{j=1}^l d_{lj}\langle\tilde{\phi}_{k\varepsilon},\tilde{\phi}_{j\varepsilon}\rangle\\
\nonumber \langle\check{\phi}_{k\varepsilon},\check{\phi}_{k\varepsilon}\rangle_a&=&1+\sum_{l=1}^{k-1}d_{kl}^2 + a\sum_{j=1}^{k-1}\sum_{l=1}^{k-1}
d_{kj}d_{kl}\langle\tilde{\phi}_{j\varepsilon},\tilde{\phi}_{l\varepsilon}\rangle
\end{eqnarray}
for any $k\neq l$. In particular, denoting $\bar{d}_k:=\max\{|d_{kl}|:k>l\}$, we can bound
\begin{eqnarray}
\nonumber \left|\langle\tilde{\phi}_{k\varepsilon},\check{\phi}_{l\varepsilon}\rangle_a\right|&\leq& a(1+(k-1)\bar{d}_k)L_{\varepsilon}^2\\
\nonumber \left|\langle\check{\phi}_{k\varepsilon},\check{\phi}_{k\varepsilon}\rangle_a-1-\sum_{l=1}^{k-1}d_{kl}^2 \right|
&\leq&aL_{\varepsilon}^2(1+(k-1)^2\bar{d}_k^2)
\end{eqnarray}
Hence for $a$ satisfying $2a^2\leq (L_{\varepsilon}^2K^2)^{-1}$, we can bound $\bar{d}_2:=|c_{21}|\leq 2aL_{\varepsilon}^2$. It then follows by induction over $k$ that $|c_{kl}|\leq 4aL_{\varepsilon}^2$ and
\[\bar{d}_k\leq 8aL_{\varepsilon}^2\] for each $k=1,\dots,K$.

To obtain functions $\bar{\phi}_{1\varepsilon},\dots,\bar{\phi}_{K\varepsilon}$ with the desired property, we furthermore need to impose the scale normalization $\langle\bar{\phi}_{k\varepsilon},\bar{\phi}_{k\varepsilon}\rangle_a = 1$. Since $\check{\phi}_{k\varepsilon}=\sum_{j=1}^kd_{kj}\tilde{\phi}_{k\varepsilon}$,
\begin{eqnarray}
\nonumber \langle\check{\phi}_{k\varepsilon},\check{\phi}_{k\varepsilon}\rangle_a &=& \langle\sum_{j=1}^k\tilde{\phi}_{j\varepsilon},T_{\mu\mu}\langle\check{\phi}_{k\varepsilon},\check{\phi}_{k\varepsilon}\rangle\tilde{\phi}_{j\varepsilon}\rangle + a\langle\check{\phi}_{k\varepsilon},\check{\phi}_{k\varepsilon}\rangle
\end{eqnarray}
Since $d_{kk}=1$, we can then bound
\[|\langle\check{\phi}_{k\varepsilon},\check{\phi}_{k\varepsilon}\rangle_a-1|\leq K\bar{d}_k^2 + aL_{\varepsilon}^2 \leq (K+8)aL_{\varepsilon}^2\]

We can therefore form
\[\bar{\phi}_k:=(\langle\check{\phi}_{k\varepsilon},\check{\phi}_{k\varepsilon}\rangle_a)^{-1/2}\check{\phi}_{k\varepsilon}\]
By construction, $\bar{\phi}_{1\varepsilon},\dots,\bar{\phi}_{K\varepsilon}$ satisfy the constraints $\langle\bar{\phi}_{k\varepsilon},\bar{\phi}_{l\varepsilon}\rangle_a=\delta_{kl}$ and, using standard bounding arguments
\[\|\bar{\phi}_{k\varepsilon}-\tilde{\phi}_{k\varepsilon}\|\leq 10KaL_{\varepsilon}^3\]
for any sufficiently small value of  $a>0$.

It remains to be shown by a continuity argument that the IMSE achieved by $\bar{\phi}_{1\varepsilon},\dots,\bar{\phi}_{K\varepsilon}$ is greater than $IMSE_K^*+\varepsilon$ for $a$ sufficiently close to zero. Using the formula from (\ref{min_mse_updated_eq}), we can write the IMSE for a given set of basis functions as
\[IMSE_K[\phi_1,\dots,\phi_K] = tr(C_K^{-1}A_K)\]
where $C_K$ and $A_K$ can be identified with $K\times K$ matrices $A_K = (A_{kl})$ and $C_K = (C_{kl})$, and
\[C_{kl}:=\langle\phi_k,T_{\mu\mu}\phi_l\rangle\]
and
\[A_{kl}:=\langle\phi_k,T_{\mu\tau}T_{\mu\tau}^*\phi_l\rangle\]
By assumption, for the basis functions $\tilde{\phi}_{1\varepsilon},\dots,\tilde{\phi}_{K\varepsilon}$ we have $C_{kl}=\delta_{kl}$ so that the corresponding matrix is the $K$-dimensional identity matrix. Furthermore, the operators $T_{\mu\mu}$ and $T_{\mu\tau}T_{\mu\tau}^*$ are compact with largest eigenvalue equal to $\nu_1<\infty$ and $\lambda_1<\infty$, respectively. Since $\|\bar{\phi}_{k\varepsilon}-\tilde{\phi}_{k\varepsilon}\|\leq aB_9K{\varepsilon}^3$, there exists a constant $\kappa_{\varepsilon}<\infty$ such that
\begin{equation}
\label{check_vs_tilde_bound}
|IMSE_K[\bar{\phi}_{1\varepsilon},\dots,\bar{\phi}_{K\varepsilon}]-IMSE_K[\tilde{\phi}_{1\varepsilon},\dots,\tilde{\phi}_{K\varepsilon}]|\leq \kappa_{\varepsilon}a
\end{equation}
for $a$ sufficiently small, which establishes the desired conclusion\qed

We also note that the convergence rate for $|IMSE_K(a)-IMSE_K^*|$ as a function of $a$ generally depends on the eigenvalues of $T_{\mu\mu}$ and $T_{\mu\tau}$, where $L_{\varepsilon}$ in the preceding proof may diverge as $\varepsilon\rightarrow0$. We leave this question for future research.

\subsection*{Proof of Proposition \ref{C1_IMSE_rate_prp}} Since $T_{\mu\mu}$ is injective on $\ker(T_{\mu\mu})^{\perp}$, any function $\phi_k^*$ solving the generalized eigenvalue problem (\ref{S_op_defn}) at eigenvalue $\lambda_k$ can be equivalently characterized by $\phi_k^*:=T_{\mu\mu}^{-1/2}\chi_k^*$, where $\chi_k^*$ is an eigenfunction
\[T_{\mu\mu}^{-1/2} T_{\mu\tau}T_{\mu\tau}^*T_{\mu\mu}^{-1/2} \chi_k^* = \lambda_k\chi_k^*\]
at the same eigenvalue $\lambda_k$. In what follows we also write $S:=T_{\mu\mu}^{-1/2} T_{\mu\tau}T_{\mu\tau}^*T_{\mu\mu}^{-1/2} $. Without loss of generality we also assume that eigenvalues are ordered according to $\lambda_1\geq\lambda_2\geq\dots$, and $\lambda_K>0$.

We next argue that under the condition (\ref{kh_bound_condition}), $S$ is compact: it follows immediately from the Karhunen-Lo\`eve expansion (\ref{mu_tau_kl}) that we can write
\begin{eqnarray}\nonumber(T_{\mu\mu}h)(x)&=&\sum_{k=1}^{\infty}\mathbb{E}[\alpha_{g^*k}^2]\langle\xi_k,h\rangle\xi_k(x)\\
\nonumber(T_{\mu\tau}h)(x)&=&\sum_{k=1}^{\infty}\sum_{l=1}^{\infty}\mathbb{E}[\alpha_{g^*k}\beta_{g^*l}]\langle\zeta_l,h\rangle\xi_k(x),
\end{eqnarray} with the analogous expression for the adjoint $T_{\mu\tau}^*$.

Since $\xi_1,\xi_2,\dots$ is a basis of $\ker(T_{\mu\mu})^{\perp}$, we can therefore evaluate the trace of $S$ in terms of that basis to obtain
\[\tr(S)=\sum_{k=1}^{\infty}\sum_{l=1}^{\infty}\frac{\mathbb{E}[\alpha_{g^*k}\beta_{g^*l}]^2}{\mathbb{E}[\alpha_{g^*k}^2]}\]
which is finite by (\ref{kh_bound_condition}). In particular, the sequence of compact operators $S_K$ defined as the analog of $S$ after truncating the respective Karhunen-Lo\`eve expansions for $\mu_{g^*}$ and $\tau_{g^*}$ after the $K$ leading principal components converges to $S$ under the trace norm, so that $S$ is also compact. Since $S$ is self-adjoint and nonnegative and therefore has a discrete spectrum of nonnegative eigenvalues $\lambda_1\geq\lambda_2\geq\dots\geq0$ with associated eigenfunctions $\chi_1^*,\chi_2^*,\dots\in L_2(\mathcal{X})$.

It remains to be shown that the norm of $\phi_k^*:=T_{\mu\mu}^{-1/2}\chi_k^*$ is also bounded. Since $\chi_k^*$ is an eigenfunction of $S$ at eigenvalue $\lambda_k>0$, we can write
\[\phi_k^*:=\lambda_k^{-1}T_{\mu\mu}^{-1/2}S\chi_k^*.\]
Hence, the norm of $\phi_k^*$ is given by
\[\|\phi_k^*\|^2 = \lambda_k^{-2}\langle T_{\mu\mu}^{-1/2}S\chi_k^*,T_{\mu\mu}^{-1/2}S\chi_k^*\rangle=
 \lambda_k^{-2}\langle \chi_k^*,ST_{\mu\mu}^{-1}S\chi_k^*\rangle\]
noting that $S$ and $T_{\mu\mu}$ are self-adjoint. Moreover, expressing $T_{\mu\mu}$ and $T_{\mu\tau}$ in terms of the Karhunen-Lo\`eve representations (\ref{mu_tau_kl}), we can use the Cauchy-Schwarz inequality to bound
\[\tr(ST_{\mu\mu}^{-1}S) \leq \left(\sum_{k=1}^{\infty}\sum_{l=1}^{\infty}\frac{\mathbb{E}[\alpha_{g^*k}\beta_{g^*l}]^2}{\mathbb{E}[\alpha_{g^*k}^2]^{3/2}}\right)^2\]
which is finite under (\ref{kh_bound_condition}). Noting that $S^rT_{\mu\mu}^{-1}S^r$ is self-adjoint, this establishes that the operator is also trace-class and therefore compact with operator norm bounded by $\tr(ST_{\mu\mu}^{-1}S)$.

In particular, since $\lambda_1\geq\dots\lambda_K>0$ by assumption, we can bound
\[\max_{k=1,\dots,K}\|\phi_k^*\|^2\leq|\lambda_K|^{-2}tr(ST_{\mu\mu}^{-1}S)\max_{k}\|\chi_k^*\|^2 \]
where $\|\chi_k^*\|$ is finite for each $k$ since $\chi_k^*\in L_2(\mathcal{X})$.

We can then apply the argument from the proof of Theorem \ref{opt_basis_thm} where we choose $\tilde{\phi}_{k\varepsilon}(x)\equiv\phi_k^*(x)$, the solutions to (\ref{S_op_defn}) corresponding to the $k$th largest eigenvalue for $k=1,\dots,K$. Since those functions are chosen independently of $\varepsilon$, the bound $B_{\varepsilon}$ is fixed at some finite value $B_0$, so that the claim follows immediately from (\ref{check_vs_tilde_bound}), noting that $\kappa_{\varepsilon}\equiv\kappa_0$ is constant.\qed

\subsection{Proof of Corollary \ref{opt_basis_cor}} By construction, the optimal basis for the regularized problem \ref{S_op_reg} satisfies the constraint $\langle \phi_k^*,(T_{\mu\mu}+a\id)\phi_l^*\rangle = \delta_{kl}$. Hence, we can rewrite the $K\times K$ matrix $P_K^*T_{\mu\mu}P_K^*=I_K-aP_K^*$, and use the Neumann series to obtain its inverse, $(P_K^*T_{\mu\mu}P_K^*)^{-1}=I_K + \frac{a}{1-a}P_K^*$. From the proof of Theorem \ref{min_mse_updated_thm}, the projection of $\tau_{g^*}$ onto the optimal basis is therefore given by
\[\tau_{g^*K}^*(x):=\sum_{k=1}^{K}\langle \mu_{g^*},\phi_k^*\rangle\frac{1+a}{1-a}\left(T_{\mu\tau}^*\phi_k^*\right)(x)\]
which establishes the formula given in the Corollary\qed

\section{Proofs for Section \ref{sec:asymptotics}}
\label{sec:app_asymptotics_proofs}


\subsection{Proof of Theorem \ref{loc_lin_consistency_thm}}

We use the main result in \cite{GNP21} which adapts Theorems 2 and 10 in \cite{Han08} to nonparametric regression for dyadic data, rather than conventional sample averages. We first apply their results separately for each clusters $g=1,\dots,G$, where we strengthen the rate conditions to ensure uniformity across clusters. We then aggregate the cluster-specific estimates to obtain the first and second conditional moments across clusters.

\subsubsection{Convergence Rate for Kernel Averages}

The local linear estimator can be expressed in terms of weighted averages of products of $X_{gi},d_{gi},Y_{gi}$. For a general notation, let $W_{sgi}=(a_{0s}+a_{1s}X_{gi})(b_{0s} + b_{1s}Y_{gi})$ be a function of $X_{gi},Y_{gi}$ that is affine in $Y_{gi}$ given fixed coefficients $a_{0s},b_{0s},a_{1s},b_{1s}$. We consider uniform convergence of conventional and dyadic kernel averages
\begin{eqnarray}
\nonumber \hat{\Psi}_g(x_1;d_1)&:=&\frac{1}{nh^d}\sum_{i=1}^{n}W_{1gi}K\left(\frac{x_1-x_{gi}}{h}\right)\dum\{d_{gi}=d_1\}\\
\nonumber\hat{\Omega}_g(x_1,x_2;d_1,d_2)&:=&\frac{1}{2\binom{n}{2}h^{2d}}\sum_{i\neq j}W_{1gi}W_{2gj}
K\left(\frac{x_1-x_{gi}}{h}\right)K\left(\frac{x_2-x_{gj}}{h}\right)\dum\{d_{gi}=d_1,d_{gj}=d_2\}
\end{eqnarray}
for each $g=1,\dots,G$ such that $R_g=1$.

\begin{lem}\label{kern_avg_conv_lem} Suppose that the bandwidth $g$ satisfies $\left(\frac{n}{\log n}\right)^{1/3} h^d\rightarrow\infty$ and $\left(\frac{\log n}{nh^d}\right)^{1/2}$. Under Assumptions \ref{loc_lin_ass} and \ref{kernel_ass}, the kernel averages $\hat{\Psi}_g$ and $\hat{\Omega}_g$ converge uniformly to their respective expectations at the rate
\begin{eqnarray}
\nonumber \max_{g=1,\dots,G}R_g\sup_{x_1\in[0,1]^d}\left|\hat{\Psi}_g(x_1;d_1)-\mathbb{E}[\hat{\Psi}_g(x_1;d_1)]\right| & = &O_P\left(\left(\frac{\log n}{nh^d}\right)^{1/2}\right)\\
\nonumber \max_{g=1,\dots,G}R_g\sup_{x_1,x_2\in[0,1]^d}\left|\hat{\Omega}_g(x_1,x_2;d_1,d_2)-\mathbb{E}[\hat{\Omega}_g(x_1,x_2;d_1,d_2)]\right| & = &O_P\left(\left(\frac{\log n}{nh^d}\right)^{1/2}\right)\\
\end{eqnarray}
for any $d_1,d_2\in\{0,1\}$.
\end{lem}

\textsc{Proof:} We prove this result using Theorem 3.2 in \cite{GNP21}. Note first that, since $W_{gi}$ are conditionally i.i.d., Assumption \ref{loc_lin_ass} implies Assumptions 3.1 and 3.3 (a) in \cite{GNP21} for $Z_{gij}:=W_{gi}W_{gj}$. Also, Assumption \ref{kernel_ass} subsumes Assumptions 3.2 and 3.3 (b) in \cite{GNP21}. Moreover, for $s\geq3$, lengthy but elementary rate calculations confirm that any bandwidth sequence with $\left(\frac{n}{\log n}\right)^{1/3} h^d\rightarrow\infty$  satisfies the additional bandwidth conditions required for their theorem.

Their argument can then be adapted to achieve uniform convergence of $\hat{\Psi}_{gs}(x_1;d_1)$ and $\hat{\Omega}_{gs}(x_1,x_2;d_1,d_2)$ with respect to $x_1,x_2$ and $g$. To that end, the grid $\mathcal{X}^*_n\{w_{n1},\dots,w_{nL_n}\}$ is chosen in a way such that the set $[0,1]^d$ is covered by the collection of balls of radius $a_nh$. We then replace the approximating grid introduced on p.19 of \cite{GNP21} with $\mathcal{X}_n^*\times\{1,\dots,G\}$, resulting in $L_n:=G\left(h^{-1}\left(\frac{\log n}{nh^d}\right)^{1/2}\right)^d$ partition elements. We can therefore conclude that
\begin{equation}\label{Omega_g_rate}\max_{g=1,\dots,G}R_g\sup_{x_1,x_2\in[0,1]^d}\left|\hat{\Omega}_{gs}(x_1,x_2;d_1,d_2)-\mathbb{E}[\hat{\Omega}_{g1}(x_1,x_2;d_1,d_2)]\right|=O_p\left(\frac{\log n}{nh^d}\right)^{1/2}\end{equation}
for $s=1,2$. The claim regarding $\hat{\Psi}(x_1;d_1)$ is proven in an analogous manner using Theorem 2 in \cite{Han08}, whose conditions are subsumed under those for Theorem 3.2 in \cite{GNP21}\qed

\subsubsection{Proof of Theorem \ref{loc_lin_consistency_thm}}

We now complete the proof of Theorem \ref{loc_lin_consistency_thm}. We consider the general case of estimating the conditional expectation
\[\Psi_g(x;d)\equiv\Psi_g(x_1,x_2;d_1,d_2):=\mathbb{E}\left[W_{1gi},W_{2gj}|X_{gi}=x,X_{gj}=x_2,D_{gi}=d_1,D_{gj}=d_2\right]\]
and $\Lambda(x_1,x_2;d_1,d_2):=\frac1G\sum_{g=1}^G\Lambda_g(x_1,x_2;d_1,d_2)$.
for general $W_{1gi},W_{2gj}$. The corresponding cluster-specific local linear estimator
\begin{eqnarray}
\nonumber \hat{\Psi}_{g}(x_1,x_2;d_1,d_2)&:=&\arg_{b_{0}^{(g)}}\min_{b_{0}^{(g)},b_{11}^{(g)},b_{12}^{(g)}}\sum_{i=1}^{n_g}\sum_{j\neq i}\Psi_{gij}(x_1,x_2;\mathbf{b})w_{gi}(x_1;d_1)w_{gj}(x_2;d_2)
\end{eqnarray}
where we denote $\Psi_{gij}(x_1,x_2;\mathbf{b}):=\left(W_{1gi}W_{2gj}-b_0^{(g)}-b_{11}^{(g)}(X_{gi}-x_1) - b_{12}^{(g)}(X_{gj}-x_2)\right)$, and as before, $w_{gi}(x_1;d_1):=K\left(\frac{X_{gi}-x_1}h\right)\dum\{d_{gi}=d_1\}$.

As in the proof of Theorem 10 in \cite{Han08}, we can write each local linear estimator as
\[\hat{\Lambda}_g(x;d) =\frac{\hat{m}_g + \hat{S}_g'\hat{M}_g^{-1}\hat{N}_g}{\hat{f}_g-\hat{S}_g'\hat{M}_g^{-1}\hat{S}_g} \]
where for greater legibility, we write $(x,d):=(x_1,x_2;d_1,d_2)$ and suppress dependence on $d_1,d_2$ wherever possible, and terms on the right-hand side are defined as follows:
\begin{eqnarray}
\nonumber \hat{m}_g&=&\hat{m}_g(x;d):=\binom{n}2^{-1}h^{-2d}\frac12\sum_{i\neq j}W_{gi}W_{gj}w_{gi}(x_1;d_1)w_{gj}(x_2;d_2)\\
\nonumber \hat{f}_g&=&\hat{f}_g(x;d):=\binom{n}2^{-1}h^{-2d}\frac12\sum_{i\neq j}w_{gi}(x_1;d_1)w_{gj}(x_2;d_2)\\
\nonumber \hat{S}_g&=&\hat{S}_g(x;d):=\binom{n}2^{-1}h^{-2d}\frac12\sum_{i\neq j}\left(\frac{x-X_{gij}}h\right)w_{gi}(x_1;d_1)w_{gj}(x_2;d_2)\\
\nonumber \hat{M}_g&=&\hat{M}_g(x;d):=\binom{n}2^{-1}h^{-2d}\frac12\sum_{i\neq j}\left(\frac{x-X_{gij}}h\right)\left(\frac{x-X_{gij}}h\right)'w_{gi}(x_1;d_1)w_{gj}(x_2;d_2)\\
\nonumber \hat{N}_g&=&\hat{N}_g(x;d):=\binom{n}2^{-1}h^{-2d}\frac12\sum_{i\neq j}\left(\frac{x-X_{gij}}h\right)W_{gij}w_{gi}(x_1;d_1)w_{gj}(x_2;d_2)
\end{eqnarray}
Applying Lemma \ref{kern_avg_conv_lem} component by component, each of these terms converges uniformly to its expectation. Specifically, denoting $b_n:=\left(\frac{\log n}{nh^d}\right)^{1/2}$, and $\Sigma:=\int uu'K(u)du$, standard calculations for conditional moment estimation using local linear regression (see also the proofs of Theorems 8 and 10 in \cite{Han08}) then yield
\begin{eqnarray}
\nonumber \hat{m}_g(x;d)&=&m_g(x;d) + O(h^2)+ O_p(b_{n})\\
\nonumber \hat{f}_g(x;d)&=&f_g(x;d) + O(h^2) + O_p(b_{n})\\
\nonumber \hat{S}_g(x;d)&=&h\Sigma\nabla_xf_g(x;d) + O(h^2)+ O_p(b_{n})\\
\nonumber \hat{M}_g(x;d)&=&\Sigma f_g(x;d) + O(h^2)+ O_p(b_{n})\\
\nonumber \hat{N}_g(x;d)&=&h\Sigma\nabla_xm_g(x;d)+ O(h^2) + O_P(b_{n})
\end{eqnarray}
uniformly in $(x;d)$, where $m_g(x;d):=\Psi_g(x;d)f_g(x;d)$.

We can now confirm that $b_n^{1/3}h\equiv\left(\frac{n}{\log n}\right)^{1/3}h^d\rightarrow\infty$ implies that \[\frac{h}{b_n}=\left(\frac{n}{\log n}\right)h^{d+1}=\left[\left(\frac{n}{\log n}\right)^{1/3}h^{d}\right]^3 h^{1-2d}\rightarrow\infty\] for any $d\geq1$. Hence, collecting terms,
\begin{eqnarray}
\nonumber \left|\hat{S}_g(x;d)'\hat{M}_g(x;d)^{-1}\hat{N}_g(x;d)-h^2\frac{\nabla_xf_g(x;d)'\Sigma\nabla_xm_g(x;d)}{f_g(x;d)}\right|
&=&O_p\left(\left\{\frac{h^2}{f_g(x;d)}+h\right\}b_n\right)\\
\nonumber&\leq&O_p\left(h^2\delta_{Gn}^{-1}b_n + hb_n\right)=hO_p\left(b_n\right)
\end{eqnarray}
where convergence is uniform in $(x;d)$. 

Hence, from standard rate calculations
\begin{eqnarray}
\nonumber\hat{\Psi}_g(x;d)&=& \frac{\hat{m}_g + \hat{S}_g'\hat{M}_g^{-1}\hat{N}_g}{\hat{f}_g-\hat{S}_g'\hat{M}_g^{-1}\hat{S}_g}\\
\nonumber&=&\frac{\Psi_g(x;d)f_g(x;d) +O(h^2)+O_p\left(b_n+hb_n\right)}{f_g(x;d) +O(h^2) + O_P\left(b_n+hb_n\right)}\\
\nonumber&=&\Psi_g(x;d) + O_P\left(b_n\right) + O\left(h^2\right)
\end{eqnarray}
Since units $i=1,\dots,n$ are sampled independently in each location, the $O_P(\cdot)$ terms are independent across locations $g=1,\dots,G$ with expectation of order $o(h^2)$.

Hence, aggregating over $g=1,\dots,G$,
\begin{eqnarray}
\nonumber\hat{\Psi}(x_1,x_2;d_1,d_2)&=&\frac1{G-1}\sum_{g=1}^GR_g\hat{\Psi}_g(x_1,x_2;d_1,d_2)
\end{eqnarray}
Hence, by the triangle inequality
\begin{eqnarray}
\nonumber \left|\hat{\Psi}(x_1,x_2;d_1,d_2)-\Psi(x_1,x_2;d_1,d_2)\right|&
\leq&\left|\frac1{G-1}\sum_{g=1}^GR_g\hat{\Psi}_g(x_1,x_2;d_1,d_2)-\Psi_g(x_1,x_2;d_1,d_2)\right|\\
\nonumber&&+\left|\frac1{G-1}\sum_{g=1}^GR_g|\Psi_g(x_1,x_2;d_1,d_2)-\Psi(x_1,x_2;d_1,d_2)\right|\\
\nonumber&=&O_p\left(\frac{b_n+h^2}{G^{1/2}}\right) + O_P\left(\frac1G\right)
\end{eqnarray}
since $R_g$ is equal to zero for a single unit $g^*$ selected at random and one otherwise, where we use unconfoundedness of location, Assumption \ref{unconf_loc_ass} and bounded conditional moments in Assumption \ref{loc_lin_ass}. By our previous arguments, convergence is also uniform with respect to the arguments $x_1,x_2;d_1,d_2$.

We can immediately verify that $\hat{\mu}(x;d)$ and $\mu(x;d)$ correspond to $\hat{\Psi}(x;d)$ and $\Psi(x;d)$, respectively, for the case $W_{1gi}=Y_{gi}$ and $W_{2gj}=1$, so that
\[\sup_{x\in[0,1]^d}|\hat{\mu}(x;d)-\mu(x;d)|=O_p\left(\frac{b_n+h^2}{G^{1/2}\delta{Gn}^{-1}}\right) + O_P\left(\frac1G\right) \]
For the covariance kernel $H(x_1,x_2;d_1,d_2)$, we can set $W_{1gi}=Y_{gi}$ and $W_{2gj}=Y_{gj}$ so that the cluster-specific local linear estimator
\begin{eqnarray}
\nonumber \hat{\Psi}_{g}(x_1,x_2;d_1,d_2)&:=&\arg_{b_{0}^{(g)}}\min_{b_{0}^{(g)},b_{11}^{(g)},b_{12}^{(g)}}\sum_{i=1}^{n_g}\sum_{j\neq i}\left(Y_{gi}Y_{gj}-b_0^{(g)}-b_{11}^{(g)}(X_{gi}-x_1) - b_{12}^{(g)}(X_{gj}-x_2)\right)\\
\nonumber&&\times \dum\{D_{gi}=d_1,D_{gj}=d_2\}w_{gi}(x_1;d_1)w_{gj}(x_2;d_2)
\end{eqnarray}
is uniformly consistent for any $g=1,\dots,G$ with $R_g=1$ so that for
\[\hat{H}(x_1,x_2;d_1,d_2)=\frac1{G-1}\sum_{g=1}^GR_g\hat{\Psi}_{d_1d_2,g}(x_1,x_2;d_1,d_2) - \bar{\mu}(x_1;d_1)\bar{\mu}(x_2;d_2)\]
we can conclude
\[\sup_{x_1,x_2\in[0,1]^d}|\hat{H}(x_1,x_2;d_1,d_2)-H(x_1,x_2;d_1,d_2)|=O_p\left(\frac{b_n+h^2}{G^{1/2}\delta{Gn}^{-1}}\right) + O_P\left(\frac1G\right) \]
establishing the convergence rates for a general choice of the bandwidth sequence subject to the rate conditions in the theorem.

Since by standard arguments the bias is of the order $h^2$, the rate of the root mean square error is minimized at bandwidth sequences such that $h^2=\left(\frac{\log n}{Gnh^d}\right)^{1/2}$ so that such a sequence must go to zero at a rate $h^*=O\left(\frac{\log n}{Gn}\right)^{\frac1{4+d}}$.\qed

\subsection{Proof of Proposition \ref{b_spline_consistency_thm}} We give the argument for estimation of $\Eta(x_1,x_2;d_1,d_2)$, the proof for the mean function $\mu(x_1;d_1)$ follows as a special case. We let $\Eta_g^*$ denote the projection of $\Eta_g$ onto the spline basis under the empirical measure. Parallel to the case of kernel estimation in \cite{GNP19}, the estimation error in $\hat{\Eta}_g(x_1,x_2;d_1,d_2)$ can be decomposed into
\begin{eqnarray}
\nonumber \hat{\Eta}_g(x_1,x_2;d_1,d_2) &=& \Eta_g^*(x_1,x_2;d_1,d_2) + \frac2{n_g}\sum_iH_{gn_gi}^{(1)} + \binom{n_g}{2}^{-1}\sum_{i< j}H_{gn_gij}^{(2)}\\
\nonumber&=:&\Eta_g^*(x_1,x_2;d_1,d_2) + T_{gn_g}{(1)} + T_{gn_g}^{(2)}
\end{eqnarray}
where
\begin{eqnarray}
\nonumber H_{gngi}^{(1)}&=&\mathbb{E}[H_{n_gij}|Y_{gi},X_{gi},\mathbf{X}_g] - \mathbb{E}[H_{n_gij},\mathbf{X}_g]\\
\nonumber H_{gn_gij}^{(2)}&=&\mathbb{E}[H_{n_gij}|Y_{gi},Y_{gj},X_{gi},X_{gj},\mathbf{X}_g]-H_{gngi}^{(1)}-H_{gngj}^{(1)}+\mathbb{E}[H_{n_gij},\mathbf{X}_g]
\end{eqnarray}
and $H_{n_gij}\equiv H_{n_gij}(x_1,x_2;d_1,d_2)$.

The variance bound can then be derived following the arguments in the proof of the i.i.d. case for Lemma 2.3 in \cite{CCh15}: By assumption, the term $T_{gn_g}{(1)}$ directly satisfies the conditions of their lemma.  For $T_{gn_g}{(2)}$, we set $h:=1/M$ and note that by assumption $\frac{\log n}{nh^d}\rightarrow0$, so that an analogous bound for the second term follows from arguments completely analogous to the proof of claim (ii) in Lemma 3.4 of \cite{GNP21}. The triangle inequality then yields
\[\var\left(\sup_{x_1,x_2,d_1,d_2}\left|\hat{\Eta}_g(x_1,x_2;d_1,d_2)-\Eta_g^*(x_1,x_2;d_1,d_2)\right|\right)\lesssim\frac{(M+r)^d\log n}{n},\]
noting that from known facts about tensor products of polynomial spline bases (see p.450 in \cite{CCh15}), $\lambda_{Kn}\lesssim O(1)$. 

Given these bounds, aggregation of the site specific estimates is completely analogous to the case of kernel-based estimation in the proof of Theorem \ref{loc_lin_consistency_thm} and yields
\[\sup_{x_1,x_2,d_1,d_2}\left|\Eta_g^*(x_1,x_2;d_1,d_2)-\Eta_g(x_1,x_2;d_1,d_2)\right|\lesssim\left(\frac{(M+r)^d\log n}{n}\right)^{1/2} +  M^{-2} +G^{-1} \]
which establishes the claim\qed

It remains to prove Theorem \ref{eig_cons_thm}, where for the remaining arguments we let $\|\cdot\|_F$ denote the trace operator norm $\|T\|_F:=\tr(T^*T)$. We first establish the following Lemma:

\begin{lem}\label{eig_cons_lem} Suppose that $S$ is a compact, self-adjoint operator with eigenvalues $\lambda_1\geq \lambda_2,\dots,$ counted by their multiplicity, and corresponding eigenfunctions $\phi_1,\phi_2,\dots$. Then for any sequence $\hat{S}$ of compact, self-adjoint operators with eigenvalues $\hat{\lambda}_1\geq\hat{\lambda}_2,\dots$ such that $\|\hat{S}-S\|_F=O_P(r_{Gn})$, we have
\[|\hat{\lambda}_k-\lambda_k| = O_p\left(r_{Gn}\right)\]
for each fixed $k$. Furthermore, if the eigenvalue $\lambda_k$ is simple,
\[\|\hat{\phi}_k-\phi_k\|=O_p\left(r_{Gn}\right)\]
If on the other hand the multiplicity of some $\lambda_k>0$ is $m+1$, i.e. $\lambda_{k-1}>\lambda_{k}=\dots=\lambda_{k+m}>\lambda_{k+m+1}$, then there exists an orthonormal basis $\phi_{k}^*,\dots,\phi_{k+m}^*$ for $\textnormal{span}\{\phi_{k},\dots,\phi_{k+m}\}$ such that
\begin{eqnarray}
\|\hat{\phi}_k-\phi_k^*\|=O_p\left(r_{Gn}\right)
\end{eqnarray}
for each fixed $k$.
\end{lem}

It is important to note that the error in $\hat{\phi}_k$ depends inversely on the distance between $\lambda_k$ and its adjacent eigenvalues, where even in the absence of multiplicities, the eigenvalues of a compact operator cannot be well-separated. The convergence rate for $\hat{\phi}_k$ is therefore component-wise for the eigenvectors at each distinct eigenvalue $\lambda_k$, but not uniform over all $k=1,2,\dots$.

\textsc{Proof:} We follow closely the proofs for Proposition 4.2 in \cite{GHR04} and Theorem 2 in \cite{YMW05}. Since $S$ is compact, existence of eigenvalues $\lambda_1\geq\lambda_2\geq\dots$ and eigenfunctions $\phi_1,\phi_2,\dots$ follows from Mercer's Theorem (see \cite{HGr18} for a multivariate generalization where $\mathcal{X}$ may be of dimension greater than one).

We define the resolvent maps of the operators $S$ and $\hat{S}$,
\[R(z):=\left(S - z\id\right)^{-1},\hspace{0.8cm}\textnormal{and }\hat{R}(z):=\left(\hat{S} - z\id\right)^{-1}\]
Defining the resolvent sets $\varrho(S)$ and $\varrho(\hat{S})$ via $\varrho(T):=\left\{z\in\mathbb{C}:T-z\id \textnormal{ is invertible}\right\}$, we have for $z\in\varrho(S)\cap\varrho(\hat{S})$,
\begin{eqnarray}
\nonumber \hat{R}(z)&=&\left(S-z\id + (\hat{S}-S)\right)^{-1}\\
\nonumber&=&\left(R(z)^{-1} + (\hat{S}-S)R(z) R(z)^{-1}\right)^{-1}\\
\nonumber&=&R(z)\left(\id + (\hat{S}-S)R(z)\right)^{-1}\\
\nonumber&=&R(z) + R(z)\sum_{s=1}^{\infty}\left((S-\hat{S})R(z)\right)^s
\end{eqnarray}
where the last equality uses a Neumann representation of the inverse. Therefore, if $\|\hat{S}-S\|_F\|R(z)\|_F<1$, we can use the triangle inequality for the (trace) operator norm to bound
\begin{equation}\label{resolvent_bound}\|\hat{R(z)}-R(z)\|_F\leq
\sum_{s=1}^{\infty}\left\|R(z)\left((S-\hat{S})R(z)\right)^s\right\|\leq\frac{\|\hat{S}-S\|_F\|R(z)\|_F^2}{1-\|\hat{S}-S\|_F
\|R(z)\|_F}\end{equation}

Now, consider the $k$th eigenvalue $\lambda_k$. Since the operator $S$ is self-adjoint and compact, its spectrum is real-valued and separated. In particular any nonzero eigenvalue $\lambda_k$ has only finite multiplicity $m_k<\infty$, and there exists $\varrho_k>0$ such that the $\varrho_k$-ball around $\lambda_k$ in the complex plane $\mathbb{C}$ does not contain any other eigenvalue different from $\lambda_k$.

We then let $\Gamma_k:[0,2\pi]\rightarrow\mathbb{C}$ be the positively oriented Jordan curve
\[\Gamma_k(t):= \lambda_k + \varrho_k/2e^{it}\]
around $\lambda_k$ with radius $\varrho_k/2$. By the Cauchy integral formula and Hilbert's resolvent equations (equations (2.4) and (2.5) in \cite{Cha83}) it can be verified that the operator
\[P_k:=-\frac{1}{2i\pi}\int_{\Gamma_k}R(z)dz\]
is the orthogonal projector onto the eigenspace of $S$ at the eigenvalue $\lambda_k$ (Theorem 2.27 in \cite{Cha83}). We can similarly define
\[\hat{P}_k:=-\frac{1}{2i\pi}\int_{\Gamma_k}\hat{R}(z)dz\]
Since the nonzero eigenvalues of $\hat{S}$ are also separated and of finite multiplicities, we can assume without loss of generality that the curve $\Gamma_k$ encloses a finite number of eigenvalues of $\hat{S}$, and that no eigenvalues of $\hat{S}$ lie on the curve (otherwise we can replace the radius $\varrho_k/2$ with any other number in the interval $(\varrho_k/4,\varrho_k)$). By Cauchy's integral formula, $\hat{P}_k$ is the sum of the orthogonal projectors onto the eigenspaces of $\hat{S}$ associated with the eigenvalues of $\hat{S}$ enclosed by $\Gamma_k$. In particular, $\hat{P}_k$ is an orthogonal projector into a linear subspace of finite dimension.

Next we define
\[M_k:=\sup\left\{\|R(z)\|:z\in\Gamma_k\right\}<\infty\]
and assume that $\varepsilon:=\|\hat{S}-S\|_F\leq\frac1{2M_k}$, so that in particular the bound (\ref{resolvent_bound}) holds, and we can use (\ref{resolvent_bound}) to bound
\begin{eqnarray}
\nonumber\|\hat{P}_k-P_k\|_F&\leq&\frac1{2\pi}\int_{\Gamma_k}\|\hat{R}(z)-R(z)\|_Fdz\\
\nonumber&\leq&\frac{\varrho_k}2\frac{\|\hat{S}-S\|_FM_k^2}{1-\|\hat{S}-S\|_FM_k}\\
\nonumber&\leq&\varrho_kM_k^2\varepsilon =:B_k\varepsilon
\end{eqnarray}

Since $\|\hat{P}_k-P_k\|_F\leq B_k<\|P_k\|_F$ for $G$ sufficiently large, $\hat{P}_k\neq 0$. In particular, the intersection of the $\varrho_k/2$ ball around $\lambda_k$ with the spectrum of $\hat{S}$ is nonempty.


Now, let $\phi_k$ be an eigenvector of $S$ associated with the eigenvalue $\lambda_k$, and let $\tilde{\phi}_k:=\hat{P}_k\phi_k$, so that
\begin{eqnarray}\nonumber \|\hat{P}_k-P_k\|^2&\geq&\|(\hat{P}_k-P_k)\phi_k\|^2=\|\hat{P}_k\phi_k - \phi_k\|^2\\
\nonumber&=& 1 - 2\langle\phi_k,\hat{P}_k\phi_k\rangle + \left\|\hat{P}_k\phi_k\right\|^2\\
\nonumber&=&1-(\langle\tilde{\phi}_k,\phi_k\rangle)^2
\end{eqnarray}
Furthermore,
\begin{eqnarray}
\nonumber \|\tilde{\phi}_k-\phi_k\|^2&\leq&2-2\langle\tilde{\phi}_k,\phi_k\rangle\leq 2(1-\langle\tilde{\phi}_k,\phi_k\rangle)(1+\langle\tilde{\phi}_k,\phi_k\rangle)\\
\nonumber&=&2\left(1-(\langle\tilde{\phi}_k,\phi_k\rangle)^2\right)\\
\label{lemma_c2_phi_bound}&\leq&2\|\hat{P}_k-P_k\|^2\leq 2B_k\varepsilon^2
\end{eqnarray}
Since $\|\phi_k\|=1$ and $\varepsilon$ can be made arbitrarily small by choosing $G$ large enough, it must in particular be true that $|1-\|\tilde{\phi}_k\||\leq \varepsilon$.  We can now choose $\hat{\phi}_k:=\|\tilde{\phi}_k\|^{-1}\tilde{\phi}_k$ so that $\|\hat{\phi}_k-\phi_k\|^2\leq 2(B_k+2)\varepsilon$ for any $\varepsilon<\frac12$. Furthermore by construction, $\hat{\phi}_k$ is an eigenvector of $\hat{S}$ at some eigenvalue $\hat{\lambda}_k\in\left[\lambda_k-\varrho_k/2,\lambda_k + \varrho_k/2\right]$. In particular $\hat{\lambda}_k$ is bounded away by a distance $\varrho_k/2$ from all eigenvalues of $S$ that are different from $\lambda_k$.

Reversing the roles of $S$ and $\hat{S}$, we can similarly find an eigenvector of $S$ at eigenvalue $\lambda_k$ for any vector $\hat{\phi}_k$ in the eigenspace of $\hat{S}$ at an eigenvalue $\hat{\lambda}_k$ with $|\hat{\lambda}_k-\lambda_k|\leq \varrho_k/2$, such that $\|\phi_k-\hat{\phi}_k\|$ satisfies the same bound.

Hence, for the case of and $(m+1)$-fold multiplicity, the eigenspace of $S$ corresponding to the eigenvalue $\lambda_k$ is approximated up to the error in (\ref{lemma_c2_phi_bound}) by the eigenspace of $\hat{S}$ corresponding to the eigenvalues $\hat{\lambda}_k,\dots,\hat{\lambda}_{k+m}$. Hence we can choose the basis $\phi_k^*,\dots,\phi_{k+m}^*$ of that eigenspace of $S$ by setting $\phi_k^*:=\|P_k\hat{\phi}_k\|^{-1}P_k\hat{\phi}_k$, and then sequentially orthonormalizing $P_k\hat{\phi}_{k+1},\dots,P_k\hat{\phi}_{k+m}$. The bound  $\|\tilde{\phi}_k-\phi_k\|^2\leq B_k\varepsilon^2$ can then be established using the same reasoning as following (\ref{lemma_c2_phi_bound}). Finally, the convergence rate for the eigenvalues follows from Slutsky's Lemma applied to the formula characterizing the $k$th eigenvalue, $\lambda_k=\langle\phi_k^*,S\phi_k^*\rangle$\qed

We now complete the proof of Theorem \ref{eig_cons_thm}:

\subsection{Proof of Theorem \ref{eig_cons_thm}} Since $\hat{\phi}_k$ are defined by the generalized eigenvalue problem
\[\hat{T}_{\mu\tau}\hat{T}_{\mu\tau}^*\hat{\phi}_k = \hat{\lambda}_k\hat{T}_{\mu\mu a}\hat{\phi}_k   \]
and $\hat{T}_{\mu\mu a}$ is injective, we can equivalently rewrite
\[\hat{\phi}_k:=\hat{T}_{\mu\mu a}^{-1/2}\hat{\chi}_k\]
where $\hat{\chi}_k$ solves the eigenvalue problem
\[\hat{S}\hat{\chi}_k = \hat{\lambda}_k\hat{\chi}_k\]
and $\hat{S}:=\hat{T}_{\mu\mu a}^{-1/2}\hat{T}_{\mu\tau}\hat{T}_{\mu\tau}^*\hat{T}_{\mu\mu a}^{-1/2}$. We therefore first derive the convergence rate for $\hat{\chi}_k$ with respect to $\chi_k$, the eigenfunction associated with the $k$th larges eigenvalue $\lambda_k$ of $S:=T_{\mu\mu a}^{-1/2}T_{\mu\tau}T_{\mu\tau}^*T_{\mu\mu a}^{-1/2}$.

Since by assumption of the theorem, the covariance functions $H_{\mu\mu}$ and $H_{\mu\tau}$ are estimated uniformly consistently at the rate $r_{Gn}$, it follows immediately that the corresponding Hilbert-Schmidt operators converge at the same rate under the trace (operator) norm, $\|\hat{T}_{\mu\mu}-T_{\mu\mu}\|_F=O_p(r_{Gn})$ and $\|\hat{T}_{\mu\tau}-T_{\mu\tau}\|_F=O_p(r_{Gn})$, follows immediately.

We furthermore confirm that the operator $S_0:=T_{\mu\mu}^{-1/2}T_{\mu\tau}T_{\mu\tau}^*T_{\mu\mu}^{-1/2}$ is compact. From the proof of Lemma \ref{min_mse_updated_thm}, $\min_{B}\|\tau_g - B\mu_g\|^2=\tr(T_{\mu\tau}^*T_{\mu\mu}^{-1}T_{\mu\tau})$ where the operator $T_{\mu\mu}$ is understood to be restricted to $\ker(T_{\mu\mu})^{\perp}$. Therefore,
\begin{eqnarray}\nonumber \tr(T_{\mu\mu}^{-1/2}T_{\mu\tau}T_{\mu\tau}^*T_{\mu\mu}^{-1/2})&=&\tr(T_{\mu\tau}^*T_{\mu\mu}^{-1}T_{\mu\tau})\\
\nonumber&=&\min_{B}\|\tau_g - B\mu_g\|^2\leq\|\tau_g\|^2<\infty
\end{eqnarray}
Since $S_0$ is self-adjoint, it follows that it is also trace class and therefore compact. Since $T_{\mu\mu a}^{-1/2}T_{\mu\mu}^{1/2}$ and its transpose are compact for any $a\geq0$, it also follows that $S\equiv S_a$ is compact for any $a\geq0$ as well. $\hat{S}$ can be shown to be compact by the same argument applied to sample analogs.

Next, define $A_a:=(T_{\mu\mu}+a\id)^{1/2}\equiv (T_{\mu\mu}^{1/2}+ a^{1/2}\id)$ and  $\hat{A}_a:=(\hat{T}_{\mu\mu}^{1/2}+ a^{1/2}\id)$. We can then check that
\[\hat{A}_a^{-1} - A_a^{-1} = \hat{A}_a^{-1}(\hat{A}_a - A_a)A_a^{-1}=\hat{A}_a^{-1}(\hat{T}_{\mu\mu}^{1/2}-T_{\mu\mu}^{1/2})A_a^{-1/2}\]
The difference between $\hat{S}$ and $S$ can be written as
\[\hat{S}-S = \hat{T}_{\mu\mu a}^{-1/2}\hat{T}_{\mu\tau}\hat{T}_{\mu\tau}^*\hat{T}_{\mu\mu a}^{-1/2}  -T_{\mu\mu a}^{-1/2}T_{\mu\tau}T_{\mu\tau}^*T_{\mu\mu a}^{-1/2} = R_1+R_2+R_3\]
where
\begin{eqnarray}
\nonumber R_1&=& \hat{T}_{\mu\mu a}^{-1/2}\hat{T}_{\mu\tau}\hat{T}_{\mu\tau^*}(\hat{T}_{\mu\mu a}^{-1/2}-T_{\mu\mu a}^{-1/2}) \\
\nonumber&=&\hat{T}_{\mu\mu a}^{-1/2}\hat{T}_{\mu\tau}\hat{T}_{\mu\tau^*}\hat{T}_{\mu\mu a}^{-1/2}(\hat{T}_{\mu\mu a}^{1/2}-T_{\mu\mu a}^{1/2})T_{\mu\mu a}^{-1/2}\\
\nonumber&=&\hat{S}(\hat{T}_{\mu\mu a}^{1/2}-T_{\mu\mu a}^{1/2})\hat{T}_{\mu\mu a}^{-1/2}\\
\nonumber R_2&=&  T_{\mu\mu a}^{-1/2}(\hat{T}_{\mu\tau}\hat{T}_{\mu\tau}^* - T_{\mu\tau}T_{\mu\tau}^*\hat{T}_{\mu\mu a}^{-1/2}\\
\nonumber R_3&=&  (\hat{T}_{\mu\mu a}^{-1/2}-T_{\mu\mu a}^{-1/2})T_{\mu\tau}T_{\mu\tau^*}T_{\mu\mu a}^{-1/2}\\
\nonumber&=& \hat{T}_{\mu\mu a}^{-1/2}(\hat{T}_{\mu\mu a}^{1/2}-T_{\mu\mu a}^{1/2})S
\end{eqnarray}
Since $\hat{T}_{\mu\mu}$ and $T_{\mu\mu}$ are nonnegative, the eigenvalues of $\hat{T}_{\mu\mu a}^{1/2}$ and $T_{\mu\mu a}^{1/2}$ are bounded from below by $a^{-1/2}$. It therefore follows that $\|R_1\|=O(a^{-1/2}r_{Gn})$, $\|R_2\|=O(a^{-1}r_{Gn})$, and $\|R_3\|=O(a^{-1/2}r_{Gn})$ under the trace norm. Hence together with the triangle inequality, Lemma \ref{eig_cons_lem} implies that $\hat{\chi}_k^*-\chi_k$ converges at a rate $O_P(a^{-1}r_{Gn})$. The conclusion of the Theorem then follows from the observation that the largest eigenvalue of $\hat{T}_{\mu\mu a}^{-1/2}$ is bounded by $a^{-1/2}$ \qed

\bibliographystyle{econometrica}
\bibliography{mybibnew}

\end{document}